\documentclass[11pt,a4paper]{article}
\usepackage{amsmath, amsfonts,amssymb,mathtools,nicefrac,nccmath,cases}
\usepackage{microtype,booktabs,multirow}
\usepackage[usenames,dvipsnames,table]{xcolor}
\usepackage{colortbl}
\usepackage{tikz}
\usetikzlibrary{shapes,arrows}
\usetikzlibrary{calc,shapes.callouts,shapes.arrows,bending,intersections}
\usetikzlibrary{shapes.geometric}
\usetikzlibrary{arrows.meta,arrows}
\usetikzlibrary{decorations.pathmorphing}
\usetikzlibrary{decorations.markings}
\usetikzlibrary{shapes.multipart}
\usetikzlibrary{tikzmark}
\usepackage{arydshln}
\usepackage{caption}
\usepackage{subcaption}
\usepackage[most]{tcolorbox}

\usepackage[nosort]{cite}
\usepackage{colortbl}

\usepackage{xcolor}
\usepackage{lipsum}

\usepackage[linktocpage=true]{hyperref}
\hypersetup{colorlinks=true,linkcolor=nicecolor,citecolor=nicecolor,urlcolor=nicecolor}
\definecolor{nicecolor}{rgb}{0.1, 0.3, 0.4}

\definecolor{blue}{rgb}{0.06, 0.3, 0.57}
\definecolor{Gray}{gray}{0.4}

\definecolor{nicecolor}{rgb}{0.1, 0.3, 0.4}
\definecolor{blue}{rgb}{0.06, 0.3, 0.57}
\definecolor{Gray}{gray}{0.4}
\colorlet{tableheadcolor}{gray!15} 
\colorlet{tablerowcolor}{gray!7} 
\newcommand{\cmmnt}[1]{}

\def\hybrid{\topmargin -20pt    \oddsidemargin 0pt
	\headheight 0pt \headsep 0pt
	\textwidth 6.5in        
	\textheight 9in         
	\textwidth 6.25in       
	\textheight 9 in       
	\marginparwidth .875in
	\parskip 5pt plus 1pt 
	\jot = 1.5ex
}
\hybrid
\numberwithin{equation}{section}
\numberwithin{table}{section}
\usepackage{float}
\usepackage{tabularx}
\newcolumntype{D}{>{\centering\arraybackslash}X}
\newcolumntype{L}{>{$}l<{$}}
\newcolumntype{R}{>{$}r<{$}}
\newcolumntype{C}{>{$}c<{$}}
\usepackage{IEEEtrantools}
\usepackage{makecell}

\newcommand{\beq}{\begin{equation}}  \newcommand{\eeq}{\end{equation}}
\newcommand{\bal}{\begin{aligned}}   \newcommand{\eal}{\end{aligned}}
\newcommand{\bea}{\begin{eqnarray}}  \newcommand{\eea}{\end{eqnarray}}
\def\beqa{\begin{eqnarray}}
\def\eeqa{\end{eqnarray}}

\newcommand{\bmat}{\left(\begin{array}}
\newcommand{\emat}{\end{array}\right)}

\newcommand{\cE}{\mathcal{E}}

\newcommand{\cK}{\mathcal{K}}
\newcommand{\cN}{\mathcal{N}}
\newcommand{\cW}{\mathcal{W}}

\usepackage{cancel}

\newcommand{\be}{\begin{equation}}
\newcommand{\ee}{\end{equation}}

\definecolor{Gray}{gray}{0.95}

\begin{document}

\baselineskip=14pt
\parskip 5pt plus 1pt

\vspace*{-1.5cm}
\begin{flushright}    
  {\small 
  
  }
\end{flushright}

\vspace{2cm}
\begin{center}        

   
    {\huge Merging the Weak Gravity and Distance\\
   [.3cm]   Conjectures Using BPS Extremal Black Holes }

\end{center}

\vspace{0.5cm}
\begin{center}        
{\large  Naomi Gendler$^{\dagger}$\footnote{Electronic address: ng434@cornell.edu} and Irene Valenzuela$^{\dagger \dagger}$\footnote{Electronic address: ivalenzuela@g.harvard.edu}}
\end{center}

\vspace{0.15cm}
\begin{center}        
\emph{$^\dagger$ Department of Physics, Cornell University \\
Ithaca, NY 14853, USA}
 \\[.3cm]
  \emph{$^{\dagger \dagger}$Jefferson Physical Laboratory, Harvard University, \\
  Cambridge, MA 02138, USA}
             \\[0.15cm]

\end{center}

\vspace{2cm}


\begin{abstract}
\noindent
We analyze the charge-to-mass structure of BPS states in general infinite-distance limits of $\mathcal{N}=2$ compactifications of Type IIB string theory on Calabi-Yau three-folds, and use the results to sharpen the formulation of the Swampland Conjectures in the presence of multiple gauge and scalar fields. We show that the BPS bound coincides with the black hole extremality bound in these infinite distance limits, and that the charge-to-mass vectors of the BPS states lie on degenerate ellipsoids with only two non-degenerate directions, regardless of the number of moduli or gauge fields. We provide the numerical value of the principal radii of the ellipsoid in terms of the classification of the singularity that is being approached. We use these findings to inform the Swampland Distance Conjecture, which states that a tower of states becomes exponentially light along geodesic trajectories towards infinite field distance. We place general bounds on the mass decay rate $\lambda$ of this tower in terms of the black hole extremality bound, which in our setup implies $\lambda \geq 1/\sqrt{6}$. We expect this framework to persist beyond $\mathcal{N}=2$ as long as a gauge coupling becomes small in the infinite field distance limit.

\end{abstract}

\thispagestyle{empty}
\clearpage

\setcounter{page}{1}


\newpage

  \tableofcontents

\newpage

\section{Introduction}
\label{sec:intro}

The goal of the Swampland program (see \cite{Brennan:2017rbf,Palti_2019} for reviews) is to provide a set of criteria that an effective field theory must satisfy in order to be a viable low-energy limit of a UV-complete theory of quantum gravity. Of all the criteria that have been proposed for this purpose, two of the most studied are the Swampland Distance Conjecture (SDC) \cite{Ooguri_2007} and the Weak Gravity Conjecture (WGC) \cite{Arkani_Hamed_2007}. On the one hand, the Swampland Distance Conjecture states that as an infinite distance point in field space is approached, an infinite tower of states must enter below the original cutoff of the effective theory and become light exponentially in the traversed geodesic field distance. The Weak Gravity Conjecture, on the other hand, says that for a theory containing at least one $U(1)$ gauge field to be compatible with a quantum gravity UV-completion, it must include a particle whose charge-to-mass ratio equals or exceeds the extremality bound for black hole solutions of that theory. Strong versions of the WGC \cite{Heidenreich_2016,Montero:2016tif,Heidenreich:2016aqi, Andriolo_2018} imply not only one, but infinitely many states forming a tower or a sublattice satisfying the WGC bound. When the gauge coupling goes to zero, all these states become light since the mass has to decrease at a rate at least as fast as the charge. 
The absence of free parameters in quantum gravity \cite{Ooguri_2007} implies that gauge couplings are parametrized by the vacuum expectation value of scalar fields, which necessitates that the point of vanishing gauge coupling is at infinite distance in field space. This is necessary to avoid restoring a global symmetry at finite distance, which should not occur in a consistent theory of quantum gravity \cite{Banks:1988yz,Banks_2011,Harlow2018,Harlow:2018tng}.


It is tantalizing that both the WGC and the SDC predict the existence of light states at weak coupling points. This suggests that they might be two faces of the same underlying quantum gravity criterion, as observed in \cite{Palti_2017,Grimm_2018,Lee:2018urn,Lee_2019}. It is the aim of this paper to continue this line of thought by making the connection as precise as possible in the context of Calabi-Yau compactifications of Type IIB string theory. Whether there exists a gauge coupling that goes to zero for every infinite field distance point requires further study, but we will argue in favor of this possibility as long as the gauge coupling can correspond to a $p$-form gauge field (as opposed to specifically a 1-form). 

If the same tower of states satisfies both the WGC and the SDC, it is possible to formulate a single, precise statement that is satisfied in the limit of weak coupling. The unification of these conjectures beautifully resolves the previously ambiguous aspects of each statement. On the one hand, the main open question regarding the SDC concerns the rate of decay of the characteristic mass scale of the tower; this rate appears as an unspecified order one parameter in the conjecture. Fixing this order one number for a given theory is essential in determining the precise phenomenological implications of the conjecture in the context of cosmology or particle physics. However, if this same tower of states also satisfies the WGC, then we can bound this factor in terms of the extremality bound for black holes in that theory. The WGC, on the other hand, suffers less ambiguities in the form of unspecified order one parameters: when it concerns particles and large black holes, all numerical factors are in principle specified. However, it does not state how many and which states should be light at small gauge coupling. Seen from this angle, the SDC suggests that infinitely many WGC-satisfying states become light, motivating the identification of the SDC as a Tower WGC in the weak coupling limits. It should be stressed that at no point in this work do we \textit{use} any of the Swampland conjectures to make statements. Rather, we are focused on analyzing the effective theories resulting from Calabi-Yau compactifications to \textit{inform} the conjectures and eliminate $\mathcal{O}(1)$ factors. 

Whether the exponential factor of the SDC can be fully determined by the extremality bound for black holes in the asymptotic limits depends on whether the WGC in the presence of scalar fields is equivalent to a repulsive force condition (i.e. the statement that there must exist a particle for which the Coulomb force is stronger than the attractive gravitational force plus scalar interactions). The latter condition was proposed in \cite{Palti_2017} as the proper generalization of the WGC in the presence of massless scalar fields, denoted in \cite{Lee_2019} as part of the Scalar WGC, and further relabeled in \cite{Heidenreich_2019} as the Repuslive Force Conjecture (RFC). We will keep the latter name in this paper to avoid confusion with the actual Scalar WGC \cite{Palti_2017} which simply requires that the scalar force acts stronger than the gravitational force on the particle. Whenever the extremality bound and the RFC coincide, the exponential factor of the SDC can be determined in terms of the scalar contribution to the extremality bound, which for the case of a single scalar field can be directly inferred from the gauge kinetic function as studied in \cite{Lee_2019}. That the extremality bound and force-cancellation condition coincide was proposed \cite{Lee_2019} to occur in the limit of weak gauge coupling. We will extend this to asymptotic limits in higher dimensional moduli spaces in which an arbitrary number of scalar fields are taken to the large field limit, and show that the exponential factor of the SDC can be bounded from above and below  by the scalar dependence of the gauge kinetic function, which is related to the extremality bound.

In order to make all these relations precise and check them explicitly in controlled compactifications of string theory, we are going to restrict ourselves to  $\cN=2$ supersymmetric theories that arise from compactifying string theory on a Calabi-Yau threefold, following the work initiated in \cite{Grimm_2018,Grimm_2019,Corvilain_2019,grimm2019infinite} (see \cite{Lee:2018urn,Lee_2019,Joshi:2019nzi,Marchesano_2019,Lee:2019tst,Lee:2019xtm,lee2019emergent,Baume:2019sry,Cecotti:2020rjq} for other works identifying the towers of states becoming light at the asymptotic limits of Calabi-Yau compactifications). 
 Once the conjectures are well understood and proven in supersymmetric setups with $\cN\geq 2$, less supersymmetric configurations, which are more useful for phenomenology, stand to be explored (see \cite{Baume_2016,Valenzuela:2016yny,Blumenhagen:2017cxt,Gonzalo:2018guu,Font:2019cxq, Bonnefoy:2020fwt,Bonnefoy_2019,Lust:2019zwm, enriquezrojo2020Swampland} for attempts in this direction). However, it is important to note that to talk about the SDC, one needs to be able to move in field space towards an infinite distance singularity. This is one reason why $\cN=2$ theories are interesting: they provide a non-trivial setup with a moduli space. One might expect the conjecture to still hold in field spaces with a scalar potential (regarded as one of the implications of the Refined SDC \cite{Klaewer_2017}) as long as there is a mass hierarchy privileging particular trajectories such that the conjecture can be applied to the bottom of the scalar potential. In that case, though, the realization of the conjecture gets conflated with which types of potentials are allowed in string theory, implying constraints on the latter (see \cite{grimm2019asymptotic} for an analysis initiating the classification of flux potentials at the asymptotic limits).


 In \cite{Grimm_2018} (see also \cite{Grimm_2019,Corvilain_2019,grimm2019infinite}), a tower of states in these $\mathcal{N}=2$ theories with the correct properties to satisfy the SDC was identified as a monodromy orbit of BPS states becoming light at the infinite field distance limits of the moduli space of vector multiplets. From a microscopic point of view, BPS states correspond to D3-branes wrapped on special Lagrangian 3-cycles that arise as particles in the 4d theory and can become massless at the singularities of the complex structure moduli space of Type IIB on a Calabi-Yau threefold. These states are charged under the 4d $U(1)$ gauge fields that arise from dimensional reduction of the Ramond-Ramond 4-form $C_4$, and are thus perfect candidates to also satisfy the WGC. We will explicitly calculate the charge-to-mass ratio of these BPS states in the 4d theory and compare with the black hole extremality bound specific to the theory in the infinite distance limit. In doing so, we will show that indeed the same set of BPS states satisfy the WGC, the RFC, and the SDC, and we will be able to place a bound on the mass decay rate of the SDC  in terms of the type of infinite distance singularity that is being approached. These precise numerical bounds are valid for multi-moduli large field limits of any Calabi-Yau and will be computed using the mathematical machinery of asymptotic Hodge theory, which allows us to determine the leading dependence of the gauge kinetic function on the scalar fields as a large field limit is approached. Notice that the generalization from one field to multiple fields is in general highly non-trivial, and this is where loopholes to the WGC usually arise. It is only thanks to the powerful theorems of asymptotic Hodge theory that we can overcome path dependent issues and give universal bounds that cannot be tricked by any type of alignment mechanism. We would also like to remark that asymptotic Hodge theory does not only provide useful tools, but we believe it in important piece in the quest of  abstractly identifying the mathematical structure distinguishing the swampland and the landscape. As postulated in \cite{Grimm_2018,Grimm_2019} and recently nicely formulated in \cite{Cecotti:2020rjq}, the difference between generic 4d $N=2$ supergravity effective theories (special K\"ahler geometries) and $N=2$ theories which can be completed to a consistent theory of quantum gravity, lies precisely in the structures provided by Hodge theory.

 It should also be noted that regardless of the implications of the Swampland conjectures for the space of possible low-energy effective theories, they can often teach us where to look for interesting structures in string theory as a whole, as already happened in \cite{Grimm_2018,Grimm_2019,Lee:2018urn,Lee:2019tst}. We will see another example of this here as we analyze the charge-to-mass ratios of states in our theory. The charge-to-mass ratio of BPS states turns out to form a degenerate ellipse with only two non-zero principal radii, regardless of the number of scalar fields. These principal radii can be determined in terms of the scalar dependence of the gauge kinetic function when written in a basis adapted to the asymptotic splitting of the charge lattice, which is guaranteed by Hodge theory. This enormously simplifies the task of determining the extremality bound in these theories, as all BPS states with electric charges become extremal in the asymptotic limits of the moduli space. Furthermore, in certain examples we study how the degenerate directions are truncated by the BPS charge restrictions, and the ellipsoid is capped off by the extremality surface associated to non-BPS black holes. It seems that at least in these examples, the set of BPS states is enough to satisfy the WGC convex hull condition \cite{Cheung_2014}, even taking into account directions in field space which do not support BPS charges.

The overall goal of this work is two-fold: first, to understand the structure of BPS states in the asymptotic regimes of moduli space, and second, to use this knowledge to put a bound on the mass decay rate of the states that become light according to the SDC. The outline of the paper is as follows. In section~\ref{swampyscalars}, we will review how the notions of extremality and force-cancellation change in the presence of massless scalars as well as review all the Swampland conjectures that we aim to test in this work. In section~\ref{bps}, we will compute the charge-to-mass spectrum of BPS states in a general setting, as well as give concrete examples of this structure in compactifications with one and two moduli. In section~\ref{asymptotic}, we will describe how to compute the gauge kinetic function in terms of the type of infinite distance singularity and thus read off the extremality bounds of electric black holes in general for any asymptotic limit. With this information, we will be able to formulate general bounds on the mass decay rate of the Distance Conjecture. In section~\ref{generalizations}, we provide some insight into how such a framework may manifest in more general settings; for example, considering non-BPS extremal solutions, and theories with $\cN<2$. Finally, we conclude in section~\ref{conclusions}.\\

\section{Swampland conjectures in the presence of scalar fields} \label{swampyscalars}

In this paper, we will test and sharpen the WGC and SDC by studying the properties of BPS states and the interrelations that appear among them. In particular, we will see how the WGC gets modified in the presence of scalar fields, yielding two possible generalizations in terms of the black hole extremality bound or a repulsive force condition. Both notions will coincide if the particles in question correspond to the asymptotic tower of states predicted by the SDC, which further allows us to determine the exponential mass decay rate in terms of the extremality bound.

Consider a quantum field theory with abelian gauge fields and massless scalar fields weakly coupled to Einstein gravity, with action given by
\beq
\label{action1}
S=M_p^{D-2}\int d^D x\sqrt{-h}\left(\frac{R}{2}-\frac12g_{ij}\partial_\mu\phi^i\partial^\mu\phi^j -\frac{1}{2}f_{IJ}(\phi)F^I_{\mu\nu}F^{J,\mu\nu}\right),
\eeq
where $g_{ij}$ is the field space metric and $f_{IJ}(\phi)$ is the gauge kinetic function, which can depend on the scalar fields.


The WGC  \cite{Arkani_Hamed_2007} states that in a quantum field theory with a weakly coupled gauge field and Einstein gravity, there must exist a particle whose charge-to-mass ratio is larger than the one associated to an extremal black hole in that theory, i.e. the theory must contain a superextremal particle. In the absence of massless scalar fields, this implies that its charge must be bigger than its mass in Planck units, so the WGC can also be formulated as the statement that gravity acts weaker than the gauge force over this particle. Though these statements are equivalent when there are only gauge forces and gravity, a crucial insight is to realize that they can differ in the presence of massless scalar fields.

Massless scalars alter the extremality bound of black hole solutions in a theory as well as the no-force condition for particles. Therefore, there are two apparently different generalizations of the weak gravity bound in the presence of scalars: 
\begin{itemize}
\item \textbf{Weak Gravity Conjecture (WGC):} There must exist a superextremal particle satisfying
\begin{align}
\frac{\mathcal{Q}}{M} \geq \left(\frac{\mathcal{Q}}{M}\right) \bigg|_{\text{extremal}},
\label{WGC}
\end{align}
where the extremality bound $\left(\frac{\mathcal{Q}}{M}\right) \bigg|_{\text{extremal}}$\cmmnt{factor $\gamma(\phi)$} can depend on the scalar fields.  Here $\mathcal{Q}^2=q_If^{IJ}q_J$ with $q_I$ the quantized charges. This bound is what is commonly assumed to be the generalization of the WGC with scalar fields \cite{Heidenreich_2019}, as it preserves one of the original motivations that extremal black holes should be allowed to decay in a theory of quantum gravity. In the presence of several gauge fields, the WGC can be phrased as the condition that the convex hull of the charge-to-mass ratio of the states in the theory contains the extremal region\footnote{The extremal region in the presence of scalar fields is not necessarily a unit ball; it needs to be determined for each theory.} \cite{Cheung_2014}.

\item \textbf{Repulsive Force Conjecture (RFC):} There must exist a particle which is self-repulsive, i.e. in which the gauge force acts stronger than gravity plus the scalar force,
\beq
\label{RFC}
\left(\frac{\mathcal{Q}}{M}\right)^2 \geq  \ \frac{D-3}{D-2}+  \frac{g^{ij}\partial_i M \partial_j M}{M^2} 
\eeq
where $D$ is the space-time dimension and $i, \ j$ run over the canonically normalized massless scalar fields. This bound was first proposed in \cite{Palti_2017} as the proper generalization of the WGC in the presence of scalar fields and renamed in \cite{Heidenreich_2019} as the Repulsive Force Conjecture. Notice that saturation of this bound corresponds to cancellation of forces, where the scalar Yukawa force arises simply because the particle mass $M(\phi)$ is parametrized by a massless scalar\footnote{When the mass of a particle $\chi$ is parametrized by a massless scalar field $\phi$, there is a Yukawa force interaction arising from \beq
\mathcal{L}\supset M^2 (\phi)\,\chi^2=2M\,(\partial_{\phi}M)\,\phi\,\chi^2+\dots
\eeq
as emphasized in \cite{Palti_2017}, where $\partial_{\phi}M$ plays the role of the scalar charge.}.
The argument for this bound stems from the expectation that black hole decay products should not be able to form gravitationally bound states.
\end{itemize}

In a supersymmetric theory, BPS particles satisfy a no-force condition and will indeed saturate \eqref{RFC}. Therefore, the RFC can be understood as an anti-BPS bound along the directions of the charge lattice that can support BPS states. This is what led \cite{Ooguri:2016pdq} to formulate a sharpening of the WGC for which only BPS states in a supersymmetric theory can saturate the WGC, yielding the striking result that any non-supersymmetric AdS vacua supported by fluxes must be unstable \cite{Ooguri:2016pdq,Freivogel:2016qwc}. However, this conclusion relies on this latter interpretation of the WGC as an anti-BPS bound and its fate becomes unclear whenever \eqref{WGC} and \eqref{RFC} do not coincide, or whenever it is possible to have extremal non-BPS states still satisfying a no-force condition as occurs if there is some fake supersymmetry \cite{Palti_2017, Galli_2011}. We will discuss more about this latter possibility in section \ref{generalizations}.

These two bounds \eqref{WGC} and \eqref{RFC} are in principle numerically different in the presence of scalars, as emphasized in \cite{Heidenreich_2019}. However, as we will see, they actually coincide in the asymptotic regimes of moduli space\footnote{This coincidence might also be used to fix the order one factor in the WGC applied to axions \cite{Arkani_Hamed_2007,Rudelius:2014wla,Rudelius:2015xta,Montero:2015ofa,Brown:2015iha,Bachlechner:2015qja,Heidenreich:2015wga}, since the notion of extremality is not well defined in the case of instantons.},
 congruent with the proposal in \cite{Lee_2019} that they should coincide in the weak coupling limits. In \cite{Lee_2019}, both bounds were shown to match explicitly for the weak coupling limits in 6D F-theory Calabi-Yau compactifications with 16 supercharges. In this paper, we will show that both bounds indeed coincide at any infinite field distance limit of the complex structure moduli space of four dimensional IIB Calabi-Yau string compactifications, i.e. for $\mathcal{N}=2$ four dimensional effective theories with 8 supercharges. Further connections between the WGC and the RFC can be found in \cite{Heidenreich_2019,HR,Ben1,Ben2,HL}\footnote{In particular, it is proved that extremal black holes (but not particles) always have vanishing self-force \cite{Ben1}, that particles which are self-repulsive everywhere in moduli space are superextremal \cite{Ben2}, and that those that have zero self-force everywhere, and nowhere vanishing mass are extremal \cite{Ben2}. }.

In the infinite field distance limits, there is another Swampland conjecture which also predicts the presence of new particles becoming light:
\begin{itemize}
\item \textbf{Swampland Distance Conjecture (SDC):} When approaching an infinite distance point in field space, there must exist an infinite tower of states becoming exponentially light with characteristic mass
\beq
M\sim M_0\exp{(-\lambda  \Delta \phi)}\ \text{ as } \ \Delta \phi \rightarrow \infty,
\label{SDC}
\eeq
where $\Delta\phi$ is the geodesic field distance. Here $\lambda$ is an unspecified parameter which is conjectured to be $\lambda\sim \mathcal{O}(1)$.
\end{itemize}

As already noticed in \cite{Grimm_2018,Lee_2019}, whenever there is a gauge coupling that goes to zero in the infinite field distance limit, it is possible to identify a tower of states becoming exponentially light according to the SDC that also satisfies the WGC. Hence, the SDC suggests a  stronger version of the WGC for which there must be not only a single particle but a sublattice or a tower of particles satisfying the WGC bound. In the past years, a lot of effort has been dedicated to rigorously identifying the tower of particles that become light in the infinite field distance limits of Calabi-Yau string compactifications \cite{Grimm_2018,Lee:2018urn,Lee_2019,Grimm_2019,Corvilain_2019,grimm2019infinite,Joshi:2019nzi,Marchesano_2019,Lee:2019tst,Lee:2019xtm,lee2019emergent,Baume:2019sry,Cecotti:2020rjq} (see also \cite{Palti_2017,Baume_2016,Valenzuela:2016yny,Blumenhagen:2017cxt,Gonzalo:2018guu,Palti:2015xra,Hebecker:2017lxm,Cicoli:2018tcq,Blumenhagen_2018,Buratti:2018xjt,Erkinger:2019umg}). It is the aim of this paper to study the relation between these conjectures in more detail, using the knowledge we have recently gained about these towers to define the Swampland conjectures in a precise way. In particular, if the same tower of particles satisfies both the SDC and the WGC, we can obtain information about the unspecified parameter $\lambda$ appearing in the SDC using the extremality bound of black holes. More concretely, it is precisely the contribution from scalar fields to the extremality bound that will determine mass scalar dependence of SDC tower, as we will explain below.

At the moment, it seems that we have three different Swampland conjectures predicting the existence of new light particles at weak coupling limits. Given the evidence from string compactifications, this seems redundant, as they all refer to the same asymptotic towers of particles. Hence, we would like to unify the above conjectures into a single statement that seems to hold at every infinite field distance limit of the moduli space of string compactifications. The statement goes as follows:


\begin{itemize}
\item In any infinite field distance limit with a vanishing gauge coupling, there exists an infinite tower of charged states satisfying $\left(\frac{\mathcal{Q}}{M}\right)^2\geq \left.\left(\frac{\mathcal{Q}}{M}\right)^2 \right|_{\text{extremal}}$
where the extremality bound coincides with the no-force condition:
\begin{align}
\left(\frac{\mathcal{Q}}{M}\right)^2 \bigg|_{\text{extremal}}= \frac{D-3}{D-2}+ \frac{g^{i j}\partial_{i} M  \partial_{j} M}{M^2} \label{ATC} \end{align}
and the gauge coupling decreases exponentially in terms of the geodesic field distance.
\end{itemize}


Note that this statement is stronger than the mild version of the WGC as it requires the existence of an infinite tower and not just a single particle. It matches, however, with stronger versions known as the sublattice or Tower WGC \cite{Heidenreich_2016,Montero:2016tif,Heidenreich:2016aqi, Andriolo_2018}. It also nicely fits with the notion that the WGC is a quantum gravity obstruction to restoring a U(1) global symmetry when $g\rightarrow 0$, as the infinite tower of states implies a reduction of the quantum gravity cutoff.\cmmnt{that invalidates the effective theory.} It also reproduces the SDC, as the fact that the gauge coupling decreases exponentially implies that the tower also becomes exponentially light in terms of the geodesic field distance. In addition, it further specifies the exponential rate of the tower, since it is determined by the extremality bound. Hence, there remains no unspecified order one parameter as in the original SDC.

The scalar fields entering in \eqref{ATC} are those parametrizing the gauge kinetic function of the gauge theory. If these scalars are massless, the extremal black hole solution involves a non-trivial profile for the scalars which modifies the charge-to-mass ratio of the extremal dilatonic Reissner-Nordstrom black hole. This extremality factor can be written as \cite{Horowitz:1991cd}
\beq
\left(\frac{\mathcal{Q}}{M}\right)_{\text{extremal}}^2 =  \frac{D-3}{D-2}+\frac14 |\vec{\alpha}|^2, 
\label{extbound}
\eeq
where the second term is the scalar (dilatonic) contribution. In the particular case in which the gauge coupling for all the gauge fields have the same dependence on the scalar fields,  $|\vec{\alpha}|$ becomes a numerical factor that can be easily read from the gauge kinetic function since the latter takes the form $f_{IJ}=f^0_{IJ}\,e^{\alpha_i \phi_i}$ in terms of the canonically normalized scalar fields, where $f^0_{IJ}$ is a  moduli-independent constant matrix. Otherwise, the extremality bound becomes a more complicated function that depends non-trivially on the scalar fields (see \cite{Duff:1996hp, Lu:1995sh, Lu:1995yn} for discussions of dilatonic black holes in more general settings). 

In principle, the second terms in \eqref{ATC} and \eqref{extbound} are numerically distinct if the WGC and the RFC do not coincide, but, as already discussed, we will show that they coincide in the regime approaching an infinite field distance point, so that
\beq
\frac{|\vec{\alpha}|^2}{4}=\frac{g^{ij}\partial_i M\partial_j M}{M^2}\ .
\eeq
The quantity on the right hand side is also known as the scalar charge-to-mass ratio and plays a dominant role in the Scalar WGC proposed in \cite{Palti_2017} (see also \cite{Gonzalo:2019gjp,Brahma:2019mdd,DallAgata:2020ino}), for which there must exist a state satisfying
\beq
\label{ScalarWGC}
\frac{g^{ij}\partial_i M\partial_j M}{M^2}>\frac{D-3}{D-2}
\eeq
for any direction in field space. The appealing feature of this conjecture is that, if satisfied by the SDC tower, it seems to imply that the exponential mass decay rate should be indeed of order one. 
In fact, for a single scalar field, we can see that it relates to the exponential factor $\lambda$ of the SDC that appears in \ref{SDC} as follows:
\begin{align}
\lambda =\frac12 \alpha=\frac{\partial_\phi M}{M}\ .
\label{LA}
\end{align}
However, for more general setups with several gauge and scalar fields, the situation is more complicated and the above identification \eqref{LA} is not valid. First, when the gauge kinetic function exhibits a different scalar dependence $|\vec\alpha|$, the scalar contribution to the extremality bound cannot be directly read from the gauge kinetic function. Secondly, in higher dimensional moduli spaces, path dependence issues come into play, invalidating the identification of $\lambda$ with the scalar charge-to-mass ratio. The exponential factor $\lambda$ will now be given by the projection of the scalar charge-to-mass ratio vector over the specific geodesic trajectory, as we will discuss in section~\ref{decayrate}, which can lower its value. Hence, the extremality factor (or equivalently, the charge-to-mass ratio) will only give an upper bound on the SDC exponential decay rate. Fortunately, we will still be able to place a lower bound on the mass decay rate of the BPS tower using the scalar dependence of the gauge kinetic function, but this lower bound will come from the individual $\alpha_i$ associated to the asymptotic splitting of charges that occurs at infinite field distance, as we will discuss in section \ref{decayrate}. In fact, these individual $\alpha_i$ will still be associated to the scalar contribution to the extremality bound of some particular black holes that satisfy \eqref{extbound}.

It is important to remark that the original SDC does not refer to any gauge coupling going to zero at infinite field distance, so the tower of states is not necessarily charged. Hence, the claim \eqref{ATC} only completely reproduces the SDC if there is a gauge coupling going to zero at every infinite field distance limit. This is a stronger requirement than the original SDC. Even though the limit $g\rightarrow 0$ is known to always be at infinite field distance, the opposite is not necessarily true. However, we believe that it is always possible to identify some p-form gauge coupling vanishing at every infinite field distance of string compactifications, which also relates to the proposal  \cite{Grimm_2018} in understanding the SDC as a quantum gravity obstruction to restore a global symmetry\footnote{There are actually two types of global symmetries that are restored at the infinite field distance limits of Calabi-Yau compactifications and that have been associated \cite{Grimm_2018} to the SDC. One one hand, we have a U(1) global symmetry coming from each gauge coupling that goes to zero. On the other hand, we have the global continuous version of a discrete symmetry of infinite order (a monodromy transformation) which is part of the duality group and that was used in \cite{Grimm_2018} to populate the infinite tower of states.}. Furthermore, if the Emergent String Proposal of \cite{lee2019emergent} is true, every infinite field distance limit would correspond either to a decompactification limit or a string perturbative limit, suggesting that it might always be possible to identify at least a KK photon or a 2-form gauge field becoming weakly coupled, supporting our statement. It has also been recently emphasized \cite{Kim:2019ths} that the SDC tower always hints a weakly coupled dual field theory description, which goes in the same spirit. We will comment more on this in section~\ref{generalizations}.
Notice, also, that we are not requiring this charged tower to be the leading tower, i.e. the one with the fastest decay mass rate. If it is not, it can still be used to give a lower bound on the exponential mass decay rate of the SDC, and therefore a precise upper bound on the scalar field range.

Although some arguments can be made in general for the unification of all these conjectures in the asymptotic regime, in the context of $\cN=2$ compactifications we can be very precise. For this reason, we will focus on $\cN=2$ four dimensional supergravity effective theories in the next two sections. 
We will calculate the charge-to-mass ratios of BPS states and compare these results to the black hole extremality bounds. This will allow us to compare the different possible generalizations of the WGC that we have reviewed in this section, as well as make progress in sharpening the SDC.  We will explicitly check \eqref{ATC} and provide a bound on $\lambda$ for every infinite field distance limit. In section \ref{generalizations}, we will discuss the realization of \eqref{ATC} in other setups beyond 4d $\cN=2$ theories.

\section{BPS states and extremal black holes in 4d $\cN=2$ effective field theories} \label{bps}
In this section, we analyze the structure of BPS charge-to-mass ratios in the asymptotic limits of the moduli space of 4d $\cN=2$ theories. We will calculate the extremality bound and show that the charge-to-mass ratio of BPS states lie on a degenerate ellipsoid with two finite principal radii, regardless of the number of moduli. We will exemplify this in two particular examples. In those examples, we will compare the BPS charge-to-mass ratios with the extremal black hole bounds of those particular theories, as well as calculate the lower bound on the mass decay rate $|\lambda|$ of the SDC. In section \ref{asymptotic}, we will compute the numerical value of these radii in terms of the type of asymptotic limit in full generality.

\subsection{Review: $\cN=2$ EFTs and BPS states in Calabi-Yau compactifications} \label{bpsreview}

Consider a Calabi-Yau threefold characterized by a set of $h^{2,1}$ complex structure moduli $T^i$ which span a special K\"ahler submanifold.
Compactifying type IIB theory on this Calabi-Yau threefold results in a four-dimensional effective theory with $\mathcal{N}=2$ supersymmetry and $n_V=h^{2,1}$ vector multiplets, yielding $n_V+1$ U(1) gauge fields of field strength $F^I_{\mu\nu}$ with $I=0,\dots ,h^{2,1}$. The theory also contains hypermultiplets involving the K\"ahler moduli deformations which we will ignore for the moment, since they are not relevant for our purposes. The low energy bosonic effective action reads
\begin{align}
S = \int d^4 x \left[ \frac{R}{2}-K_{i \bar{j}} \partial_{\mu} T^{i} \partial^{\mu} \bar{T}^{j}+\mathcal{I}_{I J} \mathcal{F}_{\mu \nu}^{I} \mathcal{F}^{J, \mu \nu}+\mathcal{R}_{I J} \mathcal{F}_{\mu \nu}^{I}(\star \mathcal{F})^{J, \mu \nu} \right],
\label{lagrangian}
\end{align}
where $K_{i\bar{j}}, \ \mathcal{I}_{I J},$ and $\mathcal{R}_{I J}$ are determined by the geometrical data of the compactification\footnote{Note that $K_{i\bar{j}}=\frac12 g_{i j}$ from the previous section.}. In particular,
\begin{align}
K_{i \bar{j}} = \partial_i \partial_{\bar{j}} \mathcal{K}, \  \ \ \ \mathcal{I}_{I J} = \text{Im}(\mathcal{N}_{IJ}), \  \ \ \ \mathcal{R}_{I J} = \text{Re}(\mathcal{N}_{IJ}) ,
\end{align}
where $\mathcal{K}$ is the K\"ahler potential
\begin{align}
\label{Kperiods}
\mathcal{K} = -\text{log} \ i\left[\overline{X}^I F_I -  X^I \overline{F}_I \right]
\end{align}
 and
\begin{align}
\mathcal{N}_{I J}=\bar{F}_{I J}+2 i \frac{\operatorname{Im} F_{I K} \operatorname{Im} F_{J L} X^{K} X^{L}}{\operatorname{Im} F_{M N} X^{M} X^{N}}.
\end{align}
Here $\{X^I,F^I\}$ are the periods of the holomorphic $(3,0)-$form of the Calabi-Yau threefold and can be written as holomorphic functions of the scalars $T^i$. They can be determined in terms of a prepotential function $F$ through $F_I=\partial_{X^I} F$ and  $F_{IJ} = \partial_I \partial_J F$.
Note that capital indices range from $0$ to $h^{2,1}$, but that in the case where a prepotential exists we can go to special coordinates where $X^I=(1,T^i)$ with $i=1,\dots,h^{2,1}$. The complex scalars have components
\beq
T^i=\theta^i+it^i
\eeq
where $\theta^i$ are the axions and $t^i$ are usually dubbed as \emph{saxions}.


The theory also contains BPS particles that arise from D3-branes wrapped on special Lagrangian 3-cycles whose volumes are parametrized by the complex structure moduli. 
The mass of a BPS state  is given by the central charge: $M=|Z|$, which can be written
\begin{align}
Z = e^{\mathcal{K}/2} (q_I X^I - p^I F_I)
\label{centralcharge}
\end{align} 
and the normalized charge in \eqref{WGC} of this BPS state is given by $\mathcal{Q}^2 =\frac12 |Q|^2$ \cite{Palti_2019}, where $|Q|^2$ is defined as
\begin{align}
|Q|^2 = -\frac{1}{2} \vec{\mathfrak{q}}^T \mathcal{M} \vec{\mathfrak{q}},
\label{qsquare}
\end{align}
where $\vec{\mathfrak{q}}$ is the vector of integrally quantized charges: $\vec{\mathfrak{q}} = \begin{pmatrix} p^I \\ q_I \end{pmatrix}$ and
\begin{align}
\mathcal{M}=\begin{pmatrix} 
\mathcal{I}+\mathcal{R} \mathcal{I}^{-1} \mathcal{R} & -\mathcal{R} \mathcal{I}^{-1} \\
-\mathcal{I}^{-1} \mathcal{R} & \mathcal{I}^{-1} 
\end{pmatrix}
\label{Mmatrix}
\end{align}
Note that we can also write the normalized charge in terms of the central charge \cite{Ceresole_1996}:
\begin{align}
\label{QN=2}
|Q|^2 = |Z|^2 + K^{i \bar{j}} D_i Z D_{\bar{j}} \overline{Z},
\end{align}
where $D_i$ is the covariant derivative, which acts on $Z$ as $D_i Z = \partial_i Z + \frac12 K_i Z$. The specific form of the prepotential, and therefore the field and gauge kinetic metrics, depend on which region of the complex structure moduli space of the Calabi-Yau threefold we are in. We will see that in the asymptotic regimes of the moduli space, there is a well-defined notion of electric and magnetic BPS states which become massless or infinitely heavy respectively at the infinite field distance limit as long as we classify the trajectories in field space into different growth sectors. Hence, a charge that is an electric BPS state which becomes massless along one path may no longer become massless within a different growth sector. By default, though, the reader can assume we are denoting the electric charges as $q_I$ unless otherwise noted. 


Note that not all possible combinations of quantized (electric and magnetic) charge are actually associated to BPS states. The central charge \eqref{Z} is only the mass that a BPS state would have if the charges $q_I,p^I$ are actually populated by a physical BPS state. In characterizing the charge-to-mass ratios of these D3-branes, it will be necessary to determine which choices of charge do correspond to BPS particles. In general, the quantized charges that can support BPS states will exhibit a conal structure (analogous to the effective cone of 4-cycles on the K\"ahler moduli side). The attractor mechanism \cite{Ferrara_1995, Ferrara_1996,Ferrara:1996um, Strominger_1996} tells us that in the presence of a BPS black hole, there is an effective potential for the moduli in the theory, and that the dynamics of the scalars in this potential are such that they all flow to a fixed point on the black hole horizon, regardless of the initial scalar profile at spatial infinity. Thus, we learn that the BPS choices of quantized charges are those which allow the scalars in the theory to flow to fixed points on the horizon of a black hole. Said differently, if the ``wrong" charges are picked for a black hole, the scalars in the theory will not exhibit an attractor flow and the black hole is not BPS. The details of using the attractor mechanism to determine which states are BPS was analyzed in \cite{Denef_1999, Denef_2000, moore1998arithmetic, moore1998attractors,Ortin:2015hya}. It is also important to note that just because a given charge site is able to support a BPS state does not mean that such a state exists in the theory. Therefore, when we refer to particular BPS states, we are referring to charge lattice sites which can in principle be populated by a BPS state. To determine if a given BPS state at a particular point in field space continues being BPS when moving to a different point, one must study the presence of walls of marginal stability \cite{Denef_2000, Douglas:2000qw, Douglas:2000gi, Denef:2001xn, Aspinwall:2001zq, Aspinwall:2002nw, Jafferis:2008uf, Aspinwall:2009qy, Andriyash:2010yf}, as also done in \cite{Grimm_2018} for the asymptotic limits. In this section, we will ignore the possible presence of these walls and just discuss the conal strucuture of charges that can in principle support BPS states, leaving a more detailed study to section \ref{asymptotic}.

\subsection{BPS and extremality bounds in the infinite distance limit} \label{bpsextsame}

A primary goal of this work is to compare the BPS spectrum arising from general compactifications of this type with the extremality bounds of black hole solutions in the resulting 4d effective theory. Having reviewed the key ingredients for calculating the charge-to-mass ratios of BPS states in section~\ref{bpsreview}, let us now begin to do this comparison. We will revisit why BPS black holes are extremal in the asymptotic limits of the moduli space, implying the identification of the above WGC and the RFC conjectures, and identify some particular solutions, which we will denote as single-charge states, that will play a crucial role throughout the paper.

%


To analyze the extremality bound of black holes in the effective theory, we first note that in the infinite distance limits of moduli space, the K\"ahler metric $K_{i \bar{j}}$ always behaves to leading order as
\beq
K_{i \bar{i}}=\frac{d_i}{4 t_i^2}+\dots
\eeq
where $d_i$ is an integer associated to the type of singular limit as we will discuss in detail in section \ref{MHS}.
 Therefore, in terms of the canonically normalized scalar fields $\phi^i=\sqrt{\frac{d_i}{2}}\log t^i$, the gauge kinetic function $\mathcal{I}_{IJ}$ will generically have exponential dependence: $\mathcal{I}_{IJ} \sim e^{\vec{\alpha} \cdot \vec{\phi}}$. Thus, the black hole solutions of this theory are Reissner-Nordstrom dilatonic black holes. For black holes charged under a single gauge field with a dilatonic coupling, the extremal charge-to-mass ratio is given by \eqref{extbound}\footnote{Note that we have switched notation  to $Q$ as defined in \eqref{qsquare} which involves an extra factor of 2.}
\begin{align}
\left( \frac{Q}{M} \right) \bigg|_{\text{extremal}} = \sqrt{1 +\frac{1}{2}|\vec{\alpha}|^2}
\label{qm_ext}
\end{align}
where we have specialized to a black hole charged under a 1-form gauge field in 4 dimensions. If the black hole is charged under several gauge fields with different scalar dependencies, the dilatonic contribution to the extremality bound cannot be simply read from the gauge kinetic function and will depend on the scalars themselves.

On the other hand, the ``no-force" condition is given by the BPS bound in our $\cN=2$ setup:
\begin{align}
\left( \frac{Q}{M} \right) \bigg|_{\text{BPS}} = \sqrt{1+4 K^{i\bar{j}} \frac{ \partial_{i} |Z| \partial_{\bar{j}} |Z|}{|Z|^2}} \ .
\end{align}
Comparing the above two equations, we can see that in order to show equality of the extremal bound and the BPS bound, it suffices to show that the gauge kinetic function $\mathcal{I}_{IJ}$ and the absolute value squared of the central charge $|Z|^2$ have the same functional dependence on the scalars, i.e. that $|Z|^2\sim e^{\vec \alpha \cdot \vec \phi}$. 

In section \ref{ctomellipse}, we will give a general proof of this correlation between the central charge and the gauge kinetic function for any number of moduli and gauge fields in any asymptotic limit by using the growth theorem of asymptotic Hodge theory. Here, however, we give some preliminary insight into why this holds in the well-studied large complex structure limit.

In theories with $\cN=2$ supersymmetry, the gauge kinetic function and the field metric are related in the large complex structure limit as
\begin{align}
\mathcal{I}_{IJ}^{-1} = -\frac{6}{\kappa} \begin{pmatrix} 1 & 0 \\ 0 & \frac{1}{4} K^{i \bar{j}} \end{pmatrix},
\end{align}
where we have set the axions to zero for the moment.
Here, $\kappa$ is related to the K\"ahler potential: $\kappa = \frac{3}{4} e^{-\mathcal{K}}$  and the K\"ahler potential can be written in terms of the periods as in \eqref{Kperiods}. Thus, we can rewrite the gauge kinetic function as
\begin{align}
\mathcal{I}_{IJ}^{-1} = \begin{pmatrix} 4 \text{Im}(\overline{X}^I F_I)^{-1} & 0 \\ 0 &  \text{Im}(\overline{X}^I F_I)^{-1} K^{i\bar{j}} \end{pmatrix} .
\end{align}
On the other hand, the mass squared is
\begin{align}
|Z|^2 = \frac{-2 |q^T \eta \Pi |^2}{\text{Im}(\overline{X}^I F_I)},
\end{align}
where we have introduced the symplectic matrix $\eta= \begin{pmatrix} 0 & -\mathbb{I} \\ \mathbb{I} & 0 \end{pmatrix}$ to denote the symplectic product in the central charge \eqref{centralcharge}. For electric states with charge $q_I$, the symplectic product selects the periods $X^I=\left(1,T^i\right)$.
If we now focus on states charged under a single gauge field, such that either $q_0\neq 0$ or $q_1\neq 0$, we find
\begin{align}
\mathcal{I}_{II}^{-1} \propto |Z_I|^2,
\end{align}
where $|Z_I|^2 =  \frac{-2 q_I^2  \Pi^I \overline{\Pi}^I}{\text{Im}(\overline{X}^J F_J)}$ with no sum in the numerator, since the product $ |q^T \eta \Pi |^2$ is proportional to $1$ for a single-charge state with $q_0$ charge or to $K^{i\bar{i}}$ for a single-charge state with $q_i$ charge.

For the above argument it is very important that we have not turned on several charges associated to a different behavior of the gauge kinetic function at the same time. This notion of \emph{single-charge state}  will be extensively used throughout the paper, so let us explain it a bit more. A single-charge state is a state which carries charge only under a single\footnote{Actually, it can be charged under several gauge fields as long as they all have the same scalar dependence on the gauge kinetic function. See section~\ref{ctomellipse} for a very precise definition of a single-charge state in terms of the charge splitting associated to the asymptotic limit.} gauge field whose gauge kinetic function can be written to leading order as single exponential $\mathcal{I}_{IJ} \sim e^{\vec{\alpha} \cdot \vec{\phi}}$ in terms of the canonically normalized fields. In other words, it corresponds to a black hole whose extremality bound is simply given by the above dilatonic extremality formula \eqref{qm_ext}, where $\alpha$ is a constant corresponding to the exponential rate of the gauge kinetic function. Though this is a basis-dependent definition, it is a well defined notion in the asymptotic limits of moduli space. This condition selects a very particular basis of charges which will be introduced in section~\ref{asymptotic}. At this stage, let us simply add that this basis is associated to an asymptotic splitting of the charge space into nearly orthogonal subspaces which is guaranteed by the $sl(2)$-orbit theorem of Hodge Theory that will be discussed in more detail in section~\ref{MHS}. As remarked in \cite{Grimm_2018,Grimm_2019,Grimm:2019bey,Cecotti:2020rjq}, the structures provided by Hodge theory are key to distinguish what $\cN=2$ supergravity theories can actually arise from string theory, so we do not expect our results to be necessarily valid for any supergravity theory but only those consistent with a quantum gravity embedding. 

The existence of these single-charge states will allow us to determine the shape that the charge-to-mass ratio of BPS states trace out in the next section. We will identify these single-charge states in examples in sections \ref{1mod} and \ref{2mod} but leave their general identification in any asymptotic limit to section \ref{ctomellipse}, where we will also provide the numerical result for their charge-to-mass ratios in terms of discrete data associated to the infinite field distance limit.
At the moment, let us just remark that the charge-to-mass ratio of these single-charge states is simply given by a numerical factor (independent of the moduli) that can be read from the gauge kinetic function or equivalently computed from the central charge since:
\beq
|\vec \alpha|^2= 8 \frac{ K^{i\bar{j}} \partial_{i} |Z| \partial_{\bar{j}} |Z|}{|Z|^2}
\label{aZ}
\eeq
%
%
%
%

The fact that the moduli dependence of the gauge kinetic function and the mass squared are identical implies that the extremality bound and the BPS bound are equal. Let us also notice that the presence of axions will only shift the identification of single-charge BPS states, as will become clear throughout the next sections, but it is always possible to find a basis of charges such that the associated single-charge BPS states satisfy the above property. 

What about BPS states that are charged under more than one gauge field? If the gauge fields have a different scalar dependence, the extremality formula \eqref{qm_ext} is no longer valid, and one has to solve the full BPS flow equations of the attractor mechanism in order to find a extremal solution. Notice that  \eqref{qm_ext} is only a particular solution of these flow equations. Any other solution of the BPS flow equations will correspond to a BPS extremal back hole as well. These flow equations have been extensively studied in the literature for $\cN=2$ setups where black holes exhibit a well-known attractor behaviour \cite{Ferrara_1995, Ferrara_1996, Ferrara:1996um, Strominger_1996, Behrndt:1996jn, Behrndt:1997ny,Galli:2012pt}, wherein the scalars flow to fixed values on the horizon of the black hole, regardless of their initial conditions at spatial infinity. This is known as the attractor mechanism. The fixed values for the scalars at the horizon can be determined by minimizing the black hole potential $V_{bh}=Q^2$. When this charge supports a BPS state so that \eqref{QN=2} is satisfied, minimization of the black hole potential is equivalent to finding a solution to $\partial_{T^i} |Z|=0$ for the values of the scalars on the horizon. We can see this by examining the form of the attractor flow equations. The most general spherically symmetric ansatz for a black hole solution is 4d is given by
\begin{align}
\label{blackhole}
ds^2= e^{2 U(r)} dt^2 - e^{-2 U(r)} \left(\frac{1}{g(r)^2} dr^2 + r^2 d\Omega_2^2 \right).
\end{align}
Assuming spherically symmetric moduli and electromagnetic fields, there is a reduced effective action \cite{Denef_1999}
\begin{align}
S=\frac{1}{2} \int_0^\infty d\tau \bigg[\dot{U}^2 + g_{a\overline{b}} \dot{T}^a \dot{\overline{T}}^{\overline{b}}-c^2 + e^{2 U} V_{bh}(T) \bigg] ,
\end{align}
where $r=c/\sinh c\tau$, $g(r) = h(\tau) \cosh c\tau$, and $V_{bh}= |Q|^2$ as defined in \eqref{QN=2}. This describes a system in which the scalars $T^i$ flow in the potential $V_{bh}$ until a minimum is reached. This flow is given by the BPS flow equations, which read:
\beqa
\label{flow1}
&\dot U=-e^U|Z|\\
&\dot T^i=-2 e^U g^{i\bar j}\partial_{\bar j}|Z|\label{flow2}.
\eeqa
Notice that electric states have a mass and charge corresponding to a monotonically decreasing runaway function towards infinite field distance, so that there is no extremum of the black hole potential at finite distance. This implies that the scalar fields flow to $\phi\rightarrow\infty$ at the black hole horizon and the black hole entropy vanishes since $Z|_{\phi\rightarrow \infty}=0$. This also occurs for the dilatonic black hole solutions in \eqref{qm_ext}. Clearly, this singularity at the horizon does not allow us to describe these small black holes within the supergravity approximation, and the typical expectation is that stringy higher derivative corrections will eventually correct the entropy to make it finite, although this is open for debate \cite{Sen:1995in,Dabholkar:2004yr,Dabholkar:2004dq,Cano:2018hut,Cano:2018qev}. Nevertheless, small black holes can always be appropriately described in the full string theory brane or worldsheet approach, in which case one obtains a non-vanishing value for the entropy \cite{Sen:1995in}. Interestingly, it is the extremality bound associated to small black holes which is the one that plays a role in the Weak Gravity Conjecture at small gauge coupling and the one that constrains the mass decay rate of the towers of particles for the Distance Conjecture. Hence, in this paper, we will investigate this extremality bound assuming that string theory will come to the rescue to regularize these black hole solutions when approaching the horizon. In any case, the only information we need is the ADM mass $M_{ADM}=|Z|$  and the charge of the black hole $Q$ for our purposes. Electric dilatonic black holes of this type have also been studied in \cite{Draper:2019utz,Loges:2019jzs,Bonnefoy:2019nzv} in the context of the WGC and SDC. In \cite{Draper:2019utz} they were used as an example to argue for a local version of the SDC, for which local excitations of an EFT cannot sample large field excursions without inducing large curvature at a horizon or instabilities.

It should be noted that not all charges allow for a physical solution to the flow equations \eqref{flow1}-\eqref{flow2}. For instance, it could happen that the central charge is driven to zero at a regular point in field space. When this occurs, the attractor flow ``breaks'' before reaching the horizon, providing a litmus test for determining which charges can, in principle, give rise to BPS states \cite{Denef_2000}. However, imposing  $Z\neq 0$ is not always enough to guarantee a BPS black hole solution, as there are other ways for the attractor flow to break down, which must be taken into account to get the full set of charge restrictions.  In section~\ref{2mod}, we will see an example in which we will fully determine what regions of the charge lattice do not correspond to BPS directions. However, this does not mean that extremal black holes do not exist in these regions. In section \ref{generalizations} we will demonstrate the existence of non-BPS extremal black holes precisely in the quadrants of the charge lattice that do not support BPS states, and show that they still satisfy a no-force condition, confirming unification of WGC and RFC at the asymptotic/weak coupling limits.

Note that the matching between WGC and RFC tells us something very interesting about the connection between the Swampland Distance Conjecture and black hole extremality bounds. Recall that the SDC tells us that at infinite distance, there should be an infinite tower of states that becomes exponentially light, where the rate is an unspecified parameter denoted as $\lambda$ in \eqref{SDC}. For the single-charge states satisfying \eqref{aZ}, the rate is now fixed in terms of the dilatonic contribution to the extremality bound
\begin{align}
\label{mexp}
M \sim M_0\ e^{-\frac12\vec \alpha \cdot \vec \phi}
\end{align}
where recall $\phi^i$ are the canonically normalized scalars and $\alpha_i$ can be read from the gauge kinetic function. The parameter $\lambda$ associated to a geodesic will be a projection of $\vec \alpha$ over the geodesic trajectory, so it will be lower bounded by the individual $\alpha_i$'s.
For the rest of this paper, and in particular in section \ref{decayrate}, we will exploit this fact, along with our knowledge of the asymptotic behavior of the gauge kinetic function, to put bounds on the exponential factor ${\lambda}$ appearing in the Swampland Distance Conjecture for any number of scalar and gauge fields.

As already remarked, only for the particular set of single-charge BPS states does the extremality factor reduce to a numerical factor that can be read from the gauge kinetic function. In general, though, the charge-to-mass ratio of BPS states (and the extremality bound) will depend on the moduli. So what is this value in general? Can it be bounded? If so, we can use it to bound the exponential factor of the SDC. Determining the general structure of BPS states and finding these bounds is the goal of the next section.

\subsection{General structure of BPS charge-to-mass spectra} \label{generalellipse}

We are interested in the structure of the charge-to-mass vectors of electrically charged BPS D3-branes in Type IIB compactifications. By ``charge-to-mass vector," we mean the vector
\begin{align}
\vec{z} = \frac{|Q|}{M} \hat{Q}
\end{align}
with $Q$ defined in \eqref{qsquare}, which is the quantity of interest when checking the Weak Gravity Conjecture in settings with multiple gauge fields \cite{Cheung_2014}. We will show in this section that the $\vec{z}$-vectors of BPS states lie on a degenerate ellipsoid. This ellipsoid has exactly two non-degenerate directions, regardless of the number of moduli, with principal radii determined by the gauge kinetic matrix. This is one the most useful results of this paper.

To see this, recall that the mass of a BPS state is given by the central charge
\begin{align}
\frac{|Z|^2}{M^2}=1
\end{align}
Let us define the quantized ``electric charge" $\vec{q}_E$ (corresponding to states that become light in the infinite distance limit) and associated ``electric periods" $\vec{\Pi}_E$. Then the above expression can be written explicitly as:
\begin{align}
\frac{\left(q^T_E \eta \Pi_E\right) \left(q^T_E \eta \bar{\Pi}_E\right)}{ -2\text{Im}\left(\overline{X}^{I} F_I\right) M^2} =1
\end{align}
where $\eta$ is an anti-symmetric intersection matrix
\begin{align}
\eta_{IJ} = \int_X \gamma_I \wedge \gamma_J = \begin{pmatrix} 0 & -\mathbb{I} \\ \mathbb{I} & 0 \end{pmatrix}
\label{intmat}
\end{align} 
with $\gamma_I$ an appropriate integral symplectic basis of 3-cycles on the Calabi-Yau, $X$.

We would like to write this expression in terms of the $\vec{z}$-vectors. To do this, let's define a symmetric matrix $G$ such that $G^TG=-\frac{1}{2}\mathcal{I}$. Then using $\vec{z} = \frac{1}{M} G^{-1} \vec{q}$, we get
\begin{align}
-2 \frac{\left(\vec{z}_E^\dagger G^T  \eta  \vec{\Pi}_E\right)\left( \vec{\Pi}_E^\dagger  \eta^T   G \vec{z}_E \right)}{ \text{Im}\left(\overline{X}^{I} F_I\right) } =1.
\end{align}
This can be written as a quadratic equation
\begin{align}
\vec{z}_E^T \mathbb{A} \vec{z}_E =1
\end{align}
where $\mathbb{A}$ is a symmetric matrix
\begin{align}
\mathbb{A} = -2 \frac{G^T \eta \ \text{Re}\left(\vec{\Pi}_E \vec{\Pi}_E^\dagger\right) \eta^T \ G}{\text{Im}\left(\overline{X}^I F_I \right)}
\label{Amatrix}
\end{align}
Note what this implies about the structure of $\mathbb{A}$: this is a block-diagonal matrix with two sub-blocks, where each sub-matrix is given by the outer product of a vector with itself. This means that $\mathbb{A}$ has exactly $2$ non-zero eigenvalues and $h^{2,1}-1$ zero eigenvalues. The non-zero eigenvalues are easily computed:
\begin{align}
\gamma_1^{-2}= \frac{ \text{Re}\left(\vec{\Pi}_E^\dagger\right) \ \eta^T \mathcal{I} \eta \ \text{Re}\left(\vec{\Pi}_E\right)}{\text{Im}\left(\overline{X}^I F_I\right)} \label{l1} \ ,\\
\gamma_2^{-2} = \frac{ \text{Im}\left(\vec{\Pi}_E^\dagger \right) \ \eta^T \mathcal{I} \eta \ \text{Im}\left(\vec{\Pi}_E\right)}{\text{Im}\left(\overline{X}^I F_I\right)} \label{l2} \ .
\end{align}
In summary, we have seen that the charge-to-mass vectors of BPS states lie on an ellipsoid with two non-degenerate radii given by the above eigenvalues. This is also consistent with the fact that not all of these states are mutually BPS (otherwise they could fragmentate to smaller BPS states when crossing a wall of marginal stability), so they are expected to form an ellipsoid in the charge-to-mass ratio plane. However, note that pairs of states lying along lines in the degenerate directions \textit{are} mutually BPS.
Let us calculate the eigenvalues of the ellipsoid more explicitly. To do this, we first note that they are invariant under shifts in the axionic variables $\theta^i$ (where $T^i = \theta^i + i t^i$). This is because the periods transform under axion shifts as:
\begin{align}
\vec{\Pi}_E  = e^{\theta^i N_i^\dagger} \vec{\Pi}_{E,0}
\end{align}
where $\vec{\Pi}_{E,0}$ are the electric periods with all axions set to zero, and $N_i$ is a nilpotent monodromy matrix. This is called the Nilpotent Orbit Theorem \cite{schmid} and is reviewed in more detail in section~\ref{MHS}.
The gauge kinetic matrix transforms in a similar way,
\begin{align}
\mathcal{I} = e^{\theta^i N_i^\dagger} \mathcal{I}_0 e^{\theta^i N_i},
\end{align}
where $\mathcal{I}_0$ is the gauge kinetic matrix with the axions set to zero. Plugging these in to Eqs.~\ref{l1} and \ref{l2}, we see that the eigenvalues can be written as
\begin{align}
\gamma^{-2}_1 = \frac{ \text{Re}(\vec{\Pi}_{E,0}^\dagger) \ \eta \ \mathcal{I}_0 \ \eta^T \ \text{Re}(\vec{\Pi}_{E,0})}{ \text{Im}(\overline{X}^I F_I)} \label{l10}\\
\gamma^{-2}_2 = \frac{ \text{Im}(\vec{\Pi}_{E,0}^\dagger) \ \eta \ \mathcal{I}_0 \ \eta^T \  \text{Im}(\vec{\Pi}_{E,0})}{ \text{Im}(\overline{X}^I F_I)} \label{l20}
\end{align}
These expressions are much easier to calculate explicitly, namely because when the axions are zero, $\mathcal{I}_0$ is diagonal\footnote{Actually, in general, the gauge kinetic matrix can be block diagonal in the infinite distance limit. However, such a scenario corresponds to a higher-dimensional asymptotic subspace splitting, as explained in section~\ref{asymptotic}, and the results will not change.} and the periods $\vec{\Pi}_{E,0}$ are either purely real or purely imaginary. To give an explicit expression for the eigenvalues, let's write them in terms of charge-to-mass ratios of the single-charge states, as defined in the previous subsection. We will use the notation $\left(\frac{Q}{M}\right)\bigg|_{q_E^I}$ to denote the charge-to-mass ratio of a single-charge state carrying charge $q_E^I$ under the $I^{\text{th}}$ electric gauge field. This charge-to-mass ratio can be computed in terms of the electric periods and gauge kinetic function:
\begin{align}
\left(\frac{Q}{M}\right)\bigg|_{q_E^I}^2= \frac{\text{Im}(\overline{X}^J F_J)}{\mathcal{I}^E_{0,II} \Pi_0^{E,I} \overline{\Pi}_0^{E,I}},
\end{align}
with no sum over $I$. So we see that the eigenvalues of $\mathbb{A}$ can be written quite simply:
\begin{align}
\gamma^{-2}_1 &= \sum_{i  |  \text{Im}(\Pi_0^{E,I})=0} \left(\frac{Q}{M}\right)\bigg|_{q_E^I}^{-2} \label{lambda1}\\
\gamma^{-2}_2 &= \sum_{i  |  \text{Im}(\Pi_0^{E,I})\neq0} \left(\frac{Q}{M}\right)\bigg|_{q_E^I}^{-2} \label{lambda2}.
\end{align}
This result is quite remarkable, since it means that we can completely determine the structure of all BPS states in charge-to-mass ratio space just by computing the value of the charge-to-mass ratio of the single-charge BPS states which, in the previous section, we showed to be determined purely by the scalar dependence $\vec\alpha$ of the gauge kinetic function. This might not seem well defined as it is basis-dependent, but the whole point is that there is a special basis associated to each asymptotic limit in the moduli space for which the above holds. This specific basis is associated to the sl(2) splitting which will be discussed in section~\ref{asymptotic}, implying that the eigenvalues $\gamma_1,\gamma_2$ can be precisely calculated in terms of a set of integers characterizing the infinite distance limits in Calabi-Yau threefolds.

The Weak Gravity Conjecture is satisfied if the convex hull of the charge-to-mass vectors of the states of the theory contain the extremal black hole region. This extremal region is sometimes simply identified with a \emph{unit ball}, assuming that the extremality bound is equal to 1 for every charge direction. In actuality, the extremal region need not be a ball at all and can change from theory to theory. In 4d $\cN=2$, since BPS black holes are extremal in the asymptotic limits of the moduli space, this region will be equivalently given by the value of the charge-to-mass ratio of BPS states along the directions supporting BPS states. Hence, the ellipsoid with principal radii given by \eqref{lambda1} and \eqref{lambda2} corresponds to the extremal region for electrically charged black holes.

As remarked in the previous section, not all regions in the charge lattice can support BPS states. In general, we expect there to be inequalities governing the which directions of the charge lattice are BPS directions of the form $p(\vec{q})>0$ with $p$ a polynomial. Furthermore, our expectation is that these charge restrictions precisely coincide with truncating the degenerate directions of the charge-to-mass ellipsoid. This expectation arises from the fact that BPS states cannot become massless at non-singular points in moduli space \cite{Denef_2000}, and so charge directions with unbounded charge-to-mass ratios should be prohibited.

Let us also remark that even though it is clear that the WGC bound is satisfied along BPS directions since BPS states themselves saturate the extremality bound, it is not obvious at all whether BPS states are enough to satisfy the convex hull condition along any direction in charge space. In section~\ref{generalizations} we will compute the extremality bound for non-BPS directions in a particular example, and show how the convex hull of BPS states contains the entire extremal region, satisfying the WGC without the need of appealing to non-BPS states.

\subsection{Example 1: one modulus case} \label{1mod}
We have seen that the general structure of the charge-to-mass vectors for D3-branes in Calabi-Yau compactifications of Type IIB is that of an ellipsoid with two non-degenerate radii, which are given in terms of the $\left(\frac{Q}{M}\right)\bigg|_{q_E^I}$. We will make this general structure concrete by analyzing the spectrum of BPS states in two specific compactifications. As a first example, we will find the spectrum of BPS states in the infinite distance limit of a compactification on a Calabi-Yau with $h^{2,1}=1$ and compute their charge-to-mass ratios. 

The necessary ingredients for computing the charge and mass of a BPS D3-brane are the gauge kinetic matrix and the central charge. These are both computed in terms of a prepotential, which must be specified. In a compactification with $h^{2,1}=1$, there are three possible asymptotic prepotentials, two of which correspond to different infinite distance limits. We will study the limit corresponding to a prepotential
\begin{align}
F= - \frac{(X^1)^3}{X^0},
\label{F1}
\end{align}
so that the periods are
\begin{align}
\begingroup
\renewcommand*{\arraystretch}{1.5}
\begin{pmatrix} X^I \\ F_I \end{pmatrix} = \begin{pmatrix} X^0 \\ X^1 \\  \frac{(X^1)^3}{(X^0)^2} \\ -\frac{3 (X^1)^2}{X^0} \end{pmatrix}.
\endgroup
\end{align}
In special coordinates ($X^0=1, \ X^1 = T$), the K\"ahler potential is
\begin{align}
\cK = -3 \log[i(\overline{T}-T)],
\end{align}
and the gauge kinetic matrix is the real part of
\begin{align}
\mathcal{N}_{IJ} = \left(
\begin{array}{cc}
- i t^3-3 i \theta ^2 t-2 \theta ^3 & 3 \theta ^2+3 i t \theta  \\
 3 \theta ^2+3 i t \theta  & -3 i t-6 \theta  \\
\end{array}
\right),
\end{align}
where we've set $T=\theta+i t$.
The central charge is
\begin{align}
Z = \frac{q_0 + q_1 T+3 p_1 T^2-p_0 T^3}{[-i (T-\overline{T})]^{3/2}}.
\end{align}

First, let's identify the electric charges and periods. In this example, the charges corresponding to states that become light in the infinite distance limit and their associated electric periods are:
\begin{align}
\begingroup
\renewcommand*{\arraystretch}{1.5}
\vec{q}_E = \begin{pmatrix} q_0 \\ q_1 \end{pmatrix}
\endgroup,
\begingroup \ \ \ \
\renewcommand*{\arraystretch}{1.5}
\vec{\Pi}_E = \begin{pmatrix} 1 \\ T \end{pmatrix}.
\endgroup
\end{align}

We can now compute the spectrum of the charge-to-mass vectors for each D3-brane. The magnitude of the charge-to-mass ratio for these states is:
\begin{align}
\bigg| \frac{Q}{M} \bigg|  = \frac{2}{\sqrt{3}} \sqrt{\frac{3 q_0^2+6   q_0 q_1\theta +q_1^2 \left(3 \theta ^2+t^2\right)}{q_0^2+2   q_0 q_1 \theta +q_1 ^2 \left(\theta ^2+t^2\right)}}.
\end{align}

\begin{figure}
\centering
\begin{subfigure}{.5\textwidth}
  \centering
  \includegraphics[width=0.8\linewidth]{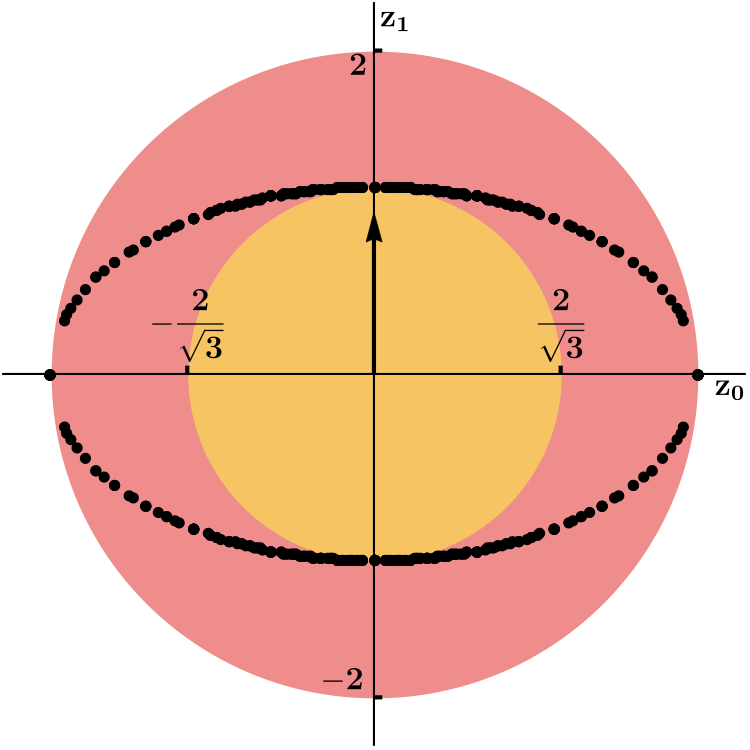}
  \caption{$(\theta, t) = (0,5)$}
  \label{bps1}
\end{subfigure}%
\begin{subfigure}{.5\textwidth}
  \centering
  \includegraphics[width=0.8\linewidth]{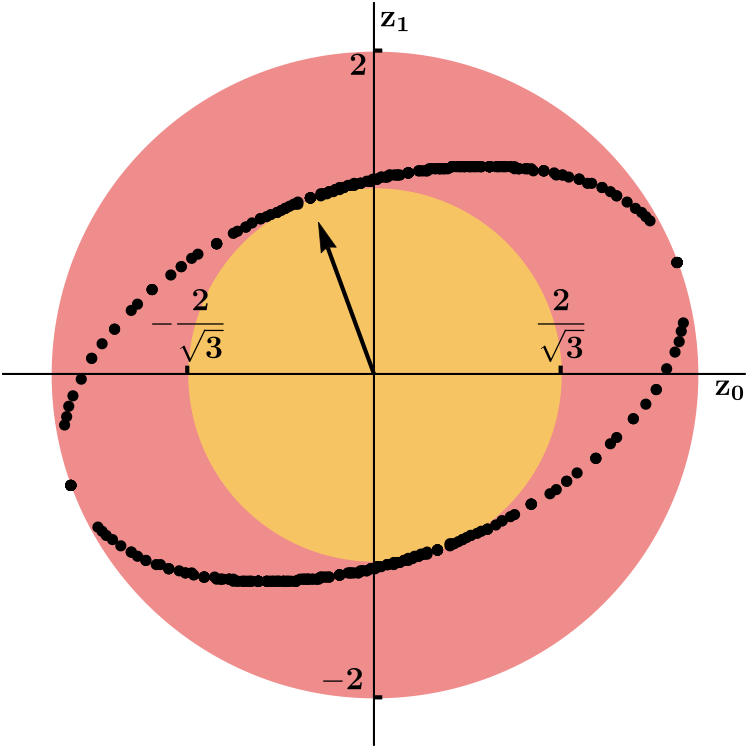}
  \caption{$(\theta, t) = (3,5)$}
  \label{bps1axion}
\end{subfigure}
\caption{The charge-to-mass ratios of BPS states in a one-modulus compactification. On the left, the axion is set to zero, while on the right, the axion is non-zero. The black arrow a principle axis of the ellipse.}
\label{elecstates1}
\end{figure}

The charge-to-mass vectors are defined as $\vec{z}=\left|\frac{Q}{M}\right| \hat{Q}$. Fig.~\ref{elecstates1} shows these vectors plotted in $\vec{z}$-space for this example for varying values of the axion. We can see that the $\vec{z}$ vectors lie on an ellipse with principal radii $\frac{2}{\sqrt{3}}$ and $2$. As the axion is shifted, the ellipse rotates and the states shift, but the values of the principal radii do not change. Additionally, we can see that as the modulus $t$ approaches $\infty$, the states move towards the $z_1$ axis, but the ellipse itself stays the same. This means that even though the charge-to-mass ratios of the individual BPS states change throughout moduli space, the shape formed by all BPS states together remains invariant!

The eigenvalues defining this ellipse can be read off from the $\mathbb{A}$ matrix defined in the previous section, which in this case is simply (setting the axions to zero)
\begin{align}
\mathbb{A} = \begin{pmatrix} \frac{1}{4} & 0 \\ 0 & \frac{3}{4} \end{pmatrix}.
\end{align}
Since the principal radii correspond to the square root of the inverse of the eigenvalues, we indeed get $\gamma_1=2$ and $\gamma_2=\frac{2}{\sqrt{3}}$, which accords with the figure.

Now we would like to compare this charge-to-mass spectrum of BPS states to the extremality bounds determined by the entries of the gauge kinetic function. In other words, we are checking that D3-branes which carry charge only under a single gauge field have charge-to-mass ratios that can be obtained from the entries of the asymptotic gauge kinetic matrix. Setting the axions to zero, the gauge kinetic matrix is
\begin{align}
\mathcal{I}_{IJ} = \begin{pmatrix} -t^3 & 0 \\ 0 & -3 t \end{pmatrix}.
\end{align}
In order to compute the extremality bound using \eqref{qm_ext}, we need to write the Lagrangian in coordinates where the scalar fields are canonically normalized. Following the conventions of \eqref{action1}, this means we want to redefine the scalar fields such that the action takes the form
\begin{align}
S = \int d^4 x \left[ \frac{R}{2}-\frac{1}{2}\partial_{\mu} \tilde{t}^{i} \partial^{\mu} \bar{\tilde{t}}^{j}+\tilde{\mathcal{I}}_{I J} \mathcal{F}_{\mu \nu}^{I} \mathcal{F}^{J, \mu \nu}+\tilde{\mathcal{R}}_{I J} \mathcal{F}_{\mu \nu}^{I}(\star \mathcal{F})^{J, \mu \nu} \right].
\label{canonicalaction}
\end{align}
To canonically normalize the scalar kinetic term in the Lagrangian, we choose new coordinates $\tilde{t} =\sqrt{\frac{3}{2}} \log{t}$. Rewriting the gauge kinetic matrix in terms of this field we get
\begin{align}
\tilde{\mathcal{I}}_{IJ} = \begin{pmatrix} -e^{\sqrt{6} \tilde{t}} & 0 \\ 0 & -3 e^{\sqrt{\frac{2}{3}}\tilde{t}} \end{pmatrix}.
\end{align}

Now we can use \eqref{qm_ext} to read off the extremality bounds associated with the single-charge black holes

\begin{align}
\left(\frac{Q}{M}\right)\bigg|_{q^0} = \sqrt{1+ \frac{1}{2} \times 6}=2 \\
\left(\frac{Q}{M}\right)\bigg|_{q^1} =  \sqrt{1+ \frac{1}{2} \times \frac23}=\frac{2}{\sqrt{3}}.
\end{align}

These values exactly correspond with the principal radii of the ellipse that the BPS states lie on. This is an example of how in the infinite distance limit, the charge-to-mass ratios of BPS states coincide with the extremality bound of black holes in that theory. Hence the ellipse in Fig.~\ref{elecstates1} can be identified with the black hole extremality bound, bounded by the above values of the principal radii.

Finally, because these two quantities coincide, we note that we can use the extremality bounds determined from the gauge kinetic function to bound the exponential factor in the Swampland Distance Conjecture
\begin{align}
\label{la1}
\frac{1}{\sqrt{6}} \leq \lambda \leq \sqrt{\frac{3}{2}}.
\end{align}
In this example, we know the exact spectrum at every point in moduli space near the infinite distance limit, so we can exactly compute the mass decay rate for a D3 brane with any given charges. However, this result can be thought of as a prototype. Here, we see how one can use the extremality bounds of black holes in the $\cN=2$ theory resulting from compactification of type IIB string theory on a Calabi-Yau to bound the mass decay rate in the Swampland Distance Conjecture. Any BPS state becoming light in this example will have an exponential mass decay rate satisfying \eqref{la1}.

\subsection{Example 2: two moduli case} \label{2mod}
Now we will repeat the above exercise for a compactification with $h^{2,1} =2$ where some of the subtleties of this exercise are more apparent. In particular, we will see that the extremal ellipsoid has a degenerate direction and that not all charge combinations give rise to BPS states.

For this example, we will study the prepotential
\begin{align}
F = -\frac{1}{6 X_0} T S^2.
\label{F2}
\end{align}

This gives rise to a complex gauge kinetic function
\begin{align}
\mathcal{N}_{IJ} = \left(
\begin{array}{ccc}
 -\frac{i (t-i \theta ) \left[(t+i \theta ) s^2+2 t \phi ^2\right]}{6 t} & \frac{i \theta  s^2+t \phi ^2}{6 t} & \frac{1}{3} (i t+\theta ) \phi  \\
 \frac{i \theta  s^2+t \phi ^2}{6 t} & -\frac{i s^2}{6 t} & -\frac{\phi }{3} \\
 \frac{1}{3} (i t+\theta ) \phi  & -\frac{\phi }{3} & -\frac{1}{3} i (t-i \theta ) \\
\end{array}
\right).
\end{align}

The central charge is (setting the axions to zero)
\begin{align}
Z =\frac{\sqrt{3}}{2\sqrt{s^2 t}} \left(\frac{1}{6} i p^0 s^2 t-\frac{p^1 s^2}{6}-\frac{p^2 s t}{3}+q_0+i q_1 t+i q_2 s \right).
\end{align}

By inspecting the above central charge, we can see that along different infinite distance limit paths, what we call ``electric" and ``magnetic" charges change. For example, in the limit $t\gg s^2 \rightarrow \infty$, the electric charges are $(q_0, \ q_2, \ p^1)$, while in the limit $s\gg t \rightarrow \infty$, the electric charges are $(q_0, \ q_1, \ q_2)$. For this example, we will choose the path $t \gg s^2 \rightarrow \infty$ so that we have

\begin{align}
\begingroup
\renewcommand*{\arraystretch}{1.5}
\vec{q}_E = \begin{pmatrix} p^1 \\ q_0 \\ q_1 \end{pmatrix}
\endgroup,
\begingroup \ \ \ \
\renewcommand*{\arraystretch}{1.5}
\vec{\Pi}_E = \begin{pmatrix} 1 \\ T \\ S^2 \end{pmatrix}.
\endgroup
\end{align}

Now we would like to know which combinations of charges can give rise to BPS states. To begin with, we compute the black hole potential $V_{bh}=Q^2$:
\begin{align}
V_{bh} = \frac{1}{12 s^2 t} \bigg(36 q_0^2 + 12 q_0 \phi(6 q_2+p^1 \phi) + 12 p^1 q_2 \phi(s^2+\phi^2) + (p^1)^2(s^2+\phi^2)^2 +18 q_2^2 (s^2+2\phi^2) \bigg).
\end{align}
BPS black hole solutions are given by solutions to the attractor flow equations \eqref{flow1}-\eqref{flow2} , which in this example are
\begin{align}
\dot{U} &= -e^U |Z| \\
\dot{t} &= -2 t e^U |Z| \\
\dot{\phi} & = - e^U \frac{q_2  (6 q_0 + p_1 s^2)}{4 t} \frac{1}{|Z|} \\
\dot{s} &= -e^U \frac{(36 q_0^2 - p_1^2 s^4)}{12 s t} \frac{1}{|Z|}.
\end{align}
We can see from the above equations that the scalars flow from their initial values at spatial infinity to fixed values on the horizon of the black hole. These fixed values are given by the critical point equations
\begin{align}
D_T Z &=0 \\
D_S Z &=0.
\end{align}
Any solution of these equations must have $t\rightarrow \infty$ while $\theta$ is unconstrained. On the other hand, the effective potential for $s$ and $\phi$ is non-trivial, and they will flow to finite minima. The physical solutions of these equations for $s$ and $\phi$ read
\begin{align}
&s=\sqrt{-\frac{6 q_0 - 6 q_2 \phi - p^1 \phi^2}{p^1}}, \ \phi \ \text{unconstrained} \\
&s= \pm \frac{\sqrt{3}}{p^1}\sqrt{-2 p^1 q_0 + 3 q_2^2} , \ \phi = -\frac{3 q_2}{p^1}.
\end{align}
The restriction that $s$ must take a positive value on the horizon of the black hole imposes non-trivial constraints on the charges. Taking the mildest restriction that still allows for a physical solution to the BPS equations, we see that we need
\begin{align}
3 q_2^2> 2 p^1 q_0.
\label{chargecondition}
\end{align}
%
The charge-to-mass ratio for the electric states is
\begin{align}
\label{QM2moduli}
\bigg| \frac{Q}{M} \bigg| = 2 \sqrt{\frac{ \left(p_1^2 \left(s^2+\phi ^2\right)^2+12 q_0 \phi  (p_1 \phi +6 q_2)+12 p_1 q_2 \phi  \left(s^2+\phi ^2\right)+36 q_0^2+18 q_2^2 \left(s^2+2 \phi ^2\right)\right)}{\left(s^2+\phi ^2\right) \left(p_1^2 \left(s^2+\phi ^2\right)+12 p_1 q_2 \phi +36 q_2^2\right)+12 q_0 \left(p_1 \left(\phi ^2-s^2\right)+6 q_2 \phi \right)+36 q_0^2}}.
\end{align}

\begin{figure}
\centering
\includegraphics[width=0.5\textwidth]{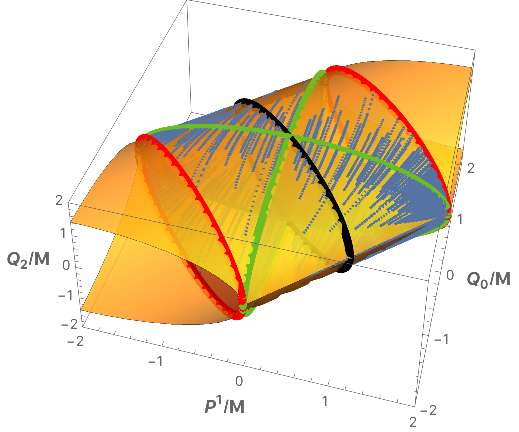}
\caption{The charge-to-mass vectors for BPS states in a compactification with $h^{2,1}=2$. The black line indicates the ellipse whose principal radii are given by Eqs.~\ref{eigenvalue1ex2} and \ref{eigenvalue2ex2}, the red ellipses indicate the outer boundaries of the BPS region, and the green ellipses indicate the boundaries $p^1 q_0 =0$ associated with the single-charge states.}
\label{bps2}
\end{figure}
We can once again plot the charge-to-mass vectors $\vec{z}$. This is shown in Fig.~\ref{bps2}, for the case where the axions are set to zero, subject to the charge restrictions derived from the attractor flow equations. We can see that the states lie on a degenerate ellipse with principal radii $\sqrt{2}$ and $\sqrt{2}$. This ellipse is defined by the matrix
\begin{align}
\mathbb{A} = \begin{pmatrix}\phantom{-} \frac{1}{4} & -\frac{1}{4} & 0 \\ -\frac{1}{4} & \phantom{-} \frac{1}{4} & 0 \\ 0 & 0 & \frac{1}{2} \end{pmatrix}.
\end{align}
In the degenerate direction, the charge-to-mass ratios appear to grow without bound. However, these are precisely the directions where the attractor flow equations forbid BPS states. Note that imposing the charge condition \eqref{chargecondition} imposes an upper bound on the charge-to-mass ratio
\begin{align}
\bigg| \frac{Q}{M} \bigg| \leq 2.
\end{align}

Now we would like to once again compare these values to the black hole extremality bounds obtained from the entries of the gauge kinetic matrix. In canonically normalized coordinates, we have (setting the axions to zero)
\begin{align}
\mathcal{I}_{IJ} = \begin{pmatrix} -\frac{1}{6} e^{2 s+ \sqrt{2} t} & 0 & 0 \\ 0 & -\frac{1}{6} e^{2 s - \sqrt{2} t} & 0 \\ 0 & 0 & -\frac{1}{3} e^{\sqrt{2}t} \end{pmatrix}.
\end{align}
Reading off the dilatonic factor $\vec \alpha$ we get:
\beq
\vec\alpha_0=(2,\sqrt{2})\ ,\quad  \vec\alpha_1=(2,-\sqrt{2})\ ,\quad \vec\alpha_2=(0,\sqrt{2}),
\eeq
yielding
\begin{align}
\left(\frac{Q}{M}\right)\bigg|_{p^1} = 2, \ \ \left(\frac{Q}{M}\right)\bigg|_{q_0} = 2, \ \ \left(\frac{Q}{M}\right)\bigg|_{q_2} = \sqrt{2},
\end{align}
where we have used the extremality bound of a dilatonic black hole given in \eqref{qm_ext}. It can be checked that the same values for $\left(\frac{Q}{M}\right)_i$ can be equivalently computed from \eqref{QM2moduli} by only turning on a charge at each time,
\begin{align}
\left(\frac{Q}{M}\right)\bigg|_{p^1}= \bigg|\frac{Q}{M}\bigg|_{p^1\neq 0}, \ \ \left(\frac{Q}{M}\right)\bigg|_{q_0}= \bigg|\frac{Q}{M}\bigg|_{q_0 \neq 0}, \ \ \left(\frac{Q}{M}\right)\bigg|_{q_1}= \bigg|\frac{Q}{M}\bigg|_{q_1\neq 0}.
\end{align}
Hence, once again, the extremality and the BPS bound coincide for these states, as expected, satisfying \eqref{aZ}.
We can now use these single-charge states to compute the principal radii of the ellipse
\begin{align}
&\gamma_1 = \left[\left(\frac{Q}{M}\right)\bigg|_{p^1}^{-2} +\left(\frac{Q}{M}\right)\bigg|_{q_0}^{-2}\right]^{-1/2} = \sqrt{2} \label{eigenvalue1ex2} \\
&\gamma_2 = \left[\left(\frac{Q}{M}\right)\bigg|_{q_2}\right]^{-1/2} = \sqrt{2},
\label{eigenvalue2ex2}
\end{align}
where we have used eqs.~\eqref{lambda1}-\eqref{lambda2}.

Thus we see that the charge-to-mass ratio is bounded
\begin{align}
\sqrt{2} \leq \frac{|Q|}{M} \leq 2.
\end{align}
Note that the bounds on $\frac{|Q|}{M}$ do \textit{not} imply a lower bound on $\lambda$. This is because moduli space now has more than one direction, so the rate at which the BPS masses change depend on the path one takes to infinite distance. We can still place a bound on $\lambda$, in fact, but we will leave this to section~\ref{asymptotic} when we revisit this example, once we have built up the machinery of asymptotic Hodge theory and described in full detail what \textit{does} constitute a bound on $\lambda$.

\subsection{Magnetic charge and dyons} \label{dyons}

Up to now, we have focused on the electrically charged BPS states, since they are the ones that become light asymptotically and fulfill the Distance Conjecture. However, we can also analyze the structure of the charge-to-mass ratio of BPS states with magnetic charges and their interplay with the WGC. In particular, let us now consider the case of dyonic BPS states.

The main qualitative difference between pure electric states and dyons is the fact that the central charge (and the black hole potential $V_{bh}$) of an electric state has runaway behavior towards infinite distance, while the central charge of a dyon has a minimum at finite distance in the moduli. 
Let us exhibit this in the one modulus example discussed in section~\ref{1mod}. Recall that this example had $h^{2,1}=1$ and $F= -\frac{(X^1)^3}{X^0}$, and let us consider now states that carry both electric and magnetic charge. The condition for such a state to be BPS is:
\begin{align}
D_T Z = \partial_T |Z| = 0.
\end{align}
There are two classes of dyonic solutions to this equation:
\begin{align}
&t=-i \sqrt{\frac{q_0}{p^1}}, \ \ q_0 \ p^1 < 0, \ (q_1, \ p^0) = (0, \ 0) \\
&t = \sqrt{\frac{q_1}{3 p^0}}, \ \ q_1 \ p^0 > 0, \ (q_0, \ p^1) =(0, \ 0). \label{dyoncharges}
\end{align}
Let us consider solutions of the second type. The charge-to-mass ratio for these BPS extremal dyonic black holes  is:
\begin{align}
\frac{|Q|}{M} = \frac{2}{\sqrt{3}} \sqrt{\frac{q_1^2+3 p_0^2 t^4}{(q_1+p_0 t^2)^2}}.
\label{dyonqm}
\end{align}
where we have set the axions to zero for simplicity, since they will not affect the discussion.
Note that in this example, the shape that the charge-to-mass vectors form is not an ellipse as in section~\ref{1mod}, but rather two straight lines (see figure \ref{convexhull_dyons}), which can, of course, be thought of as an ellipse with only one degenerate direction. This is because unlike in the case with only electric charges, the matrix $\mathbb{A}$ need not split into two non-trivial sub-blocks: it has one zero eigenvalue and one non-zero eigenvalue. This occurs because both charges $q_1,p^0$ are associated to purely imaginary periods in the central charge, so $\gamma_1^{-2}=0$ in \eqref{lambda1}. In physical terms, all of these dyonic states are \textit{mutually BPS}, since their central charges all have the same phase: none of these black holes feel any forces due to one another. This is a generic feature that can be extrapolated to other setups: whenever the states are mutually (non-mutually) BPS we expect them to lie on a plane (ellipse) in charge-to-mass ratio space (see e.g. \cite{Heidenreich_2016}). Note that as in the previous subsection, the charge restrictions \eqref{dyoncharges} have the effect of truncating the degenerate direction and bounding the charge-to-mass ratio. In section \ref{nonbps} we will discuss how to compute the extremality bound for the non-BPS directions, completing the other two quadrants in figure \ref{convexhull_dyons}.

It is also interesting to notice that unlike in the case of purely electrically charged states, the dilatonic contribution $\alpha^2$ to the extremality bound in \eqref{extbound} for dyons does not have a lower bound. Recall that since these BPS states are also extremal, $\alpha^2$ is equal to the scalar charge-to-mass ratio, which in this case is given by
\begin{align}
\frac{\alpha^2}{2} &= 4\frac{K^{T\overline{T}} \partial_T |Z| \partial_{\overline{T}} |Z|}{|Z|^2}= \frac{(q_1-3 p^0 t^2)^2}{3(q_1+p^0 t^2)^2}.
\end{align}
For electrically charged states, this quantity could never be zero, since that would mean that the mass of the state does not go to zero exponentially in the asymptotic limit. However, it can vanish for dyonic BPS sates, as we can see from the previous equation which has a zero at $t=\sqrt{\frac{q_1}{3p^0}}$. Hence, there are some particular BPS states for which $|Q|/M=1$. Notice that this solution (where $t$ takes this value at spatial infinity) corresponds to a ``double extremal black hole," wherein the values of the moduli stay constant from spatial infinity to the horizon of the black hole, as explained in \cite{Kallosh_1996}. Clearly, these states do not satisfy the scalar WGC \eqref{ScalarWGC}, which hints at the fact that this conjecture should only apply to states that become exponentially massless asymptotically, underlying the lower bound of exponential mass decay rate of the SDC. We will discuss more about this topic in the next section.

\section{General bounds in the asymptotic limit }  \label{asymptotic}

In the previous section, we saw that the charge-to-mass vectors lie on degenerate ellipsoids with exactly two non-degenerate directions for electrically charged states, which can be computed in terms of the diagonal entries of the gauge kinetic function. We calculated the numerical results in two examples, but in this section we will take a more systematic approach in providing numerical results for the charge-to-mass ratios and the bound on the SDC factor in any type of infinite field distance limit in these moduli spaces. Instead of performing a case-by-case study, we will use the mathematical machinery of asymptotic Hodge theory. In essence, this machinery provides the tools to classify all possible $\cN=2$ prepotentials that can arise near the boundaries of the complex structure moduli space of a Type II Calabi-Yau compactification. Furthermore, although calculating the exact spectrum of the charge-to-mass vectors can, in general, be cumbersome, the leading scalar dependence of the gauge kinetic function and central charge can be determined in full generality in the infinite distance limit using these techniques.

We will review these techniques in section~\ref{MHS} and apply them to the gauge kinetic function (which is necessary in order to compute the charge-to-mass ratio vectors) in section~\ref{bps}. We will also see how there is natural splitting of charges that arises asymptotically and corresponds to the special basis of charges used in section~\ref{bpsextsame}. In section~\ref{decayrate}, we will use these results to provide a general bound for the exponential factor of the Distance Conjecure for any number of moduli. Finally, since everything renders clearer in examples, we will reproduce the results of section~\ref{1mod} and \ref{2mod} using this mathematical machinery in section~\ref{MHSexamples}.

\subsection{Review: asymptotic Hodge theory techniques} \label{MHS}

We will begin with reviewing the techniques of asymptotic Hodge theory (see e.g. \cite{schmid,CKS,CattaniKaplan,Kerr2017}) applied to Calabi-Yau threefolds. These techniques have recently been applied to Calabi-Yau manifolds in \cite{Grimm_2018,Grimm_2019,Corvilain_2019,grimm2019infinite,Font:2019cxq,grimm2019asymptotic,Cecotti:2020rjq} to constrain the physics that arises in the asymptotic limits of moduli space and to check the Swampland conjectures. We refer the reader to those papers for more details.

In general, it is a tremendous task to compute the periods $\{X^I,F_I \}$ necessary to obtain the field metric and the central charge of BPS states at arbitrary points in field space. Fortunately, it is possible to give a local expansion near any infinite distance singularity of the moduli space using the theorems of asymptotic Hodge theory. These theorems allow us to determine the asymptotic form of the central charge, which is crucial for our argument, and so we review them next.

Consider the asymptotic limit given by sending $n$ scalar fields (which correspond to volumes of intersecting divisors) to large values,
\beq
T^i\rightarrow i\infty\ ,\quad i=1,\dots,n.
\eeq
It is possible to classify the types of such asymptotic limits based on the properties of the monodromy transformations around them. These monodromies are discrete transformations which are part of the duality group of the moduli space and arise from higher dimensional gauge invariance. More concretely, when circling around the singular locus, the periods undergo a monodromy transformation given by
\beq
\label{monod}
\Pi(T^i+1)=\mathbb{T}_i\ \Pi(T^i).
\eeq
From the EFT point of view, this corresponds to a discrete shift of the axion $\text{Re}(T^i)$. This is more than a symmetry---it corresponds to a redundancy, since it is part of the duality group, implying that the physics of the EFT should remain invariant. However, it becomes a continuous global symmetry in the infinite distance limit. This provides an underlying reason for the existence of a tower of light particles: it is a quantum gravity obstruction to restoring a global symmetry, as proposed in \cite{Grimm_2018}. The properties of these discrete transformations allow us to classify the different infinite distance limits and will eventually determine the exponential mass decay rate of the infinite SDC tower.

In the following, we review three key results of asymptotic Hodge theory which will be crucial in computing the charge-to-mass ratios of BPS states in theories with any number of moduli.

\begin{itemize}

\item The Nilpotent Orbit Theorem gives a universal expansion of the periods at the asymptotic limit given by
\beq
\Pi=e^{\sum_iT^iN_i}a_0(\zeta)+ \mathcal{O}(e^{2\pi i T}),
\label{nilp}
\eeq
where $a_0$ is a holomorphic function and $\zeta$ are the coordinates not sent to the limit. The nilpotent operators $N_i$ are constructed as $N_i=\log \mathbb{T}_i$, where $\mathbb{T}_i$ are the monodromy transformations in \eqref{monod}. Their nilpotent nature guarantees that the exponential $e^{TN}$ can be expanded as a polynomial with a finite number of terms, and the order of the highest term determines the singularity type. More concretely, the singularity type will be labeled as a roman numeral $X$ where $X=d+1$ with $d$ an integer defined as 
\begin{align}
N_{(i)}^{d_i} a_0 \neq 0 \ \text{and} \ N_{(i)}^{d_i+1}a_0=0 ,
\label{Nd}
\end{align} 
and upper bounded by the complex dimension of the internal space, $d\leq dim_C(CY_3)=3$. When multiple moduli approach an asymptotic limit, such a singularity type is associated to each modulus, and $N_{(i)}\equiv N_1+N_2+\dots N_i$.

By plugging the expansion \ref{nilp} of the periods into the K\"ahler potential, we find that $K$ is given to leading order by
\begin{align}
\label{K}
\cK=-\text{log}\left[t_1^{d_1} t_2^{d_2-d_1} \ .\  . \  . \  t_n^{d_n - d_{n-1}}+\dots\right]\ .
\end{align}

\item
As any singularity is approached, the third cohomology group $H^3(X,\mathbb{Z})$ degenerates into a Deligne splitting:
\begin{align}
H^3 (X,\mathbb{Z}) = \bigoplus_{p,q=0}^3 I^{p,q}.
\end{align}
This corresponds to a finer splitting of the total vector space into smaller subspaces $I^{p,q}$ adapted to the nilpotent operator $N$ such that
\beq
N I^{p,q} \subset I^{p-1,q-1}
\eeq 
and preserving certain polarization conditions \cite{Kerr2017,Grimm_2019}.

Pictorially, the dimensions of the groups of this finer splitting can be arranged into a so-called \textit{limiting Hodge diamond}:

{\centering
\begin{tikzpicture}
\node at (0+0.1,0+0.4){$i^{0,0}$};
\node at (0+0.1,1.414+0.4){$i^{1,1}$};
\node at (0+0.1,2.828+0.4){$i^{2,2}$};
\node at (0+0.1,4.243+0.3){$i^{3,3}$};
\node at (0.707+0.1,2.121+0.4){$i^{1,2}$};
\node at (-0.707+0.1,2.121+0.4){$i^{2,1}$};
\node at (-2.121+0.1,2.121+0.4){$i^{3,0}$};
\node at (2.121+0.1,2.121+0.4){$i^{0,3}$};
\node at (0.707+0.1,0.707+0.4){$i^{0,1}$};
\node at (-0.707+0.1,0.707+0.4){$i^{1,0}$};
\node at (1.414+0.1,1.414+0.4){$i^{0,2}$};
\node at (-1.414+0.1,1.414+0.4){$i^{2,0}$};
\node at (1.414+0.1,2.828+0.4){$i^{1,3}$};
\node at (-1.414+0.1,2.828+0.4){$i^{3,1}$};
\node at (0.707+0.1,2.828+0.707+0.4){$i^{2,3}$};
\node at (-0.707+0.1,2.828+0.707+0.4){$i^{3,2}$};
\draw[step=1cm,black,very thin,rotate=45] (0,0) grid (3,3);
\end{tikzpicture} \par
}

where each diagonal sum is equal to the dimension of the corresponding cohomology group:
\begin{align}
h^{p,3-p} = \sum_{i=0}^3 i^{p,i}.
\end{align}

If the manifold we are compactifying is a Calabi-Yau threefold, then the limiting Hodge diamond can be simplified. In order to fully specify the levels of the diamond, we need only specify two numbers:
\begin{itemize}
\item The value of $d$ for which $i^{3,d}=1$, which can be derived from \eqref{Nd}.
\item The value of $i^{2,2}$
\end{itemize}
To the first number we assign a capital roman numeral $X$ as explained above ($d=0 \leftrightarrow I, \ d=1 \leftrightarrow II, \ d=2 \leftrightarrow III, \ d=3 \leftrightarrow IV$), and the second we write as a subscript on the Roman numeral. Thus, when we talk about a ``singularity type," we are referring to a choice $X_{i^{2,2}}$.

When we approach a singular locus given by multiple scalar fields taking large field values ($t^i\rightarrow \infty$), we have an ``enhancement chain" specifying the enhanced singularities that arise from intersecting divisors each approaching singularites:
\beq
X_{i^{2,2}} \ \rightarrow \  X'_{i'^{2,2}} \rightarrow \dots,
\label{chain}
\eeq
with as many steps as scalar fields are sent to infinity. The order is chosen such that the first one corresponds to the coordinate $t$ that grows fastest, and so on. Note that in specifying such an enhancement chain, one has the choice between whether to specify the singularities that each individual divisor approaches, or to specify the enhanced singularity that arises from intersecting those divisors. In this paper, we choose the latter notation. For each of term in the enhancement chain we have a limiting Hodge diamond, and the fact that the polarization conditions must carry over imposes important constraints on the allowed enhancement chains. Therefore, all this machinery not only allows us to classify the individual singularity types, but also to classify the enhancement chains. This classification of singularities has been done in \cite{Kerr2017,Grimm_2019} for Calabi-Yau threefolds and in \cite{grimm2019asymptotic} for fourfolds.


\item The growth theorem provides the asymptotic behaviour of the Hodge norm of general three-forms in the infinite distance limit $v_I\in H^3 (X,\mathbb{R}) $ in terms of the singularity type and the growth sector. In particular, 
\begin{align}
 \int v_I \wedge \star v_I \sim \left(\frac{t_1}{t_2} \right)^{\ell_1^I-3} ...\left(\frac{t_{n-1}}{t_n}\right)^{\ell_{n-1}^I -3} (t_n)^{\ell_n^I-3}\ ||v_I||^2_\infty,
\label{hodgenorm}
\end{align}
where $\ell_i^I$ are the levels in which the corresponding $v_I$ lives in the limiting Hodge diamond as follows:

{\centering
\begin{tikzpicture}
\draw[step=1cm,black,very thin,rotate=45] (0,0) grid (3,3);
\draw[dashed, thick] (-3,0) -- (3,0);
\node at (3.5,0){$\ell=0$};
\draw[dashed, thick] (-3,0.707) -- (3,0.707);
\node at (3.5,0.707){$\ell=1$};
\draw[dashed, thick] (-3,1.414) -- (3,1.414);
\node at (3.5,1.414){$\ell=2$};
\draw[dashed, thick] (-3,1.414+0.707) -- (3,1.414+0.707);
\node at (3.5,1.414+0.707){$\ell=3$};
\draw[dashed, thick] (-3,1.414+2*0.707) -- (3,1.414+2*0.707);
\node at (3.5,1.414+2*0.707){$\ell=4$};
\draw[dashed, thick] (-3,1.414+3*0.707) -- (3,1.414+3*0.707);
\node at (3.5,1.414+3*0.707){$\ell=5$};
\draw[dashed, thick] (-3,1.414+4*0.707) -- (3,1.414+4*0.707);
\node at (3.5,1.414+4*0.707){$\ell=6$};
\end{tikzpicture} \par
}

Here we are assuming that $v_I$ belongs to single subspace with a definite $\ell$; otherwise, one has to sum the contribution to the Hodge norm from all of them, as we explain below.

This result is actually associated to a much deeper theorem, known as the sl(2)-orbit theorem \cite{CKS,Kerr2017}, which implies that the full vector space $H^3 (X,\mathbb{R})$ can be split into orthogonal components
\beq
\label{splitting}
H^3 (X,\mathbb{R})=\bigoplus V_\ell\ ,\quad \ell=(\ell_1,\dots,\ell_n),
\eeq
satisfying $\int_X v_\ell \wedge v_{\ell'} = 0$ unless $\ell+\ell'=6$. The Hodge norm for $v_\ell\in V_\ell$ behaves as \eqref{hodgenorm} while the non-diagonal entries $\int v_\ell \wedge \star v_{\ell'}\sim 0$ vanish to leading order up to polynomial corrections that are subleading as long as we stay within a growth sector of the form
\beq
\label{growth}
R=\left\{\frac{t^1}{t^2}>\mu, \frac{t^2}{t^3}>\mu, \dots , t^n>\mu , \quad \mu>0\right\},
\eeq 
i.e., with a specific order regarding what scalars grow fastest. Each ordered sequence of this type is associated to a different enhancement chain \eqref{chain}. Hence, for each particular enhancement chain, the Hodge norm of any general 3-form can be split asymptotically into the sum of different contributions given by \eqref{hodgenorm}.

\eqref{hodgenorm} will be the most important theorem for us, as it allows us to determine the asymptotic growth of the central charge and the gauge kinetic function, since the latter can be written in terms of the Hodge norm of a symplectic basis of 3-cycles (recall \eqref{intmat}). The above theorem implies that in the asymptotic limit we can always find a special basis such that the scalar dependence of the diagonal entries of the gauge kinetic function are simply given by \eqref{hodgenorm}, while the non-diagonal entries are suppressed.
Furthermore, the scalar dependence is simply determined in terms of a list of integers $\ell$ characterizing the singular limit. This is the special basis of charges already envisaged in section \ref{bps}.

\end{itemize}

Since we can now compute the moduli dependence of the K\"ahler potential and central charge in the infinite distance limit in terms of the singularity type, we have enough information to completely characterize the $\left(\frac{Q}{M}\right)\bigg|_{q^I}$ extremality bounds that we need to compute the principal radii of the BPS charge-to-mass ellipse, and with it to bound the exponential factor of the SDC.

\subsection{Asymptotic form of the charge-to-mass ratio} \label{ctomellipse}

Let us first use the Nilpotent Orbit Theorem \eqref{nilp} to provide the asymptotic form of the central charge for BPS states \cite{Grimm_2018,Grimm_2019},
\beq
\label{Z}
Z=e^{\cK/2}\langle q,\Pi\rangle= e^{\cK/2}\sum_{k=0}\langle q, \frac{1}{k!}\left(\sum_{i=1}^n(it^i+\theta^i)N_i\right)^ka_0\rangle +\mathcal{O}(e^{-2\pi  t}),
\eeq
where $\langle\cdot,\cdot\rangle$ denotes the symplectic product and $N_i$ is the nilpotent matrix associated to the monodromy transformations \eqref{monod}. Since these are nilpotent matrices, the sum on $k$ involves a finite number of terms, determined by the values of $d_i$ in \eqref{Nd}. For simplicity, we will set the axions to zero from now on, but their dependence can be trivially recovered by replacing $q\rightarrow \rho=e^{\theta^iN_i}q$, which is a vector of shift invariant functions that depends on the quantized charges and the axions. More specifically, we can group the terms in $Z$ as follows:
\beq
\label{Z2}
Z= e^{\cK/2}\sum_{k=0} \sum_{k_1 k_2\dots k_n} \frac{1}{k_1! \dots k_n!} \langle q,N_1^{k_1} N_2^{k_2}\dots N_n^{k_n} a_0\rangle (it_1)^{k_1} (it_2)^{k_2}\dots (it_n)^{k_n} +\mathcal{O}(e^{-2\pi  t}),
\eeq
with the constraint that $\sum_i^n k_i = k$ and $0\leq k_i\leq d_i$. The next step is to choose a basis of charges in order to make the previous result more concrete and compute the symplectic products. To this end, we will choose the basis of charges which is associated to the asymptotic splitting of the charge lattice \eqref{splitting} in the infinite distance limit. This asymptotic splitting into orthogonal subspaces is guaranteed by the sl(2)-orbit theorem and it will enormously simplify the discussion. In terms of this special basis, the central charge can be simply written as
\beq
\label{Zsl2}
Z=\sum_\ell (it_1)^{\frac{\ell_1-3}{2}} (it_2)^{\frac{\ell_2-\ell_1}2}\dots (it_n)^{\frac{\ell_n-\ell_{n-1}}{2}} \ c_\ell  \ q_\ell +\mathcal{O}(t^{i+1}/t^i)+\mathcal{O}(e^{-2\pi  t}),
\eeq
where $c_\ell$ are constants that can only depend on the scalars that are not taken to a large field limit. This result appeared already in \cite{grimm2019infinite} (or equivalently in \cite{grimm2019asymptotic} for the superpotential) and arises from rewriting the nilpotent orbit of periods in terms of the sl(2,C)-data that appears in the growth theorem \eqref{hodgenorm}. Here, $q_\ell$ are the quantized charges arising as coefficients from expanding a given charge $q$ into the sl(2) basis \eqref{splitting}. Hence, the central charge of a state with a charge $q$ that belongs to a single subspace $V_\ell$ has a well-defined leading order behaviour given by a monomial of the scalars with degrees fixed by the integer vector $\ell$. On the contrary, if the charge $q$ has components in several subspaces $V_\ell$, then we must sum over all of them, as in \eqref{Zsl2}. Notice that the nilpotent orbit result for the central charge \eqref{Z} is exact up to exponentially suppressed corrections, while the sl(2) orbit result \eqref{Zsl2} is exact up to polynomial corrections which are suppressed in the growth sector \eqref{growth}. Hence, it only provides the leading term for each charge $q_\ell$, which satisfies
\beq
\label{QM}
2 k_i-(d_i-d_{i-1})=\ell_1-\ell_{i-1}
\eeq
by comparing \eqref{Z} and \eqref{Zsl2}, where $\ell_0\equiv 3$ and $d_0\equiv 0$. This approximation of keeping only the leading term was denoted the strict asymptotic limit in \cite{grimm2019asymptotic}.

Let us now compute the charge $|Q|^2 = -\frac{1}{2} \vec{\mathfrak{q}}^T \mathcal{M} \vec{\mathfrak{q}}$, where $\mathcal{M} $ is given in terms of the gauge kinetic function in \eqref{Mmatrix}. Since $ \mathcal{M} $ can be computed in terms of the Hodge norm of a symplectic basis of 3-forms, we just need to apply the growth theorem \eqref{hodgenorm} to obtain
\beq
\label{Qgrowth}
|Q|^2=\sum_\ell  t_1^{\ell_1-3} t_2^{\ell_2-\ell_1}\dots t_n^{\ell_n-\ell_{n-1}} \ c'_\ell \ q^2_\ell,
\eeq
where $c'_\ell$ is another constant which is finite in the infinite distance limit, since it does not depend on $t^i$. For electric charges, the non-vanishing block of $\mathcal{M}$ is equal to $\mathcal{I}^{-1}$ where $\mathcal{I}=\text{Im} \mathcal{N}$ is the imaginary part of the gauge kinetic function for the electric field strengths, so \eqref{Qgrowth} provides the scalar behaviour of the gauge coupling. Notice that \eqref{Zsl2} and \eqref{Qgrowth} have the same dependence on the scalar fields, implying that both the mass and the gauge coupling exhibit the same asymptotic behaviour. 
This reflects the fact that BPS states are extremal in the asymptotic limits. In other words, it shows that the RFC bound and the extremality bound coincide at any asymptotic limit of the moduli space. 
Note that the term ``single-charge states'' introduced in section \ref{bps} has a very clear meaning here:
\begin{itemize}
\item Single-charge state: a charge vector $q$ that belongs to a single subspace $V_\ell$; i.e. it has a fixed $\ell$.
\end{itemize}
Only single-charge BPS states will exhibit a $Z$ and $Q$ corresponding to the same monomial in the scalars, satisfying \eqref{aZ} which is given by the following numerical value:
\beq
\label{alphagrowth}
\alpha^2=8 \frac{ K^{i \bar{j}} \partial_{i} |Z| \partial_{\bar{j}} |Z|}{|Z|^2}=2 \sum_{i=1}^n\frac{(\ell^I_i - \ell^I_{i-1})^2}{d_i-d_{i-1}},
\eeq
where we have used \eqref{K} to derive the asymptotic behaviour of the field metric $K_{i \bar{i}}=\frac{d_i-d_{i-1}}{4t^2_i}$ and $n$ is the number of moduli sent to the large field limit. Let us recall that $\alpha$ is defined as the dilatonic contribution to the gauge kinetic function, i.e. $\mathcal{L}\supset \frac12 e^{\alpha_i \phi_i}F^2$. Hence, the above result can also be obtained directly from the asymptotic scalar behaviour of the gauge kinetic function\footnote{Note that determining the charge-to-mass ratio by computing $|Q|$ and $M$ separately requires one to have knowledge of the currently undetermined coefficients $c_\ell$ and $c_\ell'$.} \eqref{Qgrowth} when written in terms of the canonically normalized fields $\phi^i=\sqrt{\frac{d_i-d_{i-1}}{2}} \log t_i$. 


Hence, by plugging \eqref{alphagrowth} into \eqref{extbound} we get the following charge-to-mass ratio for the single-charge BPS states,
\begin{align}
\label{QMratio}
\left(\frac{Q}{M}\right)\bigg|_{q^I}^2 = 1+\sum_{i=1}^n\frac{(\ell^I_i - \ell^I_{i-1})^2}{d_i-d_{i-1}}\ .
\end{align}
Let us summarize our findings. In general, the charge-to-mass ratio of a BPS state will depend on the value of the moduli, but as we approach an asymptotic limit in moduli space it is possible to split the lattice of charges into orthogonal subspaces such that a BPS state with charges belonging to only one of these subspaces will have a well-defined charge-to-mass ratio that is independent of the moduli and given by \eqref{QMratio}. We have denoted these states as single-charge BPS states, but one has to keep in mind that the notion \emph{single-charge} refers only to this specific basis of charges \eqref{splitting} adapted to the $sl(2)$-orbit theorem. This proves, based on the theorems of asymptotic Hodge theory, the requirement in section \ref{bps} assumed to construct the ellipse in the $Q/M$-plane. 

In section \ref{generalellipse}, we found that the charge-to-mass ratio of the states always form a degenerate ellipsoid with only two non-degenerate directions, regardless of the number of fields. The degenerate directions will be cut at some point in order to preserve the charge restrictions that allow for the existence of BPS states, so in practice the information for the charge-to-mass ratio of BPS states is encoded in a single ellipse.
Recall that to calculate the principal radii of this ellipse, which are given by the eigenvalues of $\mathbb{A}$ in \eqref{Amatrix}, we need to sum the $\left(\frac{Q}{M}\right)\bigg|_{q^I}$ corresponding to charges associated to purely imaginary and real terms in $Z$. From the expansion \eqref{Z} is clear that the terms with $k$ even (odd) will be purely real (imaginary). Hence, the question of whether a given period is purely real or purely imaginary corresponds to whether
\beq
\label{rule}
k = \sum_{i=1}^n k_i=\frac12 (\ell_n-3+ d_n)
\eeq
is even or odd. Here, $l_n-3$ precisely corresponds to the sum of all the powers of the corresponding entry of the gauge kinetic function, as can be seen in \eqref{Qgrowth}. Thus, the eigenvalues of $\mathbb{A}$ are given completely in terms of the following discrete data associated to the singular limit:
\begin{align}
\gamma_1^{-2} = \sum_{\ell \ | \ k \ \text{even}}\frac{1}{1+\sum_{i=1}^n\frac{(\ell_i-\ell_{i-1})^2}{d_i-d_{i-1}}}\label{Aeigenvalue1} \\
\gamma_2^{-2} = \sum_{\ell \ | \ k \ \text{odd}}\frac{1}{1+\sum_{i=1}^n\frac{(\ell_i-\ell_{i-1})^2}{d_i-d_{i-1}}} .\label{Aeigenvalue2}
\end{align}
These eigenvalues provide the complete structure for $|Q|/M$ of light BPS states in the asymptotic limit. All we need to know to compute them are the allowed values of $\ell$ associated to each singular limit, which have been classified in \cite{Kerr2017,Grimm_2019} for Calabi-Yau threefolds. Hence, we can explicitly give the numerical result for the extremality bound of electric black holes in $\cN=2$ in any asymptotic limit of field space! We would like to remark that even though the charge-to-mass ratio of any particular BPS state depends on the moduli, the ellipsoid representing the extremality bound in the $|Q|/M$-plane remains the same as we move towards the asymptotic limit. In other words, there is always another state which takes the place of the previous one such that even though the individual charge-to-mass ratios change, altogether they always trace out the same ellipsoid.

The eigenvalues \eqref{Aeigenvalue1} and \eqref{Aeigenvalue2} correspond to the principal radii of the smallest 2-dimensional ellipse that one can construct on the BPS ellipsoid, so they can be used to provide a lower bound for the charge-to-mass ratio of any BPS state,
\beq
\frac{|Q|}{M}
\label{boundQM}
\geq \text{min}(\gamma_1,\gamma_2) .
\eeq
There will also be an upper bound, but this will be fixed by the charge restrictions to support BPS states, which must be calculated example by example, as explained in section \ref{bps}.

For later use, notice that $\ell_i$ is bounded such as $3-d_i\leq \ell_i\leq 3+d_i$ \cite{Grimm_2018,Grimm_2019}, implying a bound on $\alpha^2$ in \eqref{alphagrowth} as follows:
\beqa
\label{boundl1}
0\leq \frac{(\ell_i - \ell_{i-1})^2}{d_i-d_{i-1}}\leq \frac{(d_i+d_{i-1})^2}{d_i-d_{i-1}} \quad \text{ for } d_i-d_{i-1}\text{ even}\\
\frac{1}{d_i-d_{i-1}}\leq \frac{(\ell_i - \ell_{i-1})^2}{d_i-d_{i-1}}\leq \frac{(d_i+d_{i-1})^2}{d_i-d_{i-1}}\quad \text{ for } d_i-d_{i-1}\text{ odd},\label{boundl2}
\eeqa
where $d_i\leq dim_C(Y_3)=3$ $\forall i$ and we have used that $\ell_i$ is even (odd) only if $ d_i-d_{i-1}$ is even (odd) as well. It is important to remark that these bounds apply to any BPS state---electric, dyonic or magnetic. If we wish to restrict ourselves to the states that become light in the asymptotic limit, we must further impose $\ell_n<3$ and $\ell_i\leq 3$ for $i<n$. This is the rule to identify a state that becomes light along any path in the growth sector \cite{Grimm_2019,Corvilain_2019}. These light states always have only electric charges since this rule also implies that the associated gauge coupling goes to zero asymptotically. Hence, even if some light state can have $\alpha_i=0$ for some modulus according to \eqref{boundl1}, it cannot vanish $\forall i$ so $\alpha^2\neq 0$.

A final comment is in order. The reader is probably wondering what happens if $d_i-d_i-1=0$, since in that case, the above formulae diverge. This is typical when there is a finite distance divisor involved in the enhancement chain in the singular limit. When this occurs, the corresponding scalar field does not appear in the leading order term of the K\"ahler potential and that is why the canonical normalization procedure fails. In that case, one should go to next to leading order in $\cK$ by analyzing the nilpotent orbit of all periods and not just the nilpotent orbit of the (3,0)-form, so the exponent will be related to the subtype of the singularity\footnote{We thank Thomas Grimm for pointing this out and useful discussions in this regard.}. We leave this for future work.

\subsection{General bounds on mass decay rate} \label{decayrate}

In the previous subsection, we described how to compute the charge-to-mass structure of light BPS states in the asymptotic limits of moduli space. We found that these charge-to-mass vectors sit on degenerate ellipsoids, and that we could compute the form of the ellipsoid using the type of infinite distance singularity that is being approached. In this section, we will use the same methods to sharpen the statement of the SDC. The Distance Conjecture predicts the existence of an infinite tower of states becoming exponentially light at every infinite field distance limit. However, the exponential mass rate of the tower is an unspecified parameter $\lambda$, supposedly of order one. Since the evidence for the SDC comes from examples in string compactifications, it seems difficult to give a universal value for this order one parameter. Nevertheless, if the tower is charged under some gauge field becoming weakly coupled, it will also satisfy the WGC, and thus we can extract the order one factor from the black hole extremality bound. The goal of this section is to provide the exact numerical value of this parameter for any infinite distance limit in which the SDC tower consists of BPS states that also satisfy the WGC. Notice that even if there are towers becoming light even faster than the BPS states in these limits, the mass decay rate for BPS states can still be used as a lower bound for $\lambda$ and provide a conservative bound on the maximum field range before the EFT breaks down. So in any case, studying the field-dependence of the BPS masses gives us very valuable information and helps us to sharpen the Distance Conjecture.

Let us remark that the exponential factor of the SDC for the case of a single field limit (only one scalar field taking large field values) in Calabi-Yau compactifications was already derived in \cite{Grimm_2018}. Here, we aim to give a bound on this factor for the case of multi-field limits in which we can have an arbitrary number of scalar fields taking large field values. Path dependence issues complicate the identification of $\lambda$, but we will still be able to give lower bounds. Using the results from the previous sections, we can see that in the asymptotic limit, the K\"ahler potential factorizes to leading order and the field space splits into a direct product of hyperbolic spaces. Any trajectory in which the axions are frozen ($\text{Re}( T^i)=0$) corresponds to a geodesic in field space. Since the metric is diagonal to leading order, we can then define a gradient of the mass given by
\beq
\vec \nabla M\equiv (\sqrt{g^{11}}\partial_1 M,\sqrt{g^{22}}\partial_2 M,\dots,\sqrt{g^{nn}}\partial_n M)
\eeq
and express the exponential factor of the mass decay rate in \eqref{SDC} as
\beq
\label{SDCfactor}
\lambda=\left|\frac{\nabla_i M}{M}u_i\right|,
\eeq
where $u_i$ is a unit vector along a geodesic trajectory. Note that $\frac{|\nabla M|}{M}$ is the scalar charge-to-mass ratio we defined in section~\ref{swampyscalars}. 

Let us first collect the information we have about the possible values for $\frac{\nabla_i M}{M}$.
In order to apply the results from the previous section, we need first to divide the space of trajectories into different growth sectors \eqref{growth}, in which an order of the growth of the scalars is specified. Associated to each growth sector there is an enhancement chain of singularities and a list of integers $\ell$ that completely determine the leading moduli dependence of the BPS masses and gauge couplings. When approaching an infinite field distance limit along a geodesic trajectory that belongs to a growth sector of the form \eqref{growth}, we have learned that single-charge BPS states satisfy
\beq
\label{scalarcharge}
\left(\frac{\nabla_i M}{M}\right)_{q^I}=\sqrt{2\frac{K^{i\bar{i}}\partial_i |Z| \partial_{\overline{i}}  |Z| }{|Z|^2}}=\frac{1}{\sqrt{2}} \frac{|\ell_i^I-\ell^I_{i-1}|}{\sqrt{d_i-d_{i-1}}},
\eeq
where no sum is implicit. This indeed corresponds to an $\mathcal{O}(1)$ parameter fixed by the discrete data associated to the singular limit. Recall that this is equal to the exponential rate of the gauge kinetic function $\alpha_i$ (see \eqref{alphagrowth}) and implies that the mass of a single-charge BPS state can be simply written as $M \sim M_0\ e^{-\frac12\sum_i\alpha_i\phi^i}$ to leading order, in terms of the canonically normalized scalars.

General BPS states with charge belonging to several subspaces $V_\ell$ in \eqref{splitting} will not exhibit such a simple expression. In this case, the mass is roughly given as a sum of exponential and there can be cancellations that reduce the value of $\frac{\nabla_i M}{M}$. Hence, it becomes very difficult to give a concrete lower bound for $\lambda$ that applies to \emph{any} BPS state in higher dimensional moduli spaces. However, it is possible to give a bound for \emph{certain} BPS states. In other words, for every geodesic trajectory in field space, we can find some BPS states (the single-charge states) becoming light and satisfying \eqref{scalarcharge}. If one can show that among this set of BPS states, there is one which corresponds to the first state of an infinite tower becoming light, then one can bound the SDC factor $\lambda$ using \eqref{scalarcharge}.

In order to properly address this issue, we need to discuss which charges are actually populated by physical BPS states.
As a reminder, \eqref{scalarcharge} gives the value of $\frac{\nabla_i M}{M}$ that a BPS state with those charges would have, but this does not mean that those charges are necessarily populated by a physical state. In general, as remarked in section \ref{bpsreview}, identifying which sites in the charge lattice are populated by physical BPS states is a hard question and an important topic of research. In \cite{Grimm_2018} it was proposed that monodromy orbits of BPS states should in fact count as stable\footnote{By studying the presence of walls of marginal stability, it was found in \cite{Grimm_2018} that the number of stable BPS states in the orbit increases exponentially as approaching the infinite field distance limit.} BPS states close enough to the singular limit, and therefore be identified with the SDC tower. These monodromy orbits are BPS towers constructed by acting successively with the monodromy transformations $\mathbb{T}_i$ over a seed charge $q_s$ \cite{Grimm_2018,Grimm_2019},
\beq
q_{\text{tower}}\equiv \exp\left(\sum_i n_i N_i\right)q_s,
\label{Qorbit}
\eeq 
 where recall $N_i=\log \mathbb{T}_i$ as defined below \eqref{nilp}\footnote{Strictly speaking, we should replace $N_i$ by $N_i^-$ which are part of the commuting $sl(2)$ algebras as explained in \cite{Grimm_2019}.}. The decay rate of the tower will be set by the first state in the tower, i.e. by the seed charge $q_s$. This seed charge alone corresponds to a single-charge state with fixed $\ell$ so it will satisfy \eqref{scalarcharge}. Its concrete value of $\ell$, though, needs to be identified case by case for each type of limit (see \cite{Grimm_2018} and \cite{Grimm_2019} for its identification in one-modulus and two-moduli limits respectively). Typically, though, it corresponds to one of the charges with the smallest possible value of \eqref{scalarcharge}, while still corresponding to a light state. Recall that the lower bounds for $\frac{\nabla_i M}{M}$ for single-charge states were given in \eqref{scalarcharge}, and small values of $\frac{\nabla_i M}{M}$ correspond to the largest values of $\ell$ that a BPS state can have while still becoming asymptotically massless. The smallest non-zero value that a single-charge BPS state can take is given by
  \beq
  \label{dim}
\left. \frac{\nabla_i M}{M}\right|_{min} = \frac{1}{\sqrt{2}} \frac{1}{\sqrt{d_i-d_{i-1}}}\left\{\begin{array}{c} 2\quad \text{ for } d_i-d_{i-1}\text{ even} \\ 1\quad \text{ for } d_i-d_{i-1}\text{ odd}\end{array}\right .
  \eeq

Once we have a bound for $\frac{\nabla_i M}{M}$, we need to discuss how much it can change when projected along a geodesic trajectory as in \eqref{SDCfactor}, i.e. how sensitive $\lambda$ is to the specific trajectory followed in field space. Notice that a state with $\frac{\nabla_i M}{M}$ nearly orthogonal to the geodesic trajectory will have an incredibly small $\lambda$. The goal now is to show there are always certain BPS states for which $\lambda$ remains of order one, and provide then a sensitive lower bound for $\lambda$.

For concreteness, let us consider a geodesic trajectory within the growth sector \eqref{growth} of the form
\beq
(t^1,t^2,\dots ,t^n)=(\mu^{n-1},\mu^{n-2},\dots ,1)
\eeq
with $\mu>1$. Then the exponential factor of the SDC \eqref{SDCfactor} reads
\beq
\lambda=\sum_{i=1}^n\left|\frac{\nabla_i M}{M}\mu^{n-i}\right|\frac{1}{\sqrt{1+\dots+\mu^{2n}}}.
\eeq
This is minimized when the trajectory has very large $\mu$ so that only the first term $i=1$ contributes in practice, so that the lower bound is given by
\begin{tcolorbox}
\beq
\label{bound}
\lambda\geq \frac{\nabla_1 M}{M}=\frac{1}{\sqrt{2}} \frac{|l_1-3|}{\sqrt{d_1}},
\eeq
\end{tcolorbox}
where we have used \eqref{scalarcharge}. This corresponds to the value of the scalar charge-to-mass ratio along the direction that grows the fastest, which is denoted as $t_1$ in the growth sector \eqref{growth}.
We can see that the concrete value will depend on the discrete data associated to the singular limit. However, using \eqref{dim} we can give a lower bound for the smallest $\lambda$ that one can find in Calabi-Yau threefolds, obtaining
\beq
\label{min}
\lambda_{min}=\frac{1}{\sqrt{2}\sqrt{d_n-d_{n-1}}}\geq \frac{1}{\sqrt{6}},
\eeq
where we have used that $d_n\leq dim_C(CY_3)=3$ and $d_{n-1}\leq d_n$. The case of $d_n=3$ corresponds to the maximally unipotency order, i.e. the large complex structure point. It can be checked that the same bound actually applies to any geodesic of the corresponding growth sector. We would like to emphasize that up to this point, in which we replaced $d\leq 3$, the discussion was general and independent of Calabi-Yaus. The growth theorem \eqref{hodgenorm} is based on asymptotic Hodge theory which is valid beyond Calabi-Yau manifolds; it is only on the specific values of $\ell,d$ that the choice of the internal manifold matters.

It is also important to note that the lower bound for $\nabla_i M/M$ might get modified if corrections that make the field metric non-diagonal become important and involve negative contributions. However, this occurs when $d_i-d_{i-1}=0$ for some of the singular divisors, which we leave for future work, as discussed at the end of section \ref{ctomellipse}. All these subtleties are particularly manifest when considering limits in higher dimensional moduli spaces in which not all moduli are sent to infinity. However, in those cases, there is something which plays in our favour, and it is that corrections to the field metric tend to make the effective result for $d_1$ smaller \cite{lee2016hodge}, increasing $\lambda$. Of course, a more refined analysis is necessary before concluding anything for these cases. Fortunately, asymptotic Hodge theory provides the tools to deal with all these path dependence issues, and it is just a matter of continuing to explore this mathematical framework.

Notice that this same bound was already given in \cite{Grimm_2018} for the case of single field limits, and only now we have been able to show its generality for multi-moduli limits. While completing this work, the evidence for this bound in Calabi-Yau manifolds \cite{Grimm_2018} was used in \cite{Andriot:2020lea} to argue that the numerical value of $1/\sqrt{6}$ should work as a general lower bound for any EFT in four dimensions, in connection with the Transplanckian Censorship Conjecture (TCC) \cite{Bedroya:2019snp}. Indeed, we can compare our lower bound 
in \eqref{bound} with the prediction for the SDC factor coming from the TCC in \cite{Bedroya:2019snp}, 
\beq
\label{TCC}
\lambda_{TCC}=\frac{2}{D\sqrt{(D-1)(D-2)}}=\frac{1}{2\sqrt{6}}
\eeq
where we have replaced the space-time dimension $D=4$ in the last step. Hence, our lower bound in \eqref{min} is only twice this TCC value. This numerical coincidence was also noticed in  \cite{Andriot:2020lea}, and it was used to propose to reduce the TCC value by a factor of $1/2$ by arguing that the potential should scale as $M^2$. However, unlike in \cite{Andriot:2020lea} 
we do not expect this correlation between \eqref{min} and \eqref{TCC} to hold necessarily in higher dimensions, since the degree of the singularity $d_n$ is not directly related to the space-time dimension, but rather to the internal dimension of the compactification space.

It is interesting to notice that the geodesic maximizing \eqref{SDCfactor} is the one parallel to the vector $\frac{\nabla_i M}{M}$, implying an exponential mass decay rate 
\beq
\label{dm2}
\lambda^2 |_{v_i=\frac{\nabla_i M}{M}}=\frac{|\nabla M|^2}{M^2}=2 \frac{K^{i\bar{j}} \partial_i |Z| \partial_{\bar{j}} |Z|}{|Z|^2}\geq \frac12(\gamma^2_{\text{min}}-1),
\eeq
where we have used \eqref{boundQM} and denoted $\gamma_{\text{min}}$ as the smaller of $\gamma_1$ and $\gamma_2$ in \eqref{Aeigenvalue1}-\eqref{Aeigenvalue2}.
Notice that this is a general bound for the charge-to-mass ratio of \textit{any} BPS state, and not just the single-charge ones. This bound also corresponds to a bound on the dilatonic factor in the extremality bound \eqref{extbound}, since $\alpha^2=4\frac{|\nabla M|^2}{M^2}$. This is a generic feature in the sense that whenever there is a gauge coupling vanishing at infinite field distance, one can use the extremality bound to provide a bound on the scalar charge-to-mass ratio \eqref{dm2} as well. However, this bound is not directly related to a lower bound on the SDC factor $\lambda$ unless there is only one scalar field. In a single field limit, the issues of path dependence disappear and the SDC factor is simply given by \eqref{dm2}. However, in higher dimensional moduli spaces, \eqref{dm2} corresponds to the exponential mass decay rate of a tower whose scalar charge-to-mass ratio is pointing in the same direction than the tangent vector of the geodesic trajectory, maximizing \eqref{SDCfactor}. Hence, unfortunately, it cannot be used as a general lower bound for the SDC factor, although the upper bound on $\frac{|\nabla M|^2}{M^2}$ that arises from imposing the BPS charge restrictions could be used as a general upper bound for the SDC factor $\lambda$.

In any case, the scalar charge-to-mass ratio is an interesting quantity in itself, as it is the protagonist of the Scalar WGC proposed in \cite{Palti_2017}.
Thus, before closing this section, we would like to comment on the realization of the scalar WGC in our setup.
Notice that, in fact, the vector \eqref{ScalarWGC} corresponds to a scalar charge under the long-range scalar force that arises from having the mass of the particle parametrized by a massless scalar field. As explained around \eqref{ScalarWGC}, it was conjectured in \cite{Palti_2017}, in analogy to the WGC, that there should always be at least one particle in which the scalar force acts stronger than the gravitational force for every massless scalar field. This implies here that some states should satisfy,
\beq
K^{ij}\nabla_j M\nabla_i M > M^2 \ \rightarrow\ \frac{|\vec\nabla M|^2}{M^2}> \frac12
\eeq
where recall $K_{ij}=\frac12 g_{ij}$. This was shown to be automatically satisfied in $\cN=2$ in the following way~\cite{Palti_2017}: the matrix $\text{Im}(F_{IJ})$ has $n_V$ positive eigenvalues and a single negative one. Therefore, the $\cN=2$ identity $Q(F)^2 =|Z|^2 - K^{i \bar{j}} D_i Z D_{\bar{j}} Z$ implies that the scalar force is greater than the gravitational force for each direction in charge space, save for one. This does not mean that every BPS state satisfies the bound but that one can always find some BPS state along that scalar direction that does. In particular, we can see that states saturating the lower bound in \eqref{boundl1}-\eqref{boundl2} will not satisfy the scalar WGC, while states saturating the upper bound in those equations will.
 It was also proposed that if this conjecture holds for some tower of states, then the SDC factor of this tower is lower bounded by 1. However, this correlation is less clear in multi-moduli limits because of the reasons explained above, since this scalar charge-to-mass ratio only gives information about a tower maximizing and not minimizing $\lambda$ in the space of trajectories. But in the presence of several towers becoming light, it might be relevant to the leading tower, i.e. the one with the largest value of $\lambda$.

In this subsection, we described the relevant aspects of asymptotic Hodge theory and showed how this machinery can be used to completely determine the BPS charge-to-mass structure in the asymptotic regimes of moduli space. In addition, we demonstrated a lower bound on the mass decay rate of certain towers of states in the infinite-distance limit and provided a numerical value for this bound in terms of the integers characterizing the type of infinite-distance singularity that is being approached. In the following subsection, we will make these structures and bounds concrete by applying these techniques to the examples described in section~\ref{bps}.

\subsection{Examples} \label{MHSexamples}

In this section, we will rederive the results for the examples in sections \ref{1mod} and \ref{2mod} using the mathematical machinery explained above, as concrete applications of the asymptotic Hodge theory techniques will help to clarify its usage. As we will see, it is much easier to get the bounds on the charge-to-mass ratios and the SDC factor this way, since all the information we need is the moduli dependence of the diagonal entries of the gauge kinetic function which can be derived from the list of integers $\ell$ characterizing each infinite field distance limit. One goal of this section is to demonstrate how useful and practical these techniques are for describing the physics near infinite-distance singularities.\\

\textbf{Example 1: One modulus case}

Let us start with the one modulus example of section \ref{1mod}.
In compactifications with a single modulus ($h^{2,1}=1$), there are two types of infinite-distance singular limits we can take: Type $II_0$ or $IV_1$. The prepotential studied in \eqref{F1} corresponds to singularity type $IV_1$, which is also known as the large complex structure limit. By definition, this limit has:
\begin{align}
d = 3.
\end{align}
The limiting Hodge diamond has the form in figure~\ref{1moddiamond} (where dots with no number are shorthand to signify a dimension of $1$), which implies that the total vector space $H^3(X,\mathbb{R})$ splits into the following possible subspaces labeled by $\ell$ in table \ref{1modtable}.

\begin{figure}[H]
\centering
\begin{tikzpicture}
\node at (0,1.75) {$1$};
\node at (0.707,2.4) {$0$};
\node at (-0.707,2.4) {$0$};
\node at (0,3.2) {$1$};
\node at (0,0)[circle,fill,inner sep=1.5pt]{};
\node at (0,1.414)[circle,fill,inner sep=1.5pt]{};
\node at (0,2.828)[circle,fill,inner sep=1.5pt]{};
\node at (0,4.243)[circle,fill,inner sep=1.5pt]{};
\node at (0.707,2.121)[circle,fill,inner sep=1.5pt]{};
\node at (-0.707,2.121)[circle,fill,inner sep=1.5pt]{};
\draw[step=1cm,black,very thin,rotate=45] (0,0) grid (3,3);
\end{tikzpicture}
\caption{}
\label{1moddiamond}
\end{figure}

\begin{table}[h!]
  \begin{center}
    \label{tab:table1}
    \begin{tabular}{|l|c|r} 
    \hline
      $\ell$ & $\text{dim}(V_{\ell})$ \\
      \hline
      0 & 1\\
      2 & 1 \\
           4 & 1 \\
                 6 & 1 \\
      \hline
    \end{tabular}
  \end{center}
  \caption{}
  \label{1modtable}
\end{table}
According to \eqref{Qgrowth}, the asymptotic behaviour of the gauge kinetic function is simply given by
\beq
Q^2=-\frac12 \vec{q}^T \mathcal{M} \vec{q} = \sum_\ell t^{\ell-3} c'_\ell q^2_\ell
\eeq
where only the components with $\ell=0,2$ (i.e. $\ell<3$) will  have a Hodge norm that vanishes in the limit $t\rightarrow \infty$, implying that the gauge kinetic function becomes large and therefore the gauge coupling goes to zero at the infinite field distance limit. We will now focus on states charged only under these gauge fields (i.e. electric charges) since they will be the ones becoming light as it is clear from \eqref{Zsl2}.
Using these values and \eqref{Mmatrix} we read off the diagonal entries of the gauge kinetic matrix:
\begin{align}
\label{coup1}
\mathcal{I}^{-1}_{00} \sim t^{\ell^0-3} = t^{-3} \\
\mathcal{I}^{-1}_{11} \sim t^{\ell^1-3} = t^{-1},
\end{align}
just as we computed in section~\ref{1mod}. The K\"ahler potential is also easily computed in terms of $d$ using \eqref{K}:
\begin{align}
\cK=-3 \text{log}(t).
\end{align}

Having pinned down the K\"ahler potential and the diagonal entries of the gauge kinetic function, we can read off the single-charge ratios $\left(\frac{Q}{M}\right)\bigg|_{q_I}$ using \eqref{QMratio}:
\begin{align}
\left(\frac{Q}{M}\right)\bigg|_{q_0} = \sqrt{1+\frac{(\ell^0-3)^2}{d}} = 2 \\
\left(\frac{Q}{M}\right)\bigg|_{q_1} = \sqrt{1+\frac{(\ell^1-3)^2}{d}} = \frac{2}{\sqrt{3}},
\end{align} 
just as we found in section~\ref{1mod}. These ratios correspond to single-charge BPS states, i.e. those with charges belonging only either to the subspace with $\ell^0=0$ or $\ell^1=2$. In this case, these values also correspond to the radii of the charge-to-mass ellipse, implying that the charge-to-mass ratio of any BPS state is bounded by
\beq
 \frac{2}{\sqrt{3}}\leq \left(\frac{Q}{M}\right)\leq 2\ .
 \eeq
 Regarding the exponential mass rate of the Distance Conjecture, in the one modulus case one simply has
 \beq
\frac{1}{\sqrt{6}}\leq\frac{ \nabla M}{M}\leq \sqrt{\frac{3}{2}},
\eeq
where we have used \eqref{scalarcharge}. The minimum and maximum values for the SDC factor $\lambda=\frac{\nabla M}{M}$ correspond then to the result for the single-charge states. The tower given by the monodromy orbit of BPS states in \cite{Grimm_2018} has charge \eqref{Qorbit} given by
\beq
q_{\text{tower}}= q_1+ nq_0,
\eeq
where we have denoted $q_0\equiv q_{\ell=0}$ and $q_1\equiv q_{\ell=2}$ to match with the notation in section \ref{bps}. Hence, the first state in the tower (the seed charge) is charged under $F_{\mu \nu}^1$ (i.e. $\ell=2$). This implies that the SDC factor for the tower, which is determined by the seed charge, is given by
\beq
\lambda_{min}=\frac{1}{\sqrt{6}}.
\eeq

\begin{figure}
\centering
\includegraphics[width=0.4\textwidth]{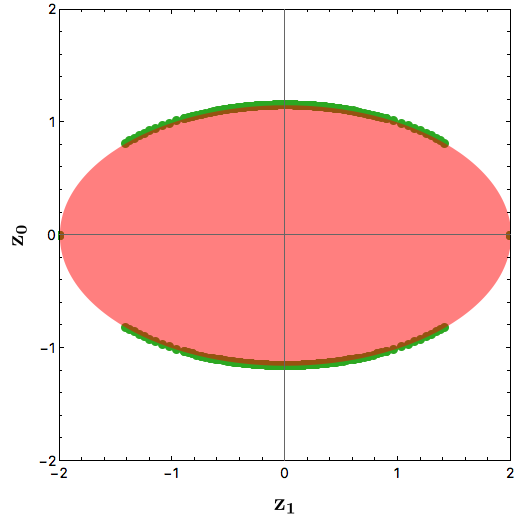}
\caption{The charge-to-mass vectors for BPS states in a single field limit with $d=3$. The red ellipse indicates the extremality region, while the green points represent only those BPS states which are stable against crossing walls of marginal stability.}
\label{mono}
\end{figure}

We are also in the position of answering whether the WGC can be satisfied in this setup with stable single-particles states or needs unstable resonances or multi-particle states. In \cite{Grimm_2018} it was argued that only part of the monodromy tower corresponds to stable BPS states, while above the species cut-off they can fragmentate to lighter BPS sates in the tower due to the presence of walls of marginal instability. It was found that the number of stable BPS states grow as $N\sim t$ as we approach the infinite field distance limit, so it actually increases exponentially in terms of the proper field distance. In figure \ref{mono} we have represented in green only the states in the tower which correspond to stable particles. The first thing to notice is that the region they occupy in the charge-to-mass ratio plane is always the same, regardless of the point in the moduli space: the bigger $t$ is, the smaller the charge-to-mass ratio of an individual state is, but the more stable states we have. This was already noted in \cite{grimm2019infinite}, and we refer the reader to this paper for an explanation of this behavior in more general terms based on asymptotic Hodge theory, valid also in multi-field limits. However, in order to conclude whether they suffice to satisfy the convex hull WGC, one also needs to determine the extremality region, which was not done in \cite{grimm2019infinite}. In view of our results, it is clear that stable BPS states are \textit{not} enough to satisfy the convex hull WGC, as they always cut part of the ellipse, so one needs to consider also unstable resonances, as usually happens in the Tower WGC. In other words, we need to take into account the full infinite tower of charges. Hence, from now on in our figures (and in all the previous figures), we will always plot the full lattice of charges that is compatible with supporting BPS states.

For completeness, we would also like to give the results for other types of infinite distance singularities associated to taking only a single modulus to be large. More generally, one gets
\beq
\sqrt{1+ \frac{1}{d}}\leq \left(\frac{Q}{M}\right)\leq \sqrt{1+d}\ 
 \eeq
and
 \beq
\frac{1}{\sqrt{2d}}\leq\frac{ \nabla m}{m}\leq \sqrt{\frac{d}{2}},
\eeq
with
\beq
\lambda_{min}=\frac{1}{\sqrt{2d}}
\eeq
where the integer can be $d=1 \ \text{or} \ 3$ in a Calabi-Yau threefold\footnote{Although in a Calabi-Yau with a single modulus there is no limit corresponding to $d=2$, in a Calabi-Yau with more moduli, it is possible to take a single field limit that has $d=2$ while keeping the rest of the moduli finite. In this case, the bounds are simply:
\beq
 \left(\frac{Q}{M}\right)=\sqrt{3}\quad \text{and}\quad \lambda=\frac{\nabla m}{m} = 1\ .
\eeq} and we have used \eqref{dim}. This matches with the results already obtained in \cite{Grimm_2018}.\\

\textbf{Example 2: two moduli case}

As a second example, let us compute the asymptotic form of the gauge kinetic function, the K\"ahler potential, the charge-to-mass ellipse, and the SDC factor for the two-modulus case studied in section~\ref{2mod}. This example corresponds to a Type $II_1$ singularity intersecting a Type $III_0$ singularity to yield a Type $IV_2$ singularity. Let us denote the scalar fields as $t,s$ such that the Type $II_1$ (Type $III_0$) singularity is located at $t\rightarrow \infty$ ($s\rightarrow \infty$). There are two possible growth sectors, depending on which modulus grows faster:

a) Growth sector: $t\gg s\gg 1$\\
This corresponds to the enhancement chain 
\beq
II_1\rightarrow IV_2
\eeq
with
\begin{align}
d_1 = 1, \ \ \ d_2 = 3,
\end{align}
and limiting Hodge diamond given by figures~\ref{sing1} and \ref{sing2}. From these diamonds, we read off the $\ell$ and $||v_\ell||^2$ values given in table~\ref{tab21}.
\begin{figure}[h!]
\centering
\begin{subfigure}{.45\textwidth}
  \centering
 \begin{tikzpicture}
\node at (0,1.75) {$1$};
\node at (0.707,2.4) {$0$};
\node at (-0.707,2.4) {$0$};
\node at (0,3.2) {$1$};
\node at (1.414,2.828)[circle,fill,inner sep=1.5pt]{};
\node at (0,1.414)[circle,fill,inner sep=1.5pt]{};
\node at (0,2.828)[circle,fill,inner sep=1.5pt]{};
\node at (-1.414,2.828)[circle,fill,inner sep=1.5pt]{};
\node at (0.707,2.121)[circle,fill,inner sep=1.5pt]{};
\node at (-0.707,2.121)[circle,fill,inner sep=1.5pt]{};
\node at (-1.414,1.414)[circle,fill,inner sep=1.5pt]{};
\node at (1.414,1.414)[circle,fill,inner sep=1.5pt]{};
\draw[step=1cm,black,very thin,rotate=45] (0,0) grid (3,3);
\end{tikzpicture}
\caption{Type $II_1$ singularity}
\label{sing1}
\end{subfigure}%
\begin{subfigure}{.45\textwidth}
  \centering
\begin{tikzpicture}
\node at (0,1.75) {$2$};
\node at (0.707,2.4) {$0$};
\node at (-0.707,2.4) {$0$};
\node at (0,3.2) {$2$};
\node at (0,0)[circle,fill,inner sep=1.5pt]{};
\node at (0,1.414)[circle,fill,inner sep=1.5pt]{};
\node at (0,2.828)[circle,fill,inner sep=1.5pt]{};
\node at (0,4.243)[circle,fill,inner sep=1.5pt]{};
\node at (0.707,2.121)[circle,fill,inner sep=1.5pt]{};
\node at (-0.707,2.121)[circle,fill,inner sep=1.5pt]{};
\draw[step=1cm,black,very thin,rotate=45] (0,0) grid (3,3);
\end{tikzpicture}
\caption{Type $IV_2$ singularity}
\label{sing2}
\end{subfigure}
\caption{}
\label{fig:test}
\end{figure}
 
\begin{table}[h!]
\begin{center}
\begin{tabular}{|c|c|c|c|c|c|c|}
\hline
&\multicolumn{2}{|c|}{$V_{\text{light}}$}&\multicolumn{2}{|c|}{$V_{\text{rest}}$}&\multicolumn{2}{|c|}{$V_{\text{heavy}}$}\\
\hline
$(\ell_1,\ell_2)$&(2,0)&(2,2)&(2,4)&(4,2)&(4,4)&(4,6)\\
\hline
$||v_\ell||^2$&$1/(ts^2)$&$1/t$&$s^2/t$&$t/s^2$&$t$&$ts^2$\\
\hline
$q_\ell$&$q_0$&$q_2$&$p^1$&$q_1$&$p^2$&$p^0$\\
\hline
\end{tabular}
\end{center}
\caption{Limit $II_1\rightarrow IV_2$}
\label{tab21}
\end{table}%

In the last row of the table, we have labeled the charges belonging to the different subspaces $V_\ell$ in order to match the notation in section \ref{bps}.
The elements in $V_{\text{light}}$ will always have a vanishing Hodge norm at the limit, $||v_\ell||^2\rightarrow 0$, regardless of the path (as long as we stay in the growth sector $t>s$). They are defined as having $\ell_i\leq 3, \ell_n<3$. Elements in $V_{\text{rest}}$ will vanish depending on the path (in this case depending on whether $t>s^2$), while elements in $V_{\text{heavy}}$ will always diverge.
If we take $t\gg s^2$ as in section \ref{2mod}, the diagonal entries of the electric gauge kinetic function are given by
\begin{align}
&\mathcal{I}_{00}^{-1} \sim t^{\ell^2_1-3} s^{\ell^2_2-\ell^2_1} = s^2/t\quad \rightarrow \quad Q^2\bigg|_{p^1}\sim (p^1)^2 s^2/t \\
&\mathcal{I}_{11}^{-1} \sim t^{\ell^0_1-3} s^{\ell^0_2-\ell^0_1} = 1/(ts^2)\quad \rightarrow \quad Q^2\bigg|_{q_0}\sim q_0^2 /(ts^2) \\
&\mathcal{I}_{22}^{-1} \sim t^{\ell^1_1-3} s^{\ell^1_2-\ell^1_1} = 1/t\quad \rightarrow \quad Q^2\bigg|_{q_2}\sim q_2^2 /t  .
\end{align}

\begin{figure}
\centering
\begin{subfigure}{.5\textwidth}
  \centering
  \includegraphics[width=0.95\linewidth]{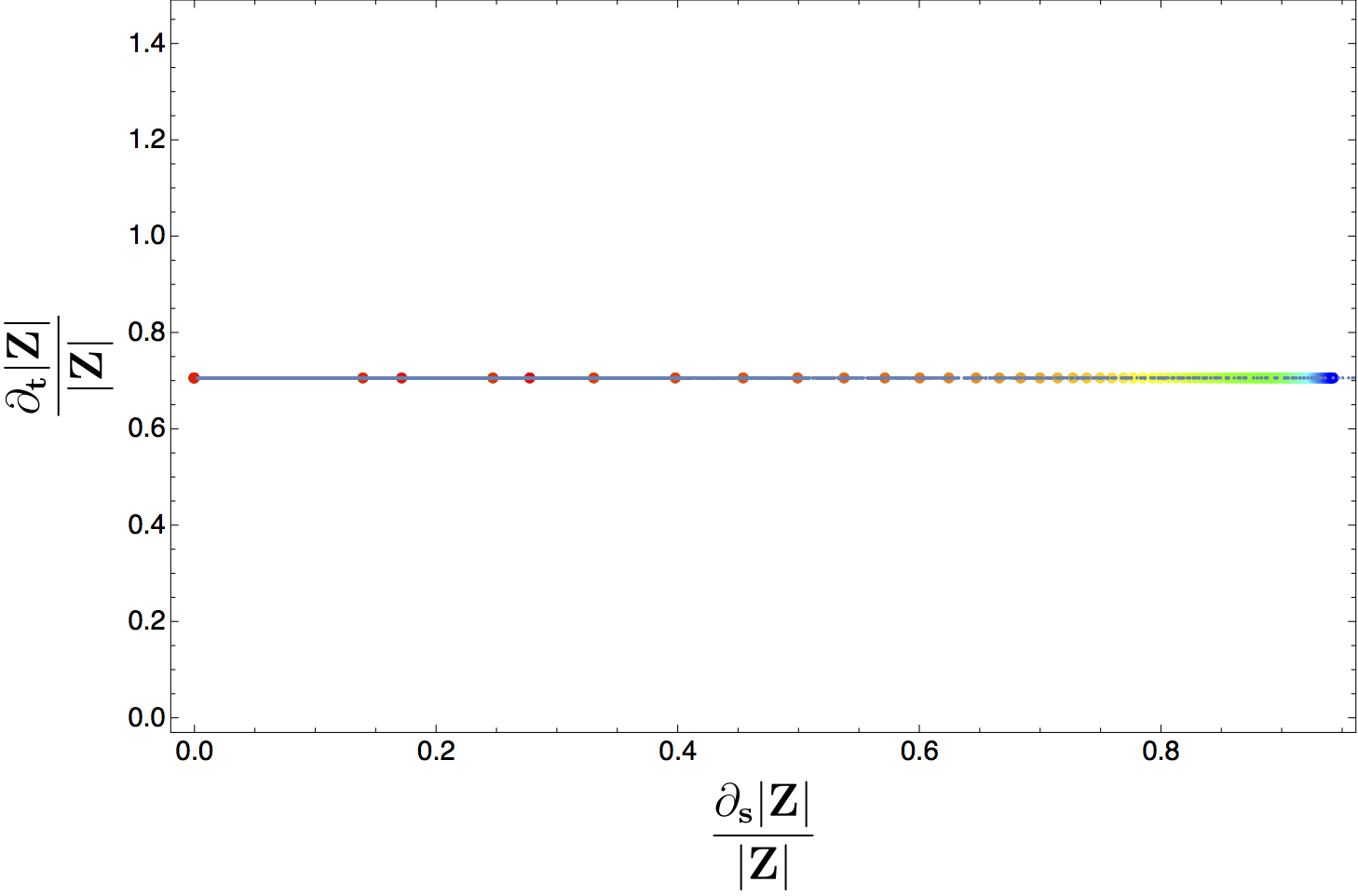}
  \caption{$t\gg s^2$}
  \label{monts}
\end{subfigure}%
\begin{subfigure}{.5\textwidth}
  \centering
  \includegraphics[width=0.95\linewidth]{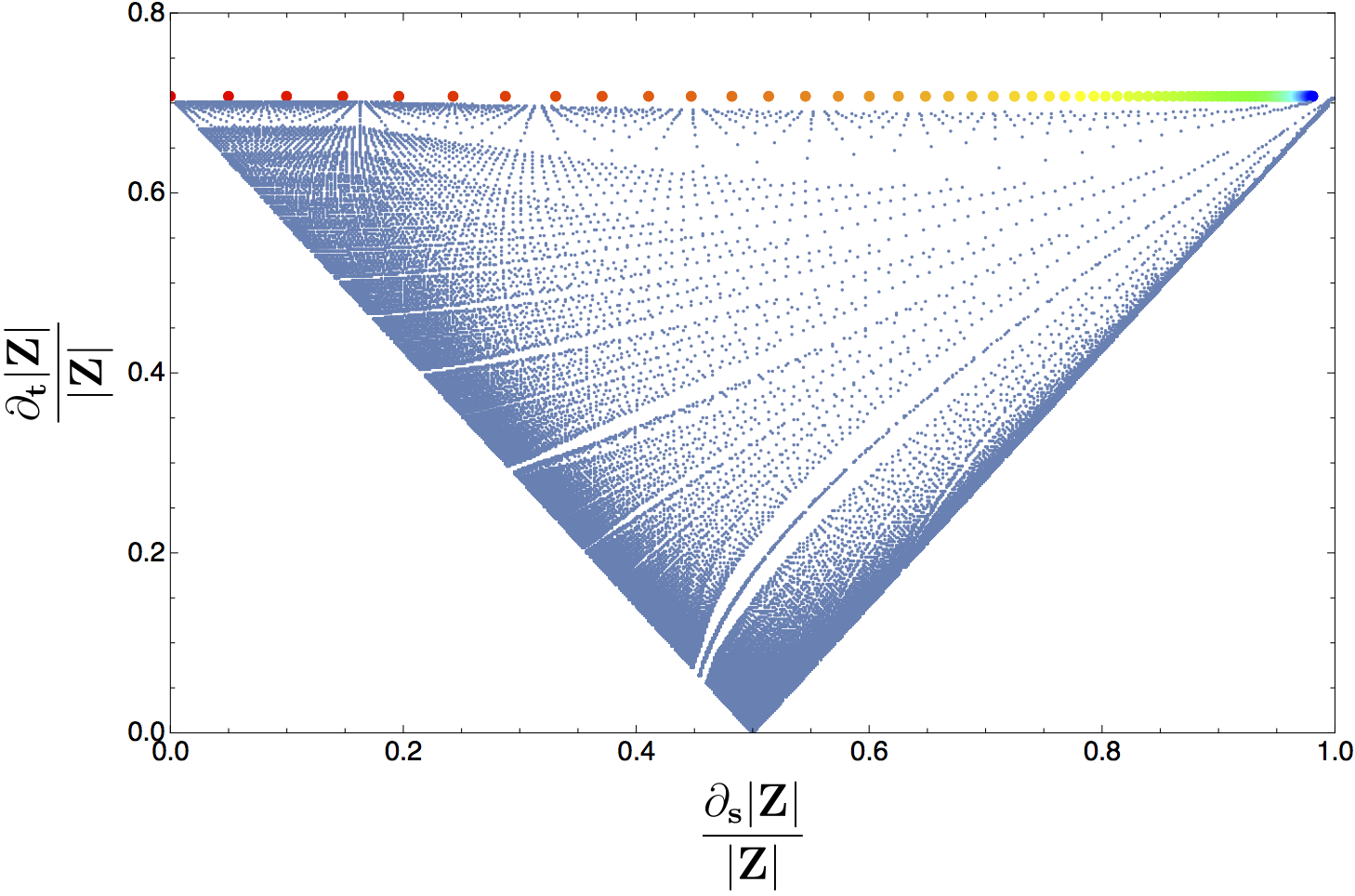}
  \caption{$s \gg t$}
  \label{monst}
\end{subfigure}
\caption{Scalar forces for the charges in the entire quantized charge lattice (in blue), along with the scalar forces for charges in the monodromy tower (in colors), for different infinite distance limits of a Calabi-Yau threefold compactification with $h^{2,1}=2$.}
\label{monodromy_orbits}
\end{figure}

Using \eqref{K} the K\"ahler potential reads $\cK=-\log(t s^2)$. The eigenvalues that determine that radii of the charge-to-mass ellipse are given by
\begin{align}
&\gamma_1^{-2} = \sum_{\ell \ | \ \kappa \ \text{even}}\frac{1}{1+\sum_{i=1}^n\frac{(\ell_i-\ell_{i-1})^2}{d_i-d_{i-1}}}= \frac12\\
&\gamma_2^{-2} = \sum_{\ell \ | \ \kappa \ \text{odd}}\frac{1}{1+\sum_{i=1}^n\frac{(\ell_i-\ell_{i-1})^2}{d_i-d_{i-1}}} =\frac14+\frac14 =\frac12,
\end{align}
where we have simply plugged the values for $(\ell_1,\ell_2)$ of table~\ref{tab21}.
The charge-to-mass ratio for any BPS state is then lower bounded by
\beq
\frac{Q}{M}\bigg|_{min}=\sqrt{2}.
\eeq
Furthermore, in order to get the SDC factor, we need to focus on the scalar charge ratio along the direction that grow fastest, in this case $t$, as explained around \eqref{bound}. In this example all single-charge states have
\beq
\frac{\nabla_t M}{M}=\frac{|\ell_1-3|}{\sqrt{2 d_1}}=\frac{1}{\sqrt{2}}.
\eeq
Since $\ell_1=2$ for all light states, the SDC factor is bounded by $\lambda\geq \frac{1}{\sqrt{2}}$. In particular, the monodromy orbit \eqref{Qorbit} with charge \cite{Grimm_2019}
\beq
q_{\text{tower}}=p^1+nq_2+\frac12 n^2 q_0
\eeq
has indeed
\beq
\lambda_{min}=\frac{1}{\sqrt{2}}.
\eeq
The values of the scalar charge-to-mass ratio for any BPS state is plotted in Fig.~\ref{monts} as blue dots while the pattern of color represents the monodromy orbit. Indeed, the first state of the tower has the smallest value of $\lambda$.\\

b) Growth sector: $s\gg t\gg 1$\\
This sector corresponds to the enhancement chain 
\beq
III_0\rightarrow IV_2
\eeq
with
\begin{align}
d_1 = 2, \ \ \ d_2 = 3,
\end{align}
and limiting Hodge diamonds given by figures~\ref{sing1_st} and \ref{sing2_st}.
\begin{figure}[h!]
\centering
\begin{subfigure}{.5\textwidth}
  \centering
 \begin{tikzpicture}
\node at (0,1.75) {$0$};
\node at (0.707,2.4) {$1$};
\node at (-0.707,2.4) {$1$};
\node at (0,3.2) {$0$};
\node at (0.707,0.707)[circle,fill,inner sep=1.5pt]{};
\node at (0.707,3.535)[circle,fill,inner sep=1.5pt]{};
\node at (-0.707,3.535)[circle,fill,inner sep=1.5pt]{};
\node at (-0.707,0.707)[circle,fill,inner sep=1.5pt]{};
\node at (0,1.414)[circle,fill,inner sep=1.5pt]{};
\node at (0,2.828)[circle,fill,inner sep=1.5pt]{};
\node at (0.707,2.121)[circle,fill,inner sep=1.5pt]{};
\node at (-0.707,2.121)[circle,fill,inner sep=1.5pt]{};
\draw[step=1cm,black,very thin,rotate=45] (0,0) grid (3,3);
\end{tikzpicture}
\caption{Type $III_0$ singularity}
\label{sing1_st}
\end{subfigure}%
\begin{subfigure}{.5\textwidth}
  \centering
\begin{tikzpicture}
\node at (0,1.75) {$2$};
\node at (0.707,2.4) {$0$};
\node at (-0.707,2.4) {$0$};
\node at (0,3.2) {$2$};
\node at (0,0)[circle,fill,inner sep=1.5pt]{};
\node at (0,1.414)[circle,fill,inner sep=1.5pt]{};
\node at (0,2.828)[circle,fill,inner sep=1.5pt]{};
\node at (0,4.243)[circle,fill,inner sep=1.5pt]{};
\node at (0.707,2.121)[circle,fill,inner sep=1.5pt]{};
\node at (-0.707,2.121)[circle,fill,inner sep=1.5pt]{};
\draw[step=1cm,black,very thin,rotate=45] (0,0) grid (3,3);
\end{tikzpicture}
\caption{Type $IV_2$ singularity}
\label{sing2_st}
\end{subfigure}
\caption{}
\label{fig:test}
\end{figure}

The K\"ahler potential is again $\cK=-\log(s^2t)$ while the asymptotic splitting is now given by table~\ref{st_splitting}.\\
\begin{table}[H]
\begin{center}
\begin{tabular}{|c|c|c|c|c|c|c|}
\hline
&\multicolumn{3}{|c|}{$\cE_{\text{light}}$}&\multicolumn{3}{|c|}{$\cE_{\text{heavy}}$}\\
\hline
$(\ell_1,\ell_2)$&(1,0)&(1,2)&(3,2)&(3,4)&(5,4)&(5,6)\\
\hline
$||v_\ell||^2$&$1/(ts^2)$&$t/s^2$&$1/t$&$t$&$s^2/t$&$s^2t$\\
\hline
$q_\ell$&$q_0$&$q_1$&$q_2$&$p^2$&$p^1$&$p^0$\\
\hline
\end{tabular}
\end{center}
\caption{Limit $III_0\rightarrow IV_2$}
\label{st_splitting}
\end{table}

This time $V_{rest}$ is empty so there is a clear distinction between electric and magnetic states for any trajectory within this growth sector.
The diagonal entries of the electric gauge kinetic function are given by

\begin{align}
&\mathcal{I}_{00}^{-1} \sim t^{l^0_1-3} s^{l^0_2-l^0_1} = 1/(ts^2)\quad \rightarrow \quad Q^2\bigg|_{q_0}\sim q_0^2 /(ts^2) \\
&\mathcal{I}_{11}^{-1} \sim t^{l^2_1-3} s^{l^2_2-l^2_1} = t/s^2\quad \rightarrow \quad Q^2\bigg|_{q_1}\sim q_1^2 t/s^2 \\
&\mathcal{I}_{22}^{-1} \sim t^{l^1_1-3} s^{l^1_2-l^1_1} = 1/t\quad \rightarrow \quad Q^2\bigg|_{q_2}\sim q_2^2 /t .
\end{align}
The eigenvalues that determine that radii of the charge-to-mass ellipse are now given by
\begin{align}
&\gamma_1^{-2} = \sum_{\ell \ | \ \kappa \ \text{even}}\frac{1}{1+\sum_{i=1}^n\frac{(\ell_i-\ell_{i-1})^2}{d_i-d_{i-1}}}= \frac14\\
&\gamma_2^{-2} = \sum_{\ell \ | \ \kappa \ \text{odd}}\frac{1}{1+\sum_{i=1}^n\frac{(\ell_i-\ell_{i-1})^2}{d_i-d_{i-1}}} =\frac14+\frac12 =\frac34
\end{align}
so the charge-to-mass for any BPS state is lower bounded by
\beq
\frac{Q}{M}\bigg|_{min}=\frac{2}{\sqrt{3}}.
\eeq
The scalar that grows the fastest is now $s$, so the SDC factor of the diagonal charge states will be bounded by
\beq
\frac{\nabla_s M}{M}=\frac{|\ell_1-3|}{\sqrt{2 d_1}}=1
\eeq
according to \eqref{bound}. 
In this case, however, there is one single-charge state (denoted as $q_2$) that has $\ell_1=3$  for which the previous result vanishes. When this occurs, the mass decay rate for that state will be fixed by the derivative with respect to the next growing scalar, in this case $t$, leading to
\beq
\label{dtmex}
 \frac{\nabla_t M}{M}=\frac{\ell_2-\ell_1}{\sqrt{2 (d_2-d_1)}}=\frac{1}{\sqrt{2}}
 \eeq
In fact, this state corresponds precisely to the seed charge of the monodromy tower,
which is given by \cite{Grimm_2019}
\beq
q_{\text{tower}}=q_2+nq_0
\eeq
Hence, the monodromy tower has indeed 
\beq
\lambda_{min}=1/\sqrt{2}.
\eeq
 The value of the scalar charge ratios are plotted in Fig.~\ref{monst} together with the monodromy orbit (the latter using the pattern of bright colors).
 
\begin{figure}[h!]
\centering
\begin{subfigure}{.5\textwidth}
  \centering
  \includegraphics[width=0.95\linewidth]{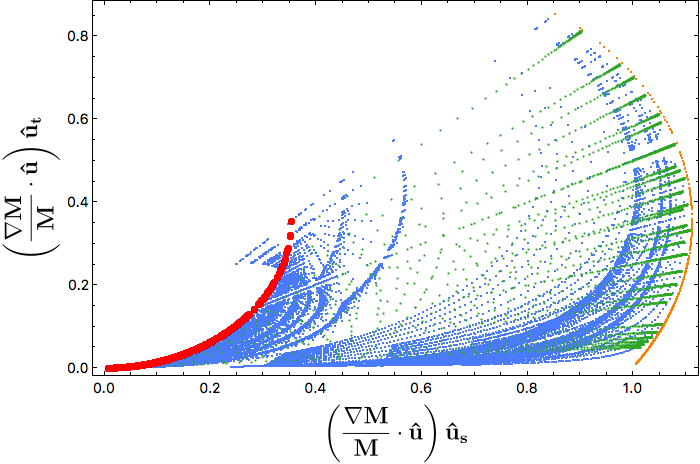}
  \caption{SDC factor in sector $s\gg t$}
  \label{plot_lambda}
\end{subfigure}%
\begin{subfigure}{.5\textwidth}
  \centering
  \includegraphics[width=0.95\linewidth]{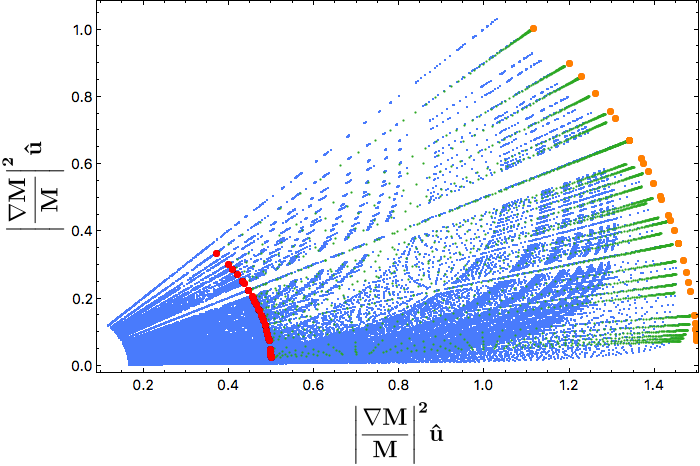}
  \caption{Scalar WGC in sector $s \gg t$}
  \label{plot_scalar}
\end{subfigure}
\centering
\begin{subfigure}{.5\textwidth}
  \centering
  \includegraphics[width=0.95\linewidth]{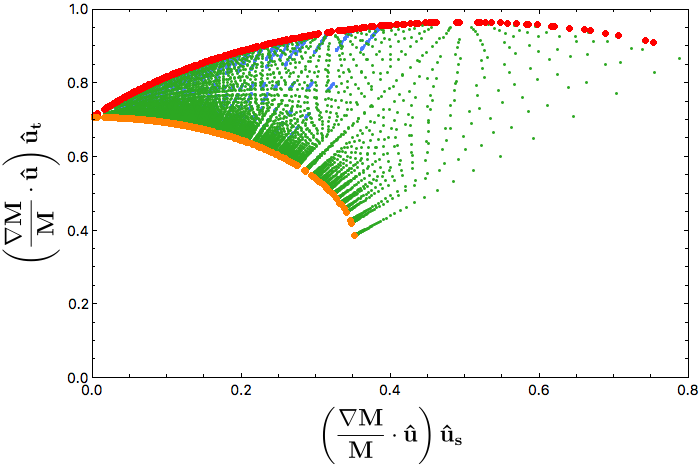}
  \caption{SDC factor in sector $t \gg s^2$}
  \label{plot_lambda_ts}
\end{subfigure}%
\begin{subfigure}{.5\textwidth}
  \centering
  \includegraphics[width=0.95\linewidth]{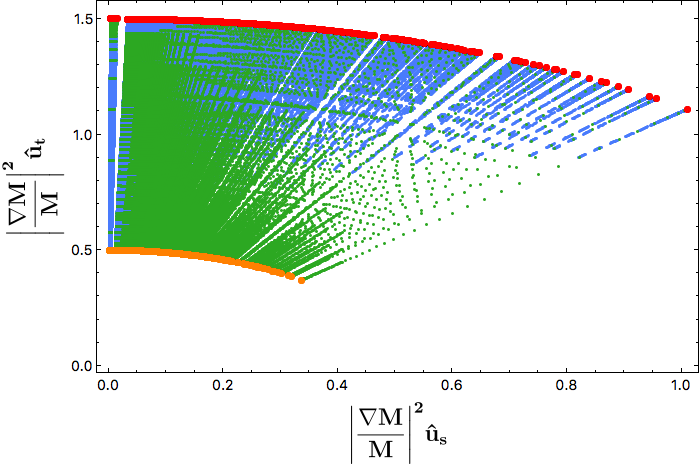}
  \caption{Scalar WGC in sector $t \gg s^2$}
  \label{plot_scalar_ts}
\end{subfigure}
\caption{The SDC factor $|\lambda|$ in each direction of field space, as well as the magnitude of the scalar WGC in each direction of field space. Blue points correspond to arbitrary sites in the quantized charge lattice. Green points correspond to points in the monodromy tower. Red points correspond to the seed charge of the monodromy tower. Orange points correspond to single-charge states that are not the seed charge. The first row shows the results for the growth sector $s \gg t$, while the second row shows the results for the growth sector $t \gg s^2$.}
\label{plot}
\end{figure}

  In order to gain a better intuition for the value of the SDC factor along every geodesic trajectory in field space, we have plotted the result of $\lambda=\left|\frac{\nabla_i M}{M} u_i\right|$ in \eqref{SDCfactor} for every BPS state in figure \ref{plot_lambda}. More concretely, we have plotted the vector $\lambda \vec u$ where $\vec u$ is a unit vector in field space, so $\vec u=(s,t)/\sqrt{s^2+t^2}$. The red dots correspond to the first state of the monodromy tower while the rest states of the tower are in green, and the blue dots refer to arbitrary BPS states in the quantized charge lattice. Notice that the bound $\lambda\geq 1/\sqrt{2}$ does not apply to all BPS states, but the whole point is that we can ensure that there are some BPS states (among them, the single-charge ones with $\lambda\geq 1$) which exhibit $\lambda\geq 1/\sqrt{2}$ along any direction in field space. The analogous figure for the $t \gg s$ limit is shown in Fig.~\ref{plot_lambda_ts} and \ref{plot_scalar_ts}.
  
  Finally, regarding the scalar WGC, something similar happens. Not all states satisfy it but there are always some that do. In particular, the monodromy tower as well as all the single-charge states satisfy it as can be seen in figure \ref{plot_scalar}, where we have represented the vector $\frac{|\nabla M|}{M} \vec u$ with $\vec u$ the unit vector in field space.

\section{Emergence and generalization to other setups} \label{generalizations}
In this section, we will discuss the generalization of our results to non-BPS states. First, we will analyze extremal non-BPS black holes still within the framework of $\cN=2$. We will describe how one can compute the extremality bound using a fake superpotential and the implications for the WGC. It seems that, at least in our examples, the convex hull of BPS states is enough to include the extremal region, even along the non-BPS directions of the charge lattice, thus satisfying the WGC without requiring the existence of non-BPS states. Secondly, we will discuss the generalization of our results beyond $\cN=2$ setups. The key ingredient for our story to hold in general is to have a gauge coupling vanishing at the infinite field distance. We believe that, indeed, one can always find a tower charged under a p-form gauge field becoming weakly coupled at infinite field distance. In such a case, one can then use the extremality bound associated to this gauge theory to bound the exponential decay rate of the Distance Conjecture.

\subsection{Non-BPS charges} \label{nonbps}
In general, obtaining the masses of non-BPS D-branes is a very complicated problem. However, extremal non-BPS black holes have been widely studied \cite{Ferrara:1997tw,Goldstein:2005hq, Tripathy:2005qp}, and we can use these findings to complete the determination of the extremality bound along the non-BPS directions in our $\cN=2$ setup. In particular, the formalism of ``fake supersymmetry"~\cite{Freedman:2003ax, Celi:2004st, Zagermann:2004ac, Skenderis:2006jq} gives a set of first-order equations analogous to the BPS flow equations which solve the Einstein equations to give an extremal black hole solution of the form \eqref{blackhole}. In this section, we will describe how to use this formalism to obtain masses of non-supersymmetric black holes, following \cite{Ceresole_2007}. 

Although the underlying explanation for fake supersymmetry is complicated and deep, our goal is very simple: find a real function $\mathcal{W}$ (the ``fake superpotential") such that $Q^2 = \mathcal{W}^2+4 K^{i\bar{j}} \partial_i \mathcal{W} \partial_{\bar{j}} \mathcal{W}$. Then everything works analogously to the BPS case but with an ADM mass  given by $M=\mathcal{W}|_{\infty}$, where the fake superpotential is evaluated at the values of the scalar fields at spatial infinity. Thus, our task is simply to find the fake superpotential\footnote{Although not relevant for our particular example, there are subtleties that arise when there are several possible fake superpotentials with a gradient flow to different attractors. In that case, one has to pick the one corresponding to the smaller value of $\mathcal{W}$. We thank Ben Heidenreich for enlightening comments on this issue.}. In general, this can be a hard problem, but there is a prescription that works in certain cases which we will exhibit here in an example.

The prescription is as follows: find a symplectic matrix $S$ that has the property that it commutes with the Hodge norm of the basis $[S,\mathcal{M}]=0$. Then the fake superpotential is:
\begin{align}
\mathcal{W} = e^{K/2} q^T S^T \eta \ \vec{\Pi},
\end{align}

If the fake superpotential can be constructed, then the extremality bound for non-BPS black holes is known. Note that if this extremality bound is known, the mildest form of the Weak Gravity Conjecture can potentially lead to very interesting constraints on geometry~\cite{Hebecker:2015zss,Heidenreich:2016jrl}. This idea was explored in \cite{demirtas2019minimal} in the context of the sublattice version of the Weak Gravity Conjecture as applied to axions and instantons. The argument is as follows~\cite{demirtas2019minimal}: the Weak Gravity Conjecture places bounds on the charge-to-mass ratios of states in a given theory. In string theory compactifications, masses of certain objects are given by volumes of cycles in the internal manifold. Then if one posits that the Weak Gravity Conjecture must be satisfied by single-particle states, there exists a bound on the masses of the non-BPS states in the theory, and hence a bound on non-holomorphic volumes of internal cycles. We point it out in this context because the formalism of the fake superpotential and the methods developed in this work give us a way to compute extremality bounds and BPS spectra precisely. This opens up an opportunity to use the mildest form of the Weak Gravity Conjecture to place bounds on the non-holomorphic aspects of the geometry of Calabi-Yau threefolds.

As an example, we can construct a fake superpotential (and hence calculate the mass) of non-BPS black holes for the one-modulus example presented in section~\ref{1mod}. We have seen that all combinations of electric charges give rise to BPS black holes, but for dyonic black holes (carrying both electric and magnetic charge) such as the ones we presented in section \ref{dyons}, there can be non-BPS extremal black hole solutions for certain charge choices. To see this, consider a black hole with charges $q_1$ and $p^0$ (exactly as was considered in section \ref{dyons}). The attractor mechanism tells us that such a black hole is BPS if on the black hole horizon
\begin{align}
\partial_i |Z| =0.
\label{bpscond}
\end{align}
In this case, we have
\begin{align}
|Z| = \sqrt{\frac{-i T \overline{T} (q_1-p^0 T^2)(q_1-p^0 \overline{T}^2)}{(2 \text{Im}(T))^3}}.
\end{align}
The condition in \eqref{bpscond} tells us that these black holes are BPS if $q_1 p^0>0$ and non-BPS if $q_1 p^0 <0$. The charge-to-mass ratio for BPS, extremal, dyonic black holes in this theory was given in \eqref{dyonqm} which we repeat here for convenience:
\begin{align}
\frac{|Q|}{M} \bigg|_{\text{BPS}}= \frac{2}{\sqrt{3}} \sqrt{\frac{q_1^2+3 p_0^2 t^4}{(q_1+p_0 t^2)^2}}.
\end{align}
As explained in section \ref{dyons}, the charge-to-mass vectors form two straight lines, representing the fact that these dyonic states are mutually BPS.

Even though the states with $q_1 p^0 <0$ are not BPS states, as discussed in \cite{Ceresole_2007}, a fake superpotential for these charges can be found:
\begin{align}
\mathcal{W}=\left|\frac{q_1 T - p^0 T^2 \overline{T}}{(2\text{Im}(T))^3} \right|,
\end{align}
which can be shown to give rise to the same $Q^2$ that is computed using the central charge.

\begin{figure}[h!]
\centering
\includegraphics[width=0.5\textwidth]{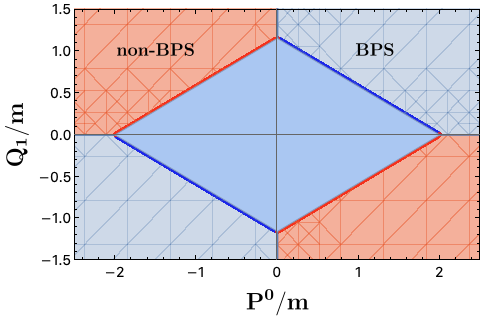}
\caption{The extremal region for both BPS and non-BPS directions in our example with $h^{2,1}=1$.}
\label{convexhull_dyons}
\end{figure}

Thus, the charge-to-mass ratio for non-BPS, extremal, dyonic black holes in this theory is (setting the axion to zero)
\begin{align}
\frac{|Q|}{M} \bigg|_{\text{non-BPS}}= \frac{2}{\sqrt{3}} \sqrt{\frac{q_1^2+3 p_0^2 t^4}{(q_1-p_0 t^2)^2}}.
\end{align}
Now we are in a position to answer the question of whether the WGC requires geometric recombination of non-trivial binding to produce non-BPS states in this theory with only dyons of the form described above. The extremal black hole region and BPS states are shown in Fig.~\ref{convexhull_dyons}. From this figure, it is clear that the convex hull of the BPS charge-to-mass ratios precisely form the entire extremal black hole region (due to the region being formed from straight lines). Thus, we conclude that in this example, BPS states alone fulfill the Convex Hull WGC, and no extra states are needed.

\begin{figure}[h!]
\centering
\includegraphics[width=\textwidth]{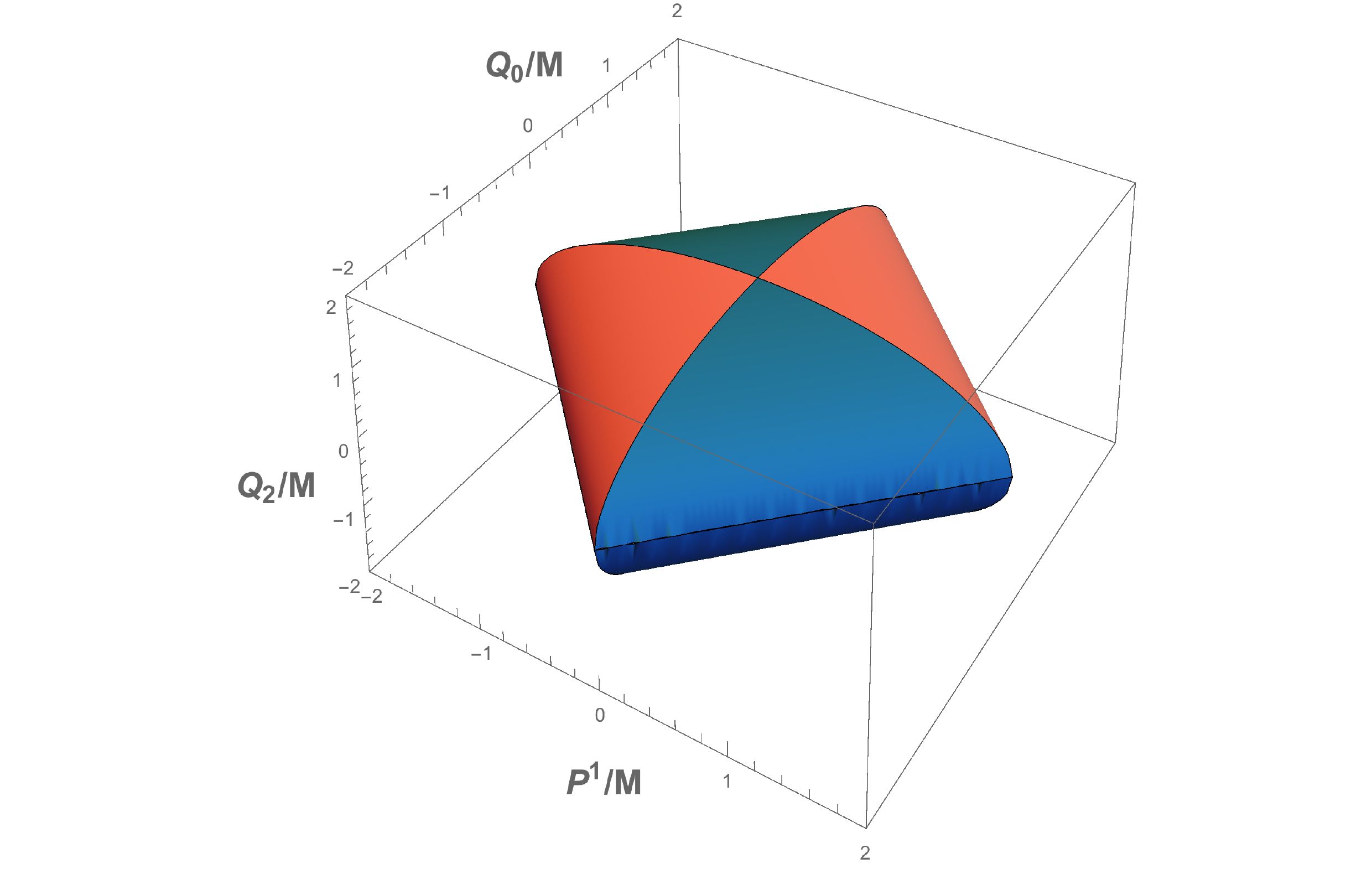}
\caption{The extremal black hole region in an example of a compactification with $h^{2,1}=2$, in the growth sector $t \gg s^2$. The blue regions correspond to extremal BPS black holes, while the red regions correspond to extremal non-BPS black holes.}
\label{extremal_2mod}
\end{figure}

Recall that in the two-moduli example of section~\ref{2mod}, there were also restrictions on which choices of quantized charges could give rise to BPS states. These restrictions had the effect of bounding the charge-to-mass ratio of states in this example from above. Now, we would like to complete the picture by determining the extremality bound for non-BPS black holes in this setup. To find this, we once again must construct a fake superpotential $\cW$ such that $Q^2 =|\cW|^2 + 4 K^{i\bar{j}} \partial_i |\cW| \partial_{\bar{j}} |\cW|$. With the axions turned off, we can find such a superpotential:
\begin{align}
\cW = \sqrt{\frac{p^1 q_0}{4 t} + \frac{3 q_2^2}{4 t} + \frac{3 q_0^2}{4 s^2 t} + \frac{(p^1)^2 s^2}{48 t}}
\end{align}
Thus, $\frac{|Q|}{\cW}$ gives the extremality bound for non-BPS black holes\footnote{This is only true under the assumption that charges satisfying $ p^1 q_0>0$ give rise to solutions for the full set of fake supersymmetry equations. Note that the axionic flow equations cannot be verified unless the axions are turned on and an appropriate fake superpotential is identified. This example serves as a proof of principle and should be studied further.}. A large black hole will evaporate until a mass of the larger of $|Z|, \ \cW$ is reached \cite{Galli_2011}. Therefore, the extremal region that should be used in verifying the Convex Hull WGC is the minimal value of $\frac{|Q|}{|Z|}$ or $\frac{|Q|}{\cW}$ for each direction in $\vec{Q}/M$ space. The extremal region that is constructed in this way is shown in Fig.~\ref{extremal_2mod}. The blue regions in the figure correspond to charges that satisfy the BPS conditions (in fact, here they satisfy $q_1 p^1<0$), while the red regions correspond to non-BPS charges. This shape is constructed by intersecting the two degenerate ellipsoids that parameterize the BPS extremality bound and the non-BPS extremality bound. Note that once again, the convex hull of the BPS extremal region precisely corresponds to the entire extremal region (BPS and non-BPS), such that no extra states beyond the BPS ones are needed in order to satisfy the Convex Hull WGC.

Note that because these states have masses which satisfy the relation $Q^2 =|\cW|^2 + 4 K^{i\bar{j}} \partial_i \cW \partial_{\bar{j}} \cW$, they satisfy (at least approximately) a force-cancellation condition, implying that the RFC still coincides with the WGC even along the non-BPS directions. In other words, extremal black holes (even if not BPS) satisfy a force-cancellation condition \cite{Heidenreich_2019,Ben1}.

Before concluding this section, let us remark that the existence of extremal non-BPS black holes might seem to be in conflict with the sharpening of the WGC \cite{Ooguri:2016pdq}, for which the WGC bound can only be saturated by BPS objects in supersymmetric theories. A possible way out is that this sharpening should be applied only to directions in the charge lattice that could support a BPS state. If, instead, a given charge direction cannot correspond to a BPS direction, then the WGC may be saturated by a non-BPS one. This is what occurs in our example, as a state with charges satisfying $q_1p^0>3 q_2^2$ cannot be BPS. However, this might change the conclusion that non-supersymmetric vacua supported by flux must be unstable \cite{Ooguri:2016pdq,Freivogel:2016qwc} (and their implications for neutrino physics in \cite{Ibanez:2017kvh,Ibanez:2017oqr,Hamada:2017yji,Gonzalo:2018dxi,Gonzalo:2018tpb}), since there could be flat domain walls satisfying the WGC and not describing an instability even in non-supersymmetric vacua. The other way out is to assume that fake supersymmetry can never be realized exactly in a UV complete theory. In other words, that we should always have higher order corrections that modify the charge-to-mass ratio of the non-BPS objects, causing them to deviate from extremality. In that case, only BPS objects protected by supersymmetry could saturate the WGC. Notice, though, that higher derivative corrections can be highly suppressed in the weak coupling limits when approaching an infinite field distance point, so it might be impossible to distinguish if the state is truly extremal from a low energy perspective.

\subsection{Kaluza-Klein tower} \label{kktower}

We have seen in our $\cN=2$ setup that the SDC tower becoming light in the infinite field distance limits also saturates the WGC, allowing for a correlation between the exponential mass decay rate and the extremality bound. This might be attributed to the fact that we are moving along the moduli space of scalars which are part of the $\cN=2$ vector multiplets, so consequently the gauge fields in the same vector multiplets become weakly coupled at the infinite field distance limit. This implies that the SDC tower, whose mass is parametrized by these scalars, is also charged under the corresponding gauge fields. However, we believe that the correlation between the SDC factor and extremality bound depends only on whether there is a gauge field that becomes weakly coupled at infinite field distance, regardless of whether it is part of a vector multiplet, and hence, valid beyond $\cN=2$. To exemplify this, let us discuss the case a KK tower of a circle compactification. Our story also holds true in this setup, since the KK tower is charged under the graviphoton whose gauge coupling indeed goes to zero at large radius.

Let us explicitly write the charge-to-mass ratio and the SDC factor associated to the KK tower in order to compare it with the general expressions we obtained in $\cN=2$. Here we will follow \cite{Heidenreich_2016,Corvilain_2019,Heidenreich_2019}. Consider a dimensional reduction of Einstein gravity in (D+1)-dimensions on a circle. The low energy D-dimensional effective theory is given by
\beq
\label{LKK}
\mathcal{L}=\frac{1}{2\kappa^2}\left( R_D-\frac{D-1}{D-2}(\nabla \phi)^2-\frac{r^2}{2}\left(\frac{r}{r_0}\right)^{\frac{2}{D-2}}e^{-2\frac{D-1}{D-2}\phi}F_2^2 \right),
\eeq
where $F_2$ is the field strength of the KK photon, $\phi=\log r$ and $r$ is the circle radius. Extremal black holes in this theory satisfy
\beq
\left(\frac{D-3}{D-2}+\frac{\alpha^2}{4}\right)m^2 = e^2q^2M_p^{D-2},
\eeq
where $\alpha$ is the coupling $e^{-\alpha\phi}$ in the gauge kinetic function in terms of the canonically normalized field. From \eqref{LKK} we get that
\beq
\alpha= 2\sqrt{\frac{D-1}{D-2}}\ ,\quad e^2=\frac{2}{r^2M_p^{D-2}}\left(\frac{r_0}{r}\right)^{\frac{2}{D-2}}
\eeq
yielding
\beq
\left(\frac{D-3}{D-2}+\frac{D-1}{D-2}\right)m^2 = e^2q^2M_p^{D-2}\quad \rightarrow\quad m^2=\frac{q^2}{r^2}\left(\frac{r_0}{r}\right)^{\frac{2}{D-2}}.
\label{WGCKK}
\eeq
Notice that, as is well known, the KK tower has precisely a mass $m^2_{KK}=m^2$ saturating the WGC bound \eqref{WGCKK}. 

On the other hand, we can write the RFC bound corresponding to the no-force cancellation condition in this setup. Since the radius is massless at this level, the KK tower feels both a gauge and a scalar force in addition to the gravitational force given by
\beq
|F|=\frac{A}{r^{D-2}vol(S^{D-2})}
\eeq
with
\beq
A_{yukawa}=\frac{\partial_i mg^{ij}\partial_j m}{M_p^{D-2}}\ ,\quad A_{grav}=\frac{(D-3)m^2}{(D-2)M_p^{D-2}},\quad A_{gauge}=e^2q^2
\eeq
leading to the following no-force condition
\beq
\frac{m^2(D-3)}{(D-2)}+\partial_i mg^{ij}\partial_j m = e^2q^2M_p^{D-2}.
\eeq
For a KK tower with $m_{KK}=\frac{q}{r}(\frac{r_0}{r})^{\frac{1}{D-2}}$ , this becomes
\beq
\frac{m^2(D-3)}{(D-2)}+ \frac{D-1}{D-2}m^2= e^2q^2M_p^{D-2},
\eeq
which precisely matches with \eqref{WGCKK}. Hence, as expected, the extremality bound and the no-force condition match for a KK tower. Furthermore, the mass of this KK tower decays exponentially with the proper field distance
\beq
m=qr_0^{\frac{1}{D-2}}\exp(-\sqrt{\frac{D-1}{D-2}}\hat \phi),
\eeq
satisfying also the SDC at large values of the radius. The SDC factor is given by 
\beq
\lambda=\frac{|\nabla m|}{m}=\sqrt{\frac{D-1}{D-2}}\ .
\eeq
 Notice that we could have directly obtained the SDC factor without knowing the explicit formula for the mass of the KK tower by using the dependence of the radius in the gauge kinetic function following \eqref{LA}, obtaining
\beq
\lambda= \frac{\alpha}{2}= \sqrt{\frac{D-1}{D-2}}.
\eeq
which is consistent with \eqref{LA}. Therefore, this works the same way as that of the $\cN=2$ setup. The fact that the extremality bound matches with the no-force condition implies that the SDC factor is given by the moduli dependence of the gauge kinetic function. The matching between extremality and no-force condition occurs because we are in a weak coupling regime, since the gauge coupling of the KK photon goes to zero at the infinite field distance. We expect this to hold in general as long as the tower of states is charged under some weakly coupled gauge theory, regardless of the amount of supersymmetry, as we will discuss in the next section.

\subsection{Emergence and generality of our results}\label{emergence}

We have seen that for towers of states which are either BPS sates or KK towers, the exponential mass decay rate is correlated with the extremality bound for black holes, so that the same order one factor that appears in the WGC can be used to provide a lower bound for the (a priori unspecified) order one factor of the Distance Conejcture. More concretely, this lower bound is given by the scalar contribution to the extremality bound, which can be directly read from the scalar dependence of the gauge kinetic function when written in some particular asymptotic basis adapted to the infinite distance limit. A very important question is to understand how general these results are, and this will be the topic of this section.

Let us assume for the moment that the tower of states of the Distance Conjecture is indeed charged under some gauge field. The above results are then completely general as long as the extremality bound \eqref{WGC} coincides with the repulsive force condition \eqref{RFC}. In that case, the scalar contribution of the extremality bound labeled by $\alpha$ becomes equal to the scalar charge-to-mass ratio $|\nabla m/m|$, which is correlated to the exponential mass decay rate of the tower\footnote{Recall that the exponential mass decay rate $\lambda$ of the Distance Conjecture is not exactly $|\nabla m/m|$ due to path dependence issues, as explained in section \ref{decayrate}. However, the SDC factor $\lambda$ can still be bounded by $|\nabla_{t^1} m/m|$ where $t^1$ is the scalar growing faster.}. Therefore, the generality of the results is linked to the question of under what circumstances the WGC and the RFC are the same condition.

The differences between the WGC and the RFC have been nicely discussed in \cite{Heidenreich_2019}. Although there are no known string theory examples which satisfy one but violate the other, a priori these are two different conditions. Furthermore, if higher derivative corrections make small black holes ``superextremal", therefore satisfying a mild version of the WGC, then we would expect them to also become self-repulsive. Otherwise, they could bind to form a single back hole of twice the charge but larger charger to mass ratio. Repeating this process, we would get black holes of increasing charge but larger and larger charge-to-mass ratio. This would be anti-intuitive since the normal expectation is that the charge-to-mass ratio should decrease when increasing the charge in order to approach the classical value of the extremality bound for large black holes, where the higher derivative corrections become negligible. It was also proposed in \cite{Heidenreich_2019} that both conjectures could actually come (only in four dimensions) from a new conjecture claiming that there must exist a multiparticle state of maximal $|Q|/M$ for every rational direction in charge space. Indeed, this seems to be satisfied in our setup, since the charge-to-mass ratio is upper bounded, at least along the BPS directions. Notice that due to the BPS condition, for a given charge there cannot be states with a larger $|Q|/M$ than the corresponding BPS state, and the possible values of $|Q|/M$ for BPS states are upper bounded by the charge restrictions. Whether this also occurs along the non-BPS directions is unclear, so we are going to explore a different approach.

In \cite{Lee_2019} it was argued that the WGC and the RFC coincide at the weak coupling limits, and this is the line of thought we have also developed in this paper. The argument in \cite{Lee_2019} goes as follows. Suppose that there is a gauge coupling going to zero at some infinite field distance point. If the asymptotically massless tower predicted by the SDC also saturates the WGC, then the scalar dependence of the gauge kinetic function must coincide\footnote{It is important here that the states saturate, and not just satisfy, the WGC. However, since large black holes are expected to saturate it, we can just scale down the charge to argue that BPS particles, along the same charge direction, have the same $Q/M$ and hence also saturate the WGC. It is not so clear, though, how to extrapolate this to non-BPS states.} with an exponential rate of the tower such that both go to zero at the infinite field distance in a way consistent with the WGC. This implies that the WGC and the RFC\footnote{The RFC was denoted in \cite{Lee_2019} as the Scalar WGC, but it should not be confused with \eqref{ScalarWGC}.} become the same in the weak coupling limit. To corroborate the argument, the authors in \cite{Lee_2019} rigorously checked it for weakly coupled limits in F-theory compactifications to six dimensions, where the tower of states correspond to the excitation modes of an asymptotically tensionless string which is dual to the heterotic perturbative string.

In this paper we have shown that the same argument holds for any infinite field distance limit of Calabi-Yau threefold compactifications of Type IIB. We have generalized the argument of \cite{Lee_2019} to any number of moduli, where the path dependent issues make it more difficult to determine the exponential mass decay rate of the tower and cannot be simply identified with the scalar dependence of the gauge kinetic function. However, we have seen that it is still possible to provide both a lower and an upper bound for this factor in terms of the properties of the infinite field distance limit. We should also note that here we are in fact interested in the opposite direction of the argument in \cite{Lee_2019}. Instead of assuming that the same tower satisfies the SDC and the WGC and get from there that the WGC and the RFC coincide, we would like to remark that whenever the WGC and the RFC coincide, then the exponential mass decay rate of the SDC tower is related to the extremality bound since the tower also satisfies the WGC. Hence, we are interested in understanding when the WGC and the RFC will coincide.

Taking into account the results of this paper as well as those of \cite{Lee_2019}, we believe that the WGC and the RFC will always coincide in the asymptotic regime approaching an infinite field distance point, as long as there is a gauge coupling vanishing at that limit. The mathematical machinery based on asymptotic Hodge theory used to show this result for Calabi-Yau string compactifications is actually very general and not restricted to Calabi-Yau manifolds or a specific dimension\footnote{For the moment, in addition to the complex structure moduli space of Type IIB Calabi-Yau compactifications, we have also applied this mathematical theorems to the K\"ahler moduli space of a Calabi-Yau threefold compactification of Type IIA to 4D and M-theory to 5D (dual to 6D F-theory) in \cite{Corvilain_2019}, as well as M-theory on a Calabi-Yau fourfold in \cite{grimm2019asymptotic}.}. Hence, we expect it to hold beyond our $\cN=2$ four-dimensional setup. But it would be interesting to find some field theoretical argument, not necessarily based on the geometry of the internal space, to explain why this should be true. The emergence proposal provides an argument in this direction, as we explain in the following.

The Emergence proposal \cite{Harlow:2015lma,Heidenreich2018a,Grimm_2018,Heidenreich:2018kpg,Corvilain_2019,Palti_2019} states that all fields are non-dynamical in the UV, and the kinetic terms emerge only in the IR from integrating out massive states up to some UV quantum gravity scale. It provides an underlying reason for the Swampland conjectures as follows: the WGC and the SDC imply light towers of states when the gauge coupling becomes small and the field distance becomes large, respectively. The emergence proposal turns this logic around and suggests that the small gauge coupling and the infinite field distance are generated from quantum corrections of integrating out the towers of states becoming light. Hence, the relations between (gauge and scalar) charges and masses that the Swampland conjectures predict would simply be a consequence of the renormalization group flow equations. The integrating out procedure must be performed up to the scale at which gravity becomes strongly coupled, also known as the species scale \cite{ArkaniHamed:2005yv,Distler:2005hi,Dimopoulos:2005ac,Dvali:2007wp,Dvali:2007hz}, and given by
\begin{align}
\label{species}
\Lambda_{UV} = \frac{M_{pl}}{\sqrt{N}}\sim (\Delta m)^{1/3}\ ,\quad N =\frac{\Lambda_{UV}}{\Delta m}\sim (\Delta m)^{-2/3},
\end{align}
where in this case $N$ is the number of species weakly coupled to gravity below $\Lambda_{UV}$.
The quantum corrected gauge coupling and the field metric from integrating out a tower of states with
\begin{align}
q_k = k, \ \
m_k = k \Delta m
\end{align}
up to the species bound are given by
\beqa
\frac{1}{g^2} \simeq  \sum_{k=1}^N q_k^2 \log \left(\frac{\Lambda_{UV}^2}{m_k^2} \right)\simeq  \frac{2}{9}\frac{1}{ (\Delta m)^2}\\
g_{\phi\phi}\simeq  \sum_{k=1}^N(\partial_\phi m_k)^2\simeq \left(\frac{\partial_\phi \Delta m}{\Delta m}\right)^2.
\eeqa
We refer the reader to \cite{Heidenreich2018a,Grimm_2018,Heidenreich:2018kpg,Corvilain_2019,Palti_2019} for more details on this one-loop computation. The interesting remark for us is that quantum corrections from integrating out a charged tower yield an IR gauge coupling and field metric given in terms of the same quantity $\Delta m$. This implies that both the gauge charge $Q=kg$ and the scalar charge $\sqrt{g^{\phi\phi}}\partial_\phi m$ become proportional to the mass $m$, which is the same behaviour we find in the asymptotic regimes of the moduli space of $\cN=2$ and that is behind the matching between the WGC and the RFC. However, in order to check this matching explicitly we would need to compute the numerical factors in the above formulae, which unfortunately seems rather impossible at the moment since the species bound \eqref{species} only gives an estimation of the order of magnitude of the UV cut-off scale. In any case, it can serve as a motivation for such a coincidence.

If the WGC and the RFC always coincide at any asymptotic limit with $g\rightarrow 0$, it still remains to be seen whether there is always a gauge coupling vanishing at every infinite field distance limit. This is linked to the very first assumption taken at the beginning of this section, for which we assumed that the tower of the Distance Conjecture is charged under some gauge field. Only if this occurs does it make sense to determine the mass decay rate in terms of the extremality bound. Clearly, such a weakly coupled gauge theory exists whenever the infinite field distance corresponds to displacing a scalar within a vector multiplet, since the gauge coupling associated to the gauge field in the same multiplet will automatically go to zero at the limit. But as we discussed, it is not restricted to the moduli space of vector multiplets. For instance, as we showed in section \ref{kktower}, it works the same way for the decompactification limit on a Kaluza-Klein circle compactification, since the KK photon becomes weakly coupled when sending the circle radius to infinity. So how general is to have a vanishing gauge couple at every infinite field distance limit? 

The first thing to notice is that this cannot be true if we require the gauge coupling to always correspond to a massless 1-form gauge field\footnote{Consider for example a compactification without isometries where there is no massless Kaluza-Klein photon, but still we have a KK tower that becomes massless in the decompactification limit.}, but it might very well be true if we allow for a general p-form gauge field. This goes along the lines of the String Emergence Conjecture in \cite{lee2019emergent}, for which the leading tower of states, i.e. the one becoming massless at the fastest rate, either corresponds to a tower of particles (dual to a KK tower in some frame) or to the excitation modes of a string becoming tensionless. In that case, we would have either a KK photon or a 2-form gauge field becoming weakly coupled. In both cases, the exponential mass rate can be determined in terms of the scalar dependence of the gauge coupling. This also matches with the expectation that the infinite tower of the SDC should admit a weakly coupled description, as emphasized in \cite{Kim:2019ths}, although we are requiring here a stronger condition by imposing that this weakly coupled description is associated to a gauge theory. A weaker condition is that there is always a gauge coupling of a p-form gauge feld which vanishes at the infinite field distance limit but this gauge field can become massive slightly away from the infinite distance point, e.g. by a Higgs mechanism. But then we would not have large charged black holes to compare the extremality bound with the RFC. However, it might still be possible to extract the SDC factor from a version of the WGC for massive gauge fields, but this goes beyond of the scope of this paper. As a final comment, notice that the existence of the tower is eventually linked to the existence of dualities, so the presence of a vanishing gauge coupling implies that every string duality should involve a weakly coupled gauge field theory in at least one of the dual descriptions.

Based on string theory examples, it seems that it is always possible to identify an infinite tower charged under a massless p-form gauge field which becomes weakly coupled at every infinite field distance point. This tower may not necessarily be the leading tower, but its exponential mass decay rate can be bounded in terms of the extremality bound associated to black holes in that gauge theory, and would serve as a lower bound for the factor $\lambda$ of the Distance Conjecture. Notice that if this tower is not the leading one, it means that the SDC factor $\lambda$ might be larger (so there is a tower which decays faster) but not smaller. Hence, we can use these charged towers to give a conservative estimate of when the effective field will break down, although it might break down even earlier.

Furthermore, we have seen that the extremality bound and the scalar dependence on the gauge kinetic function can be determined in terms of the properties of the infinite field distance limit, which can be classified. Using the classification of asymptotic limits in Calabi-Yau threefolds \cite{Grimm_2019}, we have given a lower bound for $\lambda$ for any $CY_3$. But one might envision the possibility of giving a universal lower bound for the exponential factor of the SDC based on a general classification of asymptotic limits in field space.

Needless to say, if there is some infinite field distance limit in which there is no p-form gauge field becoming weakly coupled, it would not be possible to give a universal bound for the SDC factor this way. We are not aware of any counterexamples of this type in string theory, but given the relevance of determining this numerical factor for phenomenological applications, it would be very interesting to settle this question in the future. At the moment, we can only argue that such a vanishing gauge coupling would yield a global symmetry in the infinite field distance limit, giving a motivation for the existence of the infinite tower as a quantum gravity obstruction to reach the limit and restore the global symmetry. Notice that the relation between the SDC and global symmetries was first discussed in \cite{Grimm_2018}, where in addition to this global symmetry it was noted that the monodromy transformation of infinite order populating the infinite tower (which is part of the duality group) also became global at infinite distance. So there were actually two global symmetries restored asymptotically.

\section{Conclusions} \label{conclusions}
In this paper, we have analyzed the structure of BPS charge-to-mass ratios in the asymptotic limits of compactifications of Type IIB string theory on Calabi-Yau threefolds and used the results to sharpen the Swampland Conjectures in the presence of multiple gauge and scalar fields. The results are twofold. First, we completely characterize the shape of BPS charge-to-mass ratios and the numerical values characterizing the extremality bound for any asymptotic limit. Second, we provide a lower bound on the exponential mass decay rate of the Distance Conjecture, based on the fact that the tower of states also saturates the Weak Gravity Conjecture. The latter result will hold whenever there is a gauge coupling vanishing at infinite field distance, which can, of course, also occur beyond $\cN=2$. 

We have found that the charge-to-mass ratios of BPS states lie on a degenerate ellipsoid with only up to two non-degenerate directions, regardless of the number of moduli and gauge fields. We have given a prescription to compute the exact value of the principal radii of the ellipse for any asymptotic limit in moduli space. These principal radii provide a lower bound on the charge-to-mass ratio of any BPS state in these theories, and can simply be obtained in terms of the scalar dependence of the diagonal entries of the gauge kinetic function when written in a particular basis adapted to the singular limit. This basis is associated to the asymptotic splitting of the charge space at infinite field distance, which is guaranteed by the theorems of asymptotic Hodge theory. 
We should comment that in principle, calculating the BPS charge-to-mass spectrum is straightforward if handed a prepotential, $F$ (just as we did in section~\ref{bps}). However, we have provided a general framework for obtaining the charge-to-mass ellipsoid in \textit{general}, for any infinite distance limit of any Calabi-Yau.

We have also analyzed the extremality bound for black holes in this context and shown it to match with the BPS bound, as expected from the $\cN=2$ attractor mechanism. Hence, our recipe to compute the charge-to-mass ratios can also be used to determine the extremality bound in general for the asymptotic regimes in field space. This connection implies that the Weak Gravity Conjecture and the Repulsive Force Conjecture coincide in the asymptotic regimes which, in more physical terms, correspond to the weak gauge coupling limits. It is important to remark that not all directions in the charge lattice can support BPS states, which allows us to truncate the degenerate directions of the ellipsoid, consequently implying also an upper bound for the charge-to-mass ratios. To make this more precise, we have studied the extremality bound also along the non-BPS directions of two examples using the formalism of ``fake supersymmetry," obtaining that the shape of the extremality bound mimics the one along the BPS directions. In our examples, the convex hull of BPS states is enough to include the extremal region in any direction in the entire charge space, implying that no further states beyond the BPS ones are required to satisfy the convex hull WGC. It would be interesting to give a general proof and study the implications of this result for works like \cite{demirtas2019minimal}, as we will comment below when discussing future directions.

The second main result of the paper is to determine a lower bound for the exponential mass decay rate $\lambda$ of the Swampland Distance Conjecture (SDC) in terms of some scalar contribution to the extremality bound for black holes. Whenever there is a gauge coupling vanishing at infinite field distance, as occurs in $\cN=2$ theories, there will be a tower of WGC-satisfying states which becomes exponentially light at infinite distance, thus also satisfying the SDC. If the WGC and the RFC coincide, as occurs in the asymptotic limits, the scalar contribution to the extremality bound equals the scalar charge-to-mass ratio $|\nabla M|/M$. If there is only one scalar field, this can be directly identified with the SDC factor $\lambda$. However, for higher dimensional moduli limits, there is a certain ambiguity regarding the trajectory followed in field space. More concretely, one should identify the SDC factor with $\lambda = (\vec\nabla M/M) \vec u$ where $u$ is a unit vector pointing along the geodesic trajectory. We have studied these path dependence issues and concluded that it is still possible to give a lower bound for $\lambda$ given by
\beq
\lambda\geq \frac{\nabla_1 M}{M}=\frac{1}{\sqrt{2}} \frac{|\ell_1-3|}{\sqrt{d_1}},
\eeq
where $\ell_1$ and $d_1$ are integers associated to the type of infinite distance singularity that is being approached. The subindex $1$ indicates the direction of the fastest growing scalar in the growth sector \eqref{growth}. By using the classification of singular limits in Calabi-Yau threefolds \cite{Kerr2017,Grimm_2019}, we then determined that the minimum value of $\lambda$ that can arise in a Calabi-Yau threefold compactification is
\beq
\lambda\geq \frac{1}{\sqrt{6}}\ .
\eeq
It should be stressed that the techniques of asymptotic Hodge theory used to determine the principal radii of the charge-to-mass ratio ellipsoid as well as the SDC factor are very general and do not depend on the internal manifold being Calabi-Yau. Hence, we expect some of the results to be valid beyond $\cN=2$. The main requirement for the story to hold is to have a gauge field becoming weakly coupled at the infinite field distance limit. If that occurs, there will be an infinite tower of states satisfying both the WGC and the SDC with the above properties. Notice that even if this tower is not the leading one, it can be used to give a lower bound on $\lambda$ and, therefore, a precise upper bound on the allowed field range before the EFT breaks down. Whether there is always such a vanishing gauge coupling for some p-form gauge field is an interesting question to further investigate in the future.

This work leaves several interesting directions for future research. As already mentioned, the attractor mechanism and formalism of the ``fake superpotential" open up the possibility of using the mildest form of the Weak Gravity Conjecture to say something about the presence of additional non-BPS states in the theory (either branes wrapping non-holomorphic cycles or bound states). This program has been explored in \cite{demirtas2019minimal} in the context of the sublattice Weak Gravity Conjecture and Euclidean D3-branes wrapping divisors (see also \cite{Hebecker:2015zss,Heidenreich:2016jrl}). However, in light of the current work, we can be more precise, using the mild Convex Hull WGC with precise numerical factors. For example, it would be interesting to check whether the conclusion driven from our examples that the BPS dyonic states suffice to satisfy the Convex Hull WGC without requiring the presence of additional non-BPS states, can be generalized to other setups. Additionally, it would be interesting to understand the relation between the WGC and the RFC deeper in the bulk of moduli space \cite{HR,Ben1,Ben2,HL}, where the SDC no longer applies. Finally, we believe it might be possible to find always some p-form gauge field becoming weakly coupled at infinite distance, but it would be interesting to make this more precise and find an argument motivating this in general. Notice that when this occurs, the WGC, the RFC and the SDC unify and predict the same tower of states at the asymptotic limits of field space. It is then very tantalizing to speculate that they are all just different faces of the same underlying quantum gravity principle that becomes manifest in the asymptotic regions of moduli space where we have approximate global symmetries. 

It is also imperative to extend our work to setups with less supersymmetry in order to render them applicable to situations of phenomenological interest. For instance, generating a scalar potential for the moduli constrains the trajectories that can be followed in field space. Even in the absence of a moduli space, one might expect that the bottom of the potential and trajectories selected by having a large mass hierarchy with respect to the transverse directions must still obey the SDC. Hence, in the presence of a potential, the realization of the SDC is not only tied to the kinetic structure of the scalars but also to the type of scalar potentials that can arise from string theory. In other words, the SDC actually constrains the type of potentials that can be consistently UV completed in quantum gravity. Since the scalars are no longer massless, the connection between the SDC factor and the scalar contribution to the extremality bound on the WGC disappear, but this is replaced for a connection between the SDC and constraints on the slope of the potential. The flux-induced scalar potential obeys the same rules for the scalar dependence and asymptotic behaviour as the gauge kinetic function here, as studied in detail in \cite{grimm2019asymptotic}. Analogously, the asymptotic splitting of the charge lattice described here equally well applies to flux space. And the role of the exponential rate of the gauge kinetic function $\alpha$ is played now by $|\nabla V|/V$. In fact, the same exact formulae giving the value of $\alpha$ in terms of the discrete data associated to the singular limit in \eqref{alphagrowth} can be directly applied to the different flux terms of the scalar potential. This was used in \cite{grimm2019asymptotic} to show that indeed the de Sitter conjecture claiming that $|\nabla V|/V>c$ with $c\sim\mathcal{O}(1)$ holds at the infinite field distance limits of Calabi-Yau fourfold M-theory compactifications, where the order one factor depends on the discrete data associated to the singular limit. It is not surprising, then, that the slope of the potential is correlated with the SDC factor, as proposed in \cite{Ooguri:2018wrx} (see also \cite{Andriot:2020lea}). Even if apparent from the asymptotic geometry of the field space, it is actually very astonishing that the black hole extremality bound, the exponential mass decay rate of the SDC tower, and the slope of the potential once fluxes are turned on can all be determined by the same set of integers characterizing the infinite distance limit. All these relations between the Swampland Conjectures suggest that we are only seeing the tip of the iceberg and many surprises are yet to come.

\subsubsection*{Acknowledgments}

We would like to thank Mehmet Demirtas, Thomas Grimm, Ben Heidenreich, Manki Kim, Cody Long, Liam McAllister, Miguel Montero, Jakob Moritz, and Tom Rudelius for very useful discussions and correspondence.
IV is supported by the Simons Foundation Origins
of the Universe program, and NG is supported in part by NSF grant PHY-1719877.

\bibliographystyle{jhep}
\bibliography{WGCBPS}

\end{document}